\documentstyle[11pt]{article}
\normalbaselines

\newtheorem{lem}{Lemma}[section]
\newtheorem{prop}{Proposition}[section]

\begin{document}

\def\bea*{\begin{eqnarray*}}
\def\eea*{\end{eqnarray*}}
\def\ba{\begin{array}}
\def\ea{\end{array}}
\count1=1
\def\be{\ifnum \count1=0 $$ \else \begin{equation}\fi}
\def\ee{\ifnum\count1=0 $$ \else \end{equation}\fi}
\def\ele(#1){\ifnum\count1=0 \eqno({\bf #1}) $$ \else \label{#1}\end{equation}\fi}
\def\req(#1){\ifnum\count1=0 {\bf #1}\else \ref{#1}\fi}
\def\bea(#1){\ifnum \count1=0   $$ \begin{array}{#1}
\else \begin{equation} \begin{array}{#1} \fi}
\def\eea{\ifnum \count1=0 \end{array} $$
\else  \end{array}\end{equation}\fi}
\def\elea(#1){\ifnum \count1=0 \end{array}\label{#1}\eqno({\bf #1}) $$
\else\end{array}\label{#1}\end{equation}\fi}
\def\cit(#1){
\ifnum\count1=0 {\bf #1} \cite{#1} \else 
\cite{#1}\fi}
\def\bibit(#1){\ifnum\count1=0 \bibitem{#1} [#1    ] \else \bibitem{#1}\fi}
\def\ds{\displaystyle}
\def\hb{\hfill\break}
\def\comment#1{\hb {***** {\em #1} *****}\hb }

\newcommand{\TZ}{\hbox{\bf T}}
\newcommand{\MZ}{\hbox{\bf M}}
\newcommand{\ZZ}{\hbox{\bf Z}}
\newcommand{\NZ}{\hbox{\bf N}}
\newcommand{\RZ}{\hbox{\bf R}}
\newcommand{\CZ}{\,\hbox{\bf C}}
\newcommand{\PZ}{\hbox{\bf P}}
\newcommand{\QZ}{\hbox{\rm eight}}
\newcommand{\HZ}{\hbox{\bf H}}
\newcommand{\EZ}{\hbox{\bf E}}
\newcommand{\GZ}{\,\hbox{\bf G}}

\font\germ=eufm10
\def\goth#1{\hbox{\germ #1}}
\vbox{\vspace{38mm}}

\begin{center}
{\LARGE \bf Eigenvectors of an Arbitrary Onsager Sector in Superintegrable $\tau^{(2)}$-model and Chiral Potts Model } \\[10 mm] 
Shi-shyr Roan \\
{\it Institute of Mathematics \\
Academia Sinica \\  Taipei , Taiwan \\
(email: maroan@gate.sinica.edu.tw ) } \\[25mm]
\end{center}

\begin{abstract}
We study the eigenvector problem in homogeneous superintegrable $N$-state chiral Potts model (CPM) by the symmetry principal. Using duality symmetry and (spin-)inversion in CPM, together with Onsager-algebra symmetry and $sl_2$-loop-algebra symmetry  of the superintegrable $\tau^{(2)}$-model, we construct the complete $k'$-dependent CPM-eigenvectors in the local spin basis for an arbitrary Onsager sector. In this paper, we present the complete classification of quantum numbers of superintegrable $\tau^{(2)}$-model. Accordingly, there are four types of sectors. The relationships among Onsager sectors under duality and inversion, together with their Bethe roots and CPM-eigenvectors, are explicitly found. Using algebraic-Bethe-ansatz techniques and duality of CPM, we construct the Bethe states and the Fabricius-McCoy currents of the superintegrable $\tau^{(2)}$-model through its equivalent spin-$\frac{N-1}{2}$-XXZ chain. The $\tau^{(2)}$-eigenvectors in a sector are derived from the Bethe state and the $sl_2$-product structure determined by the Fabricius-McCoy current of the sector. From those $\tau^{(2)}$-eigenvectors, the $k'$-dependence of CPM state vectors in local-spin-basis form is obtained by the Onsager-algebra symmetry of the superintegrable chiral Potts quantum chain.

\end{abstract}
\par \vspace{5mm} \noindent
{\rm 2008 PACS}: 05.50.+q, 02.20.Uw, 64.60.De, 75.10Pq \par \noindent

\setcounter{section}{0}
\section{Introduction}
\setcounter{equation}{0}
The eigenvalue spectrum of the $N$-state chiral Potts model (CPM) was solved by the method of functional relations \cite{AMP, B90, B91,BBP, MR, R0805}, where by regarding CPM as a descendant of the six-vertex  $\tau^{(2)}$-model \cite{BazS},  the chiral Potts transfer matrix can be derived as  the $Q$-operator of $\tau^{(2)}$-matrix 
in the general framework of Baxter's  $TQ$-relation \cite{Bax}.
In the superintegrable case, the degeneracy of $\tau^{(2)}$-model occurs, and the $\tau^{(2)}$-eigenspace for an eigenvalue $\tau^{(2)}({\tt t})$ forms an Onsager sector with the dimension equal to the number of chiral-Potts-eigenvalues associated with $\tau^{(2)}({\tt t})$ in the functional-relation scheme \cite{AMP, B89, B93, B94, BBP, R0805}. Much progress has been made in CPM on the study of eigenvalues, which leads to successful calculations of many important physical quantities, such as the free energy and the order parameter of the theory \cite{B88, B05a, B05b, B09a, B09b, B10,IG}. However there remain some unsolved problems significant in CPM, like correlation functions, whose solutions demand a deep understanding of eigenvectors. 
The purpose of this paper aims to provide an explicit construction of complete eigenvectors in the quantum space $V_{r, Q}$ for the (homogeneous) superintegrable CPM of a finite size $L$ with the (skewed) boundary condition $r$ and $\ZZ_N$-charge $Q$. It is known that $V_{r, Q}$ is decomposed into
(Onsager) sectors ${\cal E}_{{\tt F}, P_a, P_b}$, labeled by quantum numbers $P_a, P_b$, and the Bethe polynomial ${\tt F} (= {\tt F}({\tt t}))$ whose roots satisfy  the Bethe equation of a superintegrable $\tau^{(2)}$-model (see (\req(Bethesup)) in the paper). 
The $\tau^{(2)}$-degeneracy multiplicity of ${\cal E}_{{\tt F}, P_a, P_b}$ is  $2^{m_E}$ (a power of $2$), and the eigenvectors of CPM, depending on a temperature-like parameter $k'$, form a basis of ${\cal E}_{{\tt F}, P_a, P_b}$:
\bea(lll)
{\cal E}_{{\tt F}, P_a, P_b}= \bigoplus_s \CZ ~ \vec{v}(s ; k') & {\rm with} ~ ~ s= (s_1, \ldots, s_{m_E}) ,  s_i = \pm (: = \pm 1), &  k' \in \RZ .
\elea(Ek')
In this work, we obtain an expression of the above CPM-eigenvectors $\vec{v}(s ; k')$ in the local spin basis. First, we present the complete constraints of quantum numbers in superintegrable CPM, by which we classify all Onsager sectors into four types: $I_\pm, i_\pm$. The duality symmetry of superintegrable CPM \cite{B89,R09} interchanges sectors of type $I_\pm$ and $i_\pm$ respectively, with a precise relationship between $k'$-vectors of one sector and $k'^{-1}$-vectors of its dual sector in (\req(Ek')) under the duality correspondence of quantum spaces (see (\req(wect)), (\req(vec*)) in the paper). In particular, the duality induces an one-to-one correspondence between the state vectors in (\req(Ek')) for a $I_\pm$-sector at $k'=\infty$  and its dual $i_\pm$-sector at $k'=0$, which will serve the role of basic $\tau^{(2)}$-eigenvectors in our approach of the CPM-eigenvector problem. The basic $\tau^{(2)}$-eigenvectors can be regarded as the state vectors at $\infty$- or $0$-temperature in an Onsager sector of type $I_\pm$ or $i_\pm$ respectively. Using the $sl_2$-product algebra inherited by basic $\tau^{(2)}$-eigenvectors, the  $sl_2$-loop-algebra structure of ${\cal E}_{{\tt F}, P_a, P_b}$ is defined through the modified $\tau^{(N)}$-eigenvalue, i.e. the evaluation polynomial (\req(Ptrt)) in  the paper. There is another reflective symmetry in the theory of superintegrable CPM. Indeed, by examining the relationship between roots of the $\tau^{(2)}$-Bethe equation, the $\pm$-sectors within the same $I$- or $i$-type have shown a inversion relation with conjugate total momentum, in which a canonical identification of the $k'$-vectors  of one sector and $(-k')$-vectors of another sector in (\req(Ek')) naturally appears (see (\req(E'Eb)) in the paper).
In the discussion of $k'$-state vectors in duality and inversion, the exact correspondences are both dictated by the Onsager-algebra symmetry induced from the superintegrable chiral Potts quantum chain \cite{GR}. It is known that  there are two kinds of symmetry to describe  the $\tau^{(2)}$-degeneracy in superintegrable CPM: the Onsager-algebra symmetry \cite{R05o} and the $sl_2$-loop-algebra symmetry \cite{NiD, ND08, R06F,R09}, which play {\it different} roles in the derivation of CPM-eigenvectors.  In fact, the Onsager-algebra generators in the chiral Potts chain give rise to an irreducible Onsager-algebra representation on ${\cal E}_{{\tt F}, P_a, P_b}$, which enables us to express the state vector $\vec{v}(s ; k')$ in (\req(Ek')) using the basic $\tau^{(2)}$-eigenvectors with the $k'$-dependent coefficients determined by $({\tt F}, P_a, P_b)$ (see (\req(ci)), (\req(angle)), (\req(k'0if)) in the paper).  On the other hand, the loop-algebra symmetry is about the $sl_2$-loop-algebra structure of  basic $\tau^{(2)}$-eigenvectors in ${\cal E}_{{\tt F}, P_a, P_b}$. This $sl_2$-loop-algebra representation of ${\cal E}_{{\tt F}, P_a, P_b}$ shares the same evaluation polynomial (or Drinfeld polynomial) as the Onsager-algebra symmetry, but in essence arises from the theory of spin-$\frac{N-1}{2}$ XXZ-chains, which are in fact equivalent to  superintegrable $\tau^{(2)}$-models \cite{NiD, ND08, R06F, Tar}. By the algebraic-Bethe-ansatz of XXZ-chains \cite{Fad, KBI, KS} , we obtain the Bethe state represented by the local spin basis as the basic $\tau^{(2)}$-Bethe state of the sector.  With the help of the duality relation, the basic $\tau^{(2)}$-Bethe state is revealed as the vector with highest weight in a plus($+$)-sector, and the lowest weight in a minus($-$)-sector among basic $\tau^{(2)}$-eigenvectors, (see (\req(Betv)), (\req(BetvF)) in the paper). Furthermore, we are able to identify the explicit operator of quantum spaces which gives rise to the inversion relation in CPM through the XXZ chain equivalent to the superintegrable $\tau^{(2)}$-model. For $I_\pm$-sectors, the inversion operator is given by the reversion of all spin and site-orientation of the local spin basis, whose conjugation by duality correspondence serves the inversion operator of $i_\pm$-sectors, (see Proposition \ref{prop:inversion} in the paper).  By a similar argument in \cite{FM01, R06F}, the Fabricius-McCoy current can be successfully constructed in an arbitrary sector of the superintegrable $\tau^{(2)}$-model as a series with local-operator coefficients in the loop-algebra representation of ${\cal E}_{{\tt F}, P_a, P_b}$. When applying to the basic $\tau^{(2)}$-Bethe state, the Fabricius-McCoy current produces a local-vector form of the basic $\tau^{(2)}$-eigenvectors in ${\cal E}_{{\tt F}, P_a, P_b}$, hence follows the $k'$-dependent local-vector form of CPM-eigenvectors in (\req(Ek')) by using the Onsager-algebra symmetry.

This paper is organized as follows. Section \ref{sec:Dual} mainly reviews facts on the duality relation and quantum numbers of superintegrable CPM in \cite{R09}  which are relevant to the discussion of this work. We first in Subsection \ref{ssec.tCP} briefly survey some basic facts about the duality of homogeneous CPM and $\tau^{(2)}$-model. The detailed structures especially held for the superintegrable case are described in Subsection \ref{ssec.Supt2}. In Subsection \ref{ssec.QN}, we provide the complete constraints about quantum numbers in the superintegrable CPM, consisting of four types of Onsager sectors, $I_\pm, i_\pm$. The correspondence between $I$- and $i$-sectors under the duality relation is given here. Furthermore, by comparing solutions of the Bethe equation, we find the inversion symmetry among $\pm$-sectors of superintegrable $\tau^{(2)}$-model with the conjugate total momentum. 
In Section \ref{sec.OAlp}, we examine the degeneracy symmetries of a superintegrable $\tau^{(2)}$-sector, and show that these structures are compatible with the duality and inversion of CPM. A procedure of constructing the superintegrable CPM-eigenvectors is presented here by using  the degeneracy symmetries of CPM. 
We first in Subsection \ref{ssec.OA} re-examine in details the well-known Onsager-algebra symmetry of a superintegrable $\tau^{(2)}$-model \cite{B89, DR, Dav, GR, R09}, and show how the CPM $k'$-eigenvectors in (\req(Ek')) can be constructed from basic $\tau^{(2)}$-eigenvectors in a sector through the Onsager-algebra structure. 
The exact correspondence of CPM-eigenvectors between sectors under the duality and inversion is also established by the identification of Onsager-algebra representations.
Using the basic $\tau^{(2)}$-eigenvectors in (\req(Ek')) at $k'=\infty$ or $0$, we define $sl_2$-product-algebra and the $sl_2$-loop-algebra structures of a $\tau^{(2)}$-eigenspace in Subsection \ref{ssec:sl2t2}. This loop-algebra structure will incorporate the loop-algebra symmetry  induced from the XXZ-chain equivalent to the superintegrable $\tau^{(2)}$-model discussed in the next two sections. 
In Section \ref{sec.Degt2}, we employ the algebraic-Bethe-ansatz method as in \cite{NiD, ND08, R06F} to
investigate the basic  $\tau^{(2)}$-eigenvectors of a superintegrable $\tau^{(2)}$-model through its equivalent spin-$\frac{N-1}{2}$ XXZ chain \cite{R075}. This equivalent relation is a special case among the general equivalence between XXZ-chains with $U_q(sl_2)$-cyclic representation and arbitrary $\tau^{(2)}$-models, a result in \cite{R075, R0710}, now briefly reviewed in Subsection \ref{ssec.XXZ}. Here we assume $N (=2M+1)$ odd as in \cite{R09} for the convenience of simple notions when making the identification of local operators between XXZ chains and $\tau^{(2)}$-models. 
In Subsection \ref{ssec.XXZhw}, by the standard algebraic-Bethe-ansatz argument \cite{Fad, KBI, KS}, we obtain the Bethe state in the local spin basis and a set of operators expressed by monodromy-entries, commuting with the $\tau^{(2)}$-matrix. The Bethe state can be realized as the basic $\tau^{(2)}$-eigenvector with the highest or lowest weight. In the special case $m=r=0, L \equiv 0$,  some of these operators are corresponding to the operators of the ground-state sector in \cite{AuP7, AuP9}.
For sectors in $I_+\cap I_- (=i_+ \cap i_-)$, those operators also provide the local-operator form of algebra generators for the loop-algebra symmetry of an Onsager sector previously discussed in Subsection \ref{ssec:sl2t2}, a result generalizing those in \cite{NiD, ND08} or \cite{R06F} for the case $r=m=0$ or $r=0$, $m=M$ respectively. Furthermore, through the local operators of XXZ chains, we are able to identify the correspondence of quantum spaces which produces the inversion symmetry in superintegrable CPM. 
Section \ref{sec.loopt2} is devoted to the Fabricius-McCoy current, which is a series in the $sl_2$-loop algebra expressed by monodromy entries of the XXZ chain.  The Fabricius-McCoy current plays a crucial role in our study of CPM-eigenvector problem as an ingredient to construct the basic $\tau^{(2)}$-eigenvectors. In Subsection \ref{ssec.Ipm}, by methods in \cite{FM01, R06F}, we construct the Fabricius-McCoy currents in  sectors $I_\pm$ consistent with the inversion symmetry. In Subsection \ref{ssec.ipm}, the Fabricius-McCoy currents of $i_\pm$-sectors  are obtained by Fabricius-McCoy currents of $I_\pm$-sectors in the dual $\tau^{(2)}$-model through the duality correspondence.

Notation:  In this paper, we use the following standard notations. For a positive integer $N$ greater than one, 
$\CZ^N$ denotes the vector space of $N$-cyclic vectors with the canonical base 
$|\sigma \rangle, \sigma  \in \ZZ_N ~ (:= \ZZ/N\ZZ)$, and Weyl operators $X, Z$ satisfying  $X^N=Z^N=1, XZ= \omega^{-1}ZX$:
$$
 X |\sigma  \rangle = | \sigma  +1 \rangle , ~ \ ~ Z |\sigma  \rangle = \omega^\sigma  |\sigma  \rangle ~ ~ \ ~ ~ (\sigma  \in \ZZ_N ~ ~ \omega := {\rm e}^{\frac{2 \pi {\rm i}}{N}}).
$$
The Fourier basis $\{ \widehat{|k } \rangle \}$ of $\CZ^N$ is defined by 
\be
\widehat{|k } \rangle  = \frac{1}{\sqrt{N}} \sum_{\sigma =0}^{N-1} \omega^{-k \sigma} |\sigma \rangle , ~ ~ ~ | \sigma  \rangle = \frac{1}{\sqrt{N}} \sum_{k =0}^{N-1} \omega^{\sigma k} \widehat{|k} \rangle , ~ \sigma  \in \ZZ_N,
\ele(Fb) 
with Weyl operators, $
\widehat{X} \widehat{|k} \rangle = \widehat{|k+1} \rangle$, $\widehat{Z} \widehat{|k} \rangle = \omega^k \widehat{|k} \rangle$, satisfying the relation, 
\be
(X, Z) = (\widehat{Z}, \widehat{X}^{-1}). 
\ele(XZF)

\section{Duality and Quantum Numbers in Superintegrable Chiral Potts Model \label{sec:Dual} }
\subsection{Duality in chiral Potts model  and $\tau^{(2)}$-model \label{ssec.tCP}}
\setcounter{equation}{0}
We start with some basic notions in a homogeneous $\tau^{(2)}$-model and CPM, then state some facts in \cite{R09} about the duality of CPM. Since the duality in \cite{R09} was formulated in the form of a general inhomogeneous CPM, for easy reference in this paper, we shall summary the results in the homogeneous case, and make the modification of some conventions used in \cite{R09} for the purpose of elucidation. The correspondence between quantum spaces in duality will be essential for the later discussion of CPM-eigenvector problem. The summary will be sketchy, but also serve to establish notations.

The $L$-operator of $\tau^{(2)}$-model \cite{BBP, BazS} (see also in  \cite{R0710, R0805})  is the two-by-two matrix expressed by Weyl operators $X, Z$ or $\widehat{X}, \widehat{Z}$ in (\req(XZF)):
\bea(lll)
L ( t ) &= \left( \begin{array}{cc}
        1  -  t \frac{{\sf c}  }{\sf b' b} X ,  & (\frac{1}{\sf b }  -\omega   \frac{\sf a c }{\sf b' b} X) Z \\
       - t ( \frac{1}{\sf b'}  -  \frac{\sf a' c}{\sf b' b} X )Z^{-1}, & - t \frac{1}{\sf b' b} + \omega   \frac{\sf a' a c }{\sf b' b} X
\end{array} \right) , ~ ~ (X, Z) \leftrightarrow (\widehat{Z}, \widehat{X}^{-1}) ,
\elea(L)
with non-zero complex parameters  ${\sf a, b, a', b', c}$. It is known that the above $L$-operator satisfies the 
Yang-Baxter (YB) equation 
$$
R(t/t') (L (t) \bigotimes_{aux}1) ( 1
\bigotimes_{aux} L (t')) = (1
\bigotimes_{aux} L (t'))(L (t)
\bigotimes_{aux} 1) R(t/t')
$$
for the asymmetry $R$-matrix
$$
R(t) = \left( \begin{array}{cccc}
        t \omega - 1  & 0 & 0 & 0 \\
        0 &t-1 & \omega  - 1 &  0 \\ 
        0 & t(\omega  - 1) &( t-1)\omega & 0 \\
     0 & 0 &0 & t \omega - 1    
\end{array} \right).
$$ 
Over a chain of size $L$, we have the monodromy matrix, 
\be
L_1 (t) L_2 (t) \cdots L_L (t) = \left( \begin{array}{cc}
        A(t)  & B(t) \\
        C(t) &  D(t)
\end{array} \right)  , ~ L_\ell (t) = L (t).
\ele(Mont2) 
The $\tau^{(2)}$-model with the boundary condition 
\be
\sigma_{L+1} \equiv \sigma_1 - r \pmod{N} ,  ~ ~ ~ ( r \in \ZZ_N ),
\ele(sBy)
is the commuting family of $\tau^{(2)}$-operators defined by 
\be
\tau^{(2)}(t) = A(\omega t) + \omega^r D(\omega  t).  
\ele(tau2) 
The spin-shift operator $X (:= \prod_{\ell} X_\ell)$ commutes with (\req(tau2)), with the eigenvalue $\omega^Q$ for $Q \in \ZZ_N$. Hence $\tau^{(2)}(t)$ preserves the charge-$Q$ subspace, denoted by
\be
V_{r, Q} = \{ v=  \sum_{\sigma_1, \ldots \sigma_L} v_{\sigma_1, \ldots \sigma_L} |\sigma_1, \ldots \sigma_L \rangle \in \bigotimes^L \CZ^N| ~ \sigma_{L+1} \equiv \sigma_1 - r , ~ ~ X (v) = \omega^Q v \} ,
\ele(VrQ)
with the basis: 
\bea(lll)
V_{r, Q} & = \bigoplus_{n_\ell} \CZ |Q; n_1, \ldots n_L \rangle & (\sum_{\ell=1}^L n_\ell \equiv r, ~ ~ |n_{L+1} \rangle \rangle = \omega^{-Q n_1} | n_1 \rangle \rangle ), \\
&  = \bigoplus_{n'_\ell } \CZ |\widehat{n'_1}, \ldots \widehat{n'_L} \rangle & (\sum_{\ell=1}^L n'_\ell \equiv Q , ~ ~ |\widehat{n'_{L+1}}\rangle = \omega^{-rn'_1} |\widehat{n'_1}),
\elea(Vbasis)
where $| Q; n_1, \ldots n_L \rangle :=  N^{-1/2} \sum_{\sigma_1=0}^{N-1} \omega^{-Q \sigma_1} |\sigma_1, \ldots \sigma_L \rangle$ with $\sigma_\ell - \sigma_{\ell +1} = n_\ell$ are the basis introduced in \cite{AuP, AuP7} for cyclic boundary condition, i.e. $r=0$, case. Note that $V_{r, Q}$ is subspace of the quantum space $\bigotimes^L \CZ^N$ with a Hermitian  inner-product induced from the local spin orthonormal basis (\req(Fb)). The basis in (\req(Vbasis)) are indeed (Hermitian) orthonormal basis of $V_{r, Q}$.

The parameters of $\tau^{(2)}$-model (\req(L)) in CPM over a square lattice (rotated by $45^\circ$-degree) $\Gamma$  are 
\bea(ll)
({\sf a}',  {\sf b}', {\sf a}, {\sf b}, {\sf c})= ( x_p, 
y_p , x_p , y_p , \mu_p^2 ), & p \in {\goth W}_{k'}
\elea(pp)
where ${\goth W}_{k'}$ is the (CPM) rapidity $k'$-curve defined by 
\bea(ll)
{\goth W}_{k'}&: k x^N  = 1 -  k'\mu^{-N},   k  y^N  = 1 -  k'\mu^N, \ ( k'^2 \neq 0, 1,   k^2 + k'^2 = 1 ) 
\elea(cpmC)
(see, e.g. \cite{AMP, BBP}). The $\tau^{(2)}$-model (\req(tau2)) with parameters (\req(pp)) will also be denoted by 
\be
\tau^{(2)}(t) = \tau^{(2)}(t ; p).
\ele(t2pp')
Here the spectral parameter $t$ is identified with $x_q y_q$ for a generic rapidity $q$ in (\req(cpmC)):
\bea(ll)
t ~ ( = t_q) = x_q y_q , & q \in {\goth W}_{k'}.
\elea(tdef)
Then  $t^N$ is related to $ \mu^N$ by the equation of a genus-$(N-1)$ hyperelliptic curve $W_{k'}$:
\bea(l)
W_{k'} :  t^N = \frac{(1- k' \lambda  )( 1 - k' \lambda^{-1}) }{k^2 } ,  ~ \lambda = \mu^N .
\elea(hW)
As in \cite{B93} (3.11)-(3.13),  \cite{B90} (10), \cite{BBP} (4.27a) (4.27b), \cite{R06F} Proposition 2.1, (2.31) and \cite{R0710} (2.25), one can construct $\tau^{(j)}$-matrices from (\req(L)), with $\tau^{(0)}=0, \tau^{(1)}= I$ and $\tau^{(2)}$ in (\req(t2pp')), so that the fusion relation holds:
\bea(l)
\tau^{(2)}(\omega^{j-1} t) \tau^{(j)}(t) =  \omega^r X z( \omega^{j-1} t) \tau^{(j-1)}(t)  + \tau^{(j+1)}(t) , \ \ j \geq 1 ; \\
\tau^{(N+1)}(t) = \omega^r X z(t) \tau^{(N-1)}(\omega t) + u(t) I ,
\elea(fus)
where $z(t)=(\frac{\omega \mu_p^2 (t_p-t)^2}{y_p^4})^L $, $u(t)= \alpha_q  + \overline{\alpha}_q$ with $
\alpha_q = (\frac{\mu^N (y_p^N-x^N) (y_p^N-x^N)}{k' y_p^N y_p^N})^L$,$\overline{\alpha}_q   = ( \frac{\mu^{-N} (y_p^N-y^N) (y_p^N-y^N)}{k' y_p^N y_p^N})^L$. 

The $Q$-operator of $\tau^{(2)}$-matrix (\req(t2pp')) is the chiral Potts transfer matrices (of size $L$ and boundary condition $r$) \cite{B93, BBP,  R0710}, which are $\stackrel{L}{\otimes} \CZ^N$-operators defined by 
\bea(lll)
T (q)_{\{\sigma \}, \{\sigma'\}} &(= T (q ; p)_{\{\sigma \}, \{\sigma'\}}) = \prod_{\ell =1}^L W_{p q}(\sigma_\ell - \sigma'_\ell ) \overline{W}_{p q}(\sigma_{\ell+1} - \sigma'_\ell),& \\
\widehat{T} (q)_{\{\sigma' \}, \{\sigma'' \}}&(=\widehat{T} (q ; p )_{\{\sigma' \}, \{\sigma''\}}) = \prod_{\ell =1}^L \overline{W}_{p q}(\sigma'_\ell - \sigma''_\ell) W_{p q}(\sigma'_\ell - \sigma''_{\ell+1}),& (p, q \in {\goth W}_{k'}).
\elea(ThatT)
Here $W_{p q}, \overline{W}_{p q}$ are the Boltzmann weights in CPM \cite{BPA}:
\be
\frac{W_{p q}(\sigma)}{W_{p q}(0)}  = (\frac{\mu_p}{\mu_q})^\sigma \prod_{j=1}^\sigma
\frac{y_q-\omega^j x_p}{y_p- \omega^j x_q }  , \ ~ \
\frac{\overline{W}_{p q}(\sigma)}{\overline{W}_{p q}(0)}  = ( \mu_p\mu_q)^\sigma \prod_{j=1}^\sigma \frac{\omega x_p - \omega^j x_q }{ y_q- \omega^j y_p } 
\ele(WW)
with $\sigma \in \ZZ_N$ and $W_{p,q}(0)= \overline{W}_{p,q}(0)=1$, and they are uniquely determined by their Fourier transform: $
\overline{W}^{(f)}_{pq}(k)= \frac{1}{\sqrt{N}} \sum_{\sigma = 0}^{N-1} \omega^{k \sigma }\overline{W}_{pq}(\sigma) $, $
W^{(f)}_{pq}(k) = \frac{1}{\sqrt{N}} \sum_{\sigma = 0}^{N-1} \omega^{k \sigma }W_{pq}(\sigma)$. 
Then $T, \widehat{T}$ in (\req(ThatT)) commute with $X$, and $\widehat{T} (q)  = T (q) S_R  = S_R T (q)$
where $S_R$ is the spatial translation operator $S_R | j_1, \ldots, j_L \rangle = | j_2,  \ldots, j_{L+1} \rangle$.
By the star-triangle relation of Boltzmann weights \cite{AMPT, AuP, BPA,  MaS, MPTS}, 
$$
~ [T (q), T (q')]= [\widehat{T} (q), \widehat{T} (q')] = 0 ~ ~ ~ (q, q' \in {\goth W}_{k'}),
$$
hence  $T, \widehat{T}$ in (\req(ThatT) preserve the charge-$Q$ subspace, and are decomposed into automorphisms of  $V_{r, Q}$ in (\req(VrQ)).

Consider the $\tau^{(2)}$-model and CPM over the dual lattice $\Gamma^*$, but replacing $k'$-rapidities in (\req(pp)), (\req(tdef)) and (\req(ThatT)) by $k'^{-1}$-rapidities, denoted by $p^*, t^*, q^*$ for $p^*, q^* \in {\goth W}_{1/k'}$. 
The dual 
$\tau^{(2) *}(t^*) (= \tau^{(2) *}(t^* ; p^*))$, and CPM matrices,  $T^* (q^*)(= T^* (q^* ; p^*)), 
\widehat{T}^* (q^*)(= \widehat{T}^* (q^* ; p^*))$, are of size $L$ with the boundary condition $r^*$, defined in terms of Weyl-operators $X^*, Z^*$ of the local (face-)quantum space ${\tt C}^N = \sum_{\sigma^* \in \ZZ_N} \CZ |\sigma^* \rangle^* $ over $\Gamma^*$. The Fourier basis of $|\sigma^* \rangle^*$'s in (\req(Fb)) and (\req(XZF)) are denoted by $|n \rangle \rangle$'s with the Weyl-operators ${\tt X}, {\tt Z}$ satisfying the relation $(X^*, Z^*) = ({\tt Z}, {\tt X}^{-1})$. The matrices, $\tau^{(2) *}(t^*), T^* (q^*), \widehat{T}^* (q^*)$, all commute with  the spin-shift operator $X^* (= \prod_{\ell} X^*_\ell = \prod_{\ell} {\tt Z}_\ell = {\tt Z})$ with the dual-charge $Q^*$, hence preserve the charge-$Q^*$ subspace\footnote{The $X^*, Z^*, r^*, Q^*, V^*_{r^*, Q^*}, T^* (q^*; p^*), \widehat{T}^* (q^*; p^*) $ here are denoted by ${\tt X}^*, {\tt Z}^*, {\tt r}, {\tt Q}, W_{{\tt r}, {\tt Q}}, {\tt T}^* (q^*; p^*), \widehat{\tt T}^* (q^*; p^*)$ respectively in \cite{R09}.}: 
\be
V^*_{r^*, Q^*} = \bigoplus_{\sum_{\ell} n_\ell \equiv Q^*,  } \CZ |n_1, \ldots n_L \rangle \rangle ~ ~ ~ 
~ ( |n_{L+1} \rangle \rangle = \omega^{-{r^*}n_1} | n_1 \rangle \rangle).   
\ele(VrQ*)
Under the lattice-identification, $\Gamma^* \leftrightarrow \Gamma$, so that local quantum vector spaces at the $\ell$th position of $\Gamma^*$ and $\Gamma$ are identified via the isomorphism,
\be
\Phi : {\tt C}^N \rightarrow \CZ^N, ~ ~ |n \rangle \rangle \mapsto \widehat{|n} \rangle ~ (\Longleftrightarrow ~ |\sigma \rangle^* \mapsto |\sigma \rangle ),
\ele(Phi) 
$\tau^{(2) *}(t^*)$ and $T^* (q^*), \widehat{T}^* (q^*)$ in  the (dual-charge) $Q^*$-sector can be canonically identified with the $\tau^{(2)}$-model and CPM over $\Gamma$ using the vertical rapidity $p^* \in {\goth W}_{1/k'}$, the boundary condition $r^*$ and charge $Q^*$, denoted by $\tau^{(2) \dagger}(t^*) (= \tau^{(2) \dagger}(t^*; p^*))$ and $T^\dagger (q^*) (= T^\dagger (q^*; p^*))$, $\widehat{T}^\dagger (q^*) (= \widehat{T}^\dagger (q^*; p^*))$. On the other hand, when $(r, Q)= (Q^*, r^*)$, the quantum space $V_{r, Q}$ in (\req(VrQ)) and $V^*_{r^*, Q^*}$ in (\req(VrQ*)) are isomorphic under isometric linear transformation:
\be
\Theta : V_{r, Q} \cong V^*_{r^*, Q^*} , ~ ~ |Q; n_1, \ldots n_L \rangle \mapsto |n_1, \ldots n_L \rangle \rangle, ~ ~ (r, Q)= (Q^*, r^*).
\ele(Theta)
Consider the duality of rapidity curves between ${\goth W}_{k'}$ and ${\goth W}_{1/k'}$ (\cite{R09} (3.9)), 
\be
{\goth W}_{k'} \stackrel{\sim}{\longrightarrow} {\goth W}_{1/k'} , ~ p= (x_p, y_p, \mu_p)  \longrightarrow  p^* = (x_{p*}, y_{p*}, \mu_{p*}):= ( {\rm i}^\frac{1}{N} x_p \mu_p, {\rm i}^\frac{1}{N} y_p \mu_p^{-1}, \mu_p^{-1}), 
\ele(pp*)
so that the spectral parameter $t$ is sent to $t^* = (-1)^\frac{1}{N} t$. Then the Boltzmann weights (\req(WW)) with rapidities in  ${\goth W}_{k'}$ and ${\goth W}_{1/k'}$ are related by $
\frac{\overline{W}^{(f)}_{p q }(k)}{\overline{W}^{(f)}_{pq}(0)} = W_{p^*q^*}(k)$ , $\frac{W^{(f)}_{p q}(k)}{W^{(f)}_{p q}(0)} =  \overline{W}_{p^*q^*}(N-k)$.  (\cite{R09} (3.17)).
Under the duality identification of quantum spaces (\req(Theta)) and rapidities (\req(pp*)),
the $\tau^{(2)}$-model $\tau^{(2)}(t)$ and $T (q), \widehat{T} (q)$ of CPM  are equivalent to the dual models, $\tau^{(2) *}(t^*)$ and $T^* (q^*), \widehat{T}^* (q^*)$. Indeed, the following similar relations hold between $\tau^{(2)}$-models and CPMs when $(r, Q)= (Q^*, r^*)$, under the identification (\req(pp*)) for rapidities  and (\req(Phi)) (\req(Theta)) for quantum spaces:
$$
\begin{array}{cccc}
{\rm Quantum ~ space}:&V_{r, Q} & V^*_{r^*, Q^*}  & V_{r^*, Q^*} ; \\
\tau^{(2)}-{\rm model}:&\tau^{(2)}(t), & \tau^{(2) *}(t^*) = \Theta \tau^{(2)}(t) \Theta^{-1} & \tau^{(2) \dagger}(t^*)= \ \Phi  \tau^{(2) *}(t^*) \Phi^{-1};  \\
{\rm CPM}: &{}^{T (q)}_{\widehat{T} (q)}, &  {}^{T^* (q^*)}_{\widehat{T}^* (q^*)}  =  \bigg( \frac{W_{p^* q^*}^{(f)}(0) }{W_{p q}^{(f)}(0)} \bigg)^L \Theta {}^{T (q)}_{\widehat{T} (q)} \Theta^{-1},  & {}^{T^\dagger (q^*)}_{\widehat{T}^\dagger (q^*)} =  \Phi   {}^{T^* (q^*)}_{\widehat{T}^* (q^*)} \Phi^{-1}.
\end{array}
$$
Hence one obtains the duality between $\tau^{(2)}$-models and CPMs with $k'$ and $k'^{-1}$-rapidities: 
\bea(ll)
\tau^{(2) \dagger }(t^*) = \Psi  \tau^{(2)}(t) \Psi^{-1} , & 
{}^{T^\dagger (q^*)}_{\widehat{T}^\dagger (q^*)}  =  \bigg( \frac{W_{p^* q^*}^{(f)}(0)}{W_{p q}^{(f)}(0)}\bigg)^L  {}^{\Psi T (q) \Psi^{-1}}_{\Psi \widehat{T} (q) \Psi^{-1}} , 
\elea(Dual)
where $\Psi (:= \Phi \Theta)$ is the duality correspondence which preserves the Hermitian metric of quantum spaces: 
\be
\Psi : V_{r, Q} \longrightarrow V_{r^*, Q^*}, ~ ~ |Q; n_1, \ldots n_L \rangle \mapsto |\widehat{n_1}, \ldots \widehat{n_L} \rangle, ~ ~ (\sum_{\ell=1}^L n_\ell \equiv r) ~ ~ (r, Q)= (Q^*, r^*),
\ele(Psi)
(\cite{R09} (3.12) (3.15)\footnote{The $\tau^{(2)}_F(t; p, p')$ in \cite{R09} is equal to $\tau^{(2) *}(t^*; p'^*, p^*)$ here.} (3.16) (3.19) (3.22)).

\subsection{Duality in superintegrable chiral Potts model  and $\tau^{(2)}$-model \label{ssec.Supt2}}
For the rest of this paper, we consider the superintegrable homogeneous $\tau^{(2)}$-model and CPM with $p$ in (\req(pp)) defined by\footnote{Here we consider only the case $n_0=0$ for the homogeneous superintegrable CPM in \cite{R09} where the vertical rapidity is defined by $(x_p, y_p, \mu_p)= (\eta^{\frac{1}{2}}\omega^m,  \eta^{\frac{1}{2}}, \omega^{n_0})$. By applying the similar transformation $Z^{2n_0}$ on the local quantum space $\CZ^N$ which changes $X$ to $\omega^{2n_0}X$, the discussion for the case of an arbitrary $n_0$ can be reduced to  that of $n_0=0$ by the similarity relation of chiral Potts transfer matrices at $p=(x_p,y_p, \mu_p)$ and $p(i):= (x_p,y_p, \omega^i \mu_p)$: $T(q(i); p(i)) =Z^{2i} T(q; p) Z^{-2i} \omega^{-2ri}$, $\widehat{T}(q(i); p(i)) = Z^{2i} \widehat{T}(q; p) Z^{-2i}$ for $i=n_0$. }
\bea(ll)
p ~ (= p (k')): (x_p, y_p, \mu_p)= (\eta^{\frac{1}{2}}\omega^m,  \eta^{\frac{1}{2}}, 1) \in {\goth W}_{k'}, & \eta (=\eta (k')) := (\frac{1-k'}{1+k'})^{\frac{1}{N}}, 
\elea(pcood)
for $0\leq m \leq N-1$, where $k' \neq 0 , \pm 1$ (\cite{AMP, AuP7, AuP9, R05o} for the case $m=r=0$, \cite{B89, B93, B94} for $m=0, r \in \ZZ_N$, \cite{R075} for $r=0, m= \frac{N-1}{2}$, and \cite{R09} for $r, m \in \ZZ_N$.) The duality of superintegrable CPM has been discussed years ago in \cite{B89} for $m=0$ case. In this subsection, we summarize the results in \cite{R075} Section 3.5 about the superintegrable CPM for arbitrary $r$ and $m$, but stated only in the homogeneous case.

The $L$-operators $L_\ell (t)$ of (\req(Mont2)) for $p$ in (\req(pcood)) are gauge-equivalent to 
\be
{\tt L}_\ell ( {\tt t} ) = \left( \begin{array}{cc}
        1  -    {\tt t} X   &  (1  -\omega^{1+m}  X) Z \\
       - {\tt t} ( 1  -  \omega^{m}  X )Z^{-1} & - {\tt t} +  \omega^{1+2m}  X
\end{array} \right) ~ ~ ~ ( {\tt t} := \omega^m t_p^{-1} t ) , 
\ele(hsupL)
by the diagonal matrix ${\rm dia}[\eta^\frac{1}{2}, 1]$, i.e. ${\rm dia}[\eta^\frac{1}{2}, 1]\cdot L_\ell (t)\cdot {\rm dia}[\eta^\frac{-1}{2}, 1] = {\tt L}_\ell ( {\tt t} )$ for all $\ell$. Hence the $\tau^{(2)}$-models for all $k'$ are the same when using the normalized spectral parameter ${\tt t}$, a property which enables us to discuss the $\tau^{(2)}(t, p(k'))$ simultaneous for all $k'$. It is well-known that the $\tau^{(2)}$-eigenvalues can be solved by the Bethe ansatz equation \cite{AMP, B93, B94, R0805, R09}. Express the $\tau^{(2)}$-eigenvalue (\req(t2pp')) in the form
\be
\tau^{(2)}(t; p) = \tau^{(2)}({\tt t}) = \omega^{-P_a} (1- \omega^{-m} {\tt t})^L \frac{ {\tt F} ({\tt t})}{{\tt F}( \omega {\tt t})} + \omega^{P_b}(1- \omega^{1-m} {\tt t})^L \frac{{\tt F} (\omega^2 {\tt t})}{{\tt F} ( \omega {\tt t})}  
\ele(stauev)
where $P_a, P_b$ are integers with $0 \leq P_a + P_b \leq N-1$, and the roots of polynomial  ${\tt F}( {\tt t} ) = \prod_{j=1}^J (1+ \omega {\tt v}_j {\tt t})$ satisfy the Bethe equation (\cite{AMP} (4.11), \cite{B93} (6.22), \cite{B94} (16), \cite{R0805} (4.31) (4.32)):
\be
 ( \frac{{\tt v}_i  + \omega^{-1-m}  }{{\tt v}_i  + \omega^{-2-m} })^L   = - \omega^{-P_a- P_b} \prod_{j=1}^J \frac{{\tt v}_i -  \omega^{-1}   {\tt v}_j }{ {\tt v}_i -\omega  {\tt v}_j } , \ \ i= 1, \ldots, J.
\ele(Bethesup)
Note that a $\tau^{(2)}$-eigenvalue is uniquely determined by a triple $({\tt F}, P_a, P_b)$ satisfying (\req(Bethesup)), and the corresponding $\tau^{(2)}$-eigenspace will be denoted by ${\cal E}_{{\tt F}, P_a, P_b}$.

Using (\req(fus)), one finds the eigenvalue-expression of $\tau^{(j)}$-model, $\tau^{(j)}(t; p) = \tau^{(j)}({\tt t})$, in terms of  $P_a, P_b$ and the Bethe polynomial ${\tt F}({\tt t})$. In particular, one obtains the polynomial ${\tt P}({\tt t})$ from $\tau^{(N)}({\tt t})$:
\bea(ll)
{\tt P}( {\tt t} )= {\tt t}^{-P_a-P_b} \frac{\tau^{(N)}({\tt t})}{{\tt F}({\tt t})^2} , &{\tt P}({\tt t}) = \omega^{-P_b}  \sum_{k=0}^{N-1} \frac{(1-  {\tt t}^N )^L (\omega^{k} {\tt t})^{-(P_a+P_b)}}{(1- \omega^{-m+k} {\tt t})^L {\tt F} (\omega^k {\tt t}) {\tt F}( \omega^{k+1} {\tt t}) } 
\elea(Pt)
(\cite{AMP} (2.37) for the case $N=3, {\tt F}=1$, \cite{B93} (6.25), \cite{B94} (17) and \cite{R09} (2.23) (2.26) (2.33)). The Bethe relation (\req(Bethesup)) is indeed the polynomial criterion of the function ${\tt P}( {\tt t} )$ in (\req(Pt)), which can be regarded as a ${\tt t}^N$-polynomial with ${\tt t}^N$-degree $m_E$, 
\bea(ll)
m_E:= [\frac{(N-1)L-P_a-P_b-2J}{N}]  & \Leftrightarrow \\
 N m_E = (N-1)L-P_a-P_b-2J-d_E  & {\rm for ~ some} ~ 0 \leq d_E \leq N-1 ,
\elea(ME)
and ${\tt P}(0) \neq 0$. Write  ${\tt P}( {\tt t} )$ in terms of its roots ${\tt t}_i^N$'s, and define the (evaluation) polynomial ${\tt P}_{\rm ev} (\xi )$ of degree $m_E$ by 
\be
{\tt P}_{\rm ev}({\tt t}^N) := \frac{{\tt P}( {\tt t} )}{{\tt P}(0)} = \frac{(-1)^{m_E}}{\prod_{i=1}^{m_E} {\tt t}_i^N} \prod_{i=1}^{m_E} ({\tt t}^N   - {\tt t}_i^N ) = \prod_{i=1}^{m_E} (1 - {\tt a}_i {\tt t}^N)   , ~ ~ {\tt a}_i := {\tt t}_i^{-N}.
\ele(Ptrt)
Denote
\be
\cos \theta_i := \frac{{\tt t}_i^N + 1}{{\tt t}_i^N - 1} = \frac{ 1+ {\tt a}_i}{1 - {\tt a}_i} , ~ ~ i=1, \ldots, m_E ,
\ele(ci)
and define the "normalized" reciprocal polynomial of ${\tt P}_{\rm ev} (\xi )$:
\bea(l)
{\tt P}'_{\rm ev} (\xi ) = \frac{(-\xi)^{m_E} }{\prod_{i=1}^{m_E} {\tt a}_i } {\tt P}_{\rm ev} (\frac{1}{\xi }) = \prod_{i=1}^m (1- {\tt a}^{-1}_i \xi ). 
\elea(P'ev)
By the method of functional relations in CPM \cite{BBP}, one can solve the $T , \widehat{T}$-eigenvalues through the polynomial ${\tt P}( {\tt t} )$ in (\req(Pt)).
The explicit form of $T, \widehat{T}$-eigenvalues are expressed by (\cite{AMP} (2.22) for the case $r=m=0$, \cite{B93} (6.14), \cite{B94} (21) for $m=0, r \in \ZZ_N$ , and \cite{R09} (2.36) for $m, r \in \ZZ_N$)
\bea(ll)
T(q) = \alpha_1 N^L \frac{R_m ( {\tt x} )^L ( 1-  {\tt x} )^L}{R_m ({\tt y})^L( 1-   {\tt x}^N )^L}  {\tt x}^{P_a }{\tt y}^{P_b }  \mu^{-P_\mu} \frac{{\tt F}({\tt t} )}{\omega^{P_b+m(P_b +P_a)}{\tt F}( \omega^{m+1}) } {\cal G}(\lambda) ,  \\
\widehat{T} (q) 
= \alpha_1^{-1} N^L \frac{R_m ( {\tt x})^L ( 1- {\tt x} )^L}{R_m ({\tt y})^L( 1- {\tt x}^N )^L}  {\tt x}^{P_a } {\tt y}^{P_b }  \mu^{-P_\mu} \frac{{\tt F}({\tt t} )}{{\tt F}(\omega^m) } {\cal G}(\lambda),  
\elea(TTform)
where $\alpha_1 := (-1)^{mL} \omega^{\frac{m(m+1) L+ 2m P_a  }{2}}$, $R_m (z):= \frac{( 1- z^N )}{\prod_{j=0}^{N-1-m} (1- \omega^j z )}$, $\mu:= \mu_q, ~ \lambda:= \mu^N$, with the total momentum 
\be
S_R = \omega^{- m(m+1) L +m(P_b -P_a)} \frac{\omega^{P_b} {\tt F}( \omega^{1+m})}{{\tt F}(\omega^m)}.
\ele(tmom)
Here the variables ${\tt x}, {\tt y}, {\tt t}$ are the normalized coordinates of $x_q , y_q, t_q$:
\be
{\tt x} := \omega^m x_p^{-1} x_q , ~ ~ {\tt y} := y_p^{-1} y_q , ~ ~ {\tt t} := \omega^m t_p^{-1} t_q , 
\ele(xyt)
and ${\cal G}(\lambda)$ is the factor-function of ${\tt P}({\tt t})$ in (\req(Pt)):  ${\cal G} (\lambda) {\cal G} (\lambda^{-1}) = \frac{{\tt P}({\tt t})}{{\tt P}(\omega^m )}$, with the expression 
\be
{\cal G}(\lambda)= \prod_{i=1}^{m_E} \frac{(\lambda +1) - (\lambda -1) w_i}{2 \lambda } ,
\ele(cG)
where $w_i$'s are solutions for ${\tt t}^N= {\tt t}^N_i$ in the following equivalent form of $W_{k'}$ in (\req(hW)), 
\bea(lll)
W_{k'}&:  \frac{(1-k')^2 }{4} w^2 = \frac{(1-k')^2 }{4} + \frac{k'}{1- {\tt t}^N }, &   (w:= \frac{\lambda + 1}{\lambda- 1}).
\elea(hWw)
Let $\overline{w}_i $ be the $w$-value in (\req(hWw)) for a zero ${\tt t}^N_i$ of ${\tt P}({\tt t})$ with 
$$
{\rm Re} ((1-k') \overline{w}_i) > 0 ~ ~ ~ {\rm for } ~ k' \in \RZ . 
$$
Note that ${\rm Re} ((1-k') \overline{w}_i) \longrightarrow  1$ as $k' \longrightarrow 0$, and ${\rm Re} (\frac{1-k'}{k'}  \overline{w}_i) \longrightarrow \pm 1 $ as $k' \longrightarrow \pm \infty$. Any choice of $w_i=  s_i \overline{w}_i ~ (1 \leq i \leq m_E)$ with $s_i = \pm$ in (\req(cG)) gives rise to a $T$-(or $\widehat{T}$)-eigenvalue (\req(TTform)) with the norm-one eigenvector, denoted by $\vec{v}(s_1, \ldots, s_{m_E} ; k')$. All such vectors form a basis of ${\cal E}_{{\tt F}, P_a, P_b}$ in (\req(Ek')),
(when $s_E=0$, $\vec{v}(s ; k')=$ the norm-one base element in the 1-dimensional space ${\cal E}_{{\tt F}, P_a, P_b}$). 
For rapidities $q$ near $p$ in ${\goth W}_{k'}$, up to first order of  small $\epsilon$, we have 
$$
\begin{array}{lll} 
x_q = \omega^m \eta^{\frac{1}{2}} (1- 2k' \epsilon ), & y_q = \eta^{\frac{1}{2}}(1 + 2k' \epsilon ) , & \mu_q =  1+ 2(k'-1) \epsilon . 
\end{array}
$$
Then  $\widehat{T} (q)$ is expressed by ( \cite{AMP} (1.11)-(1.17), \cite{R09} (2.43) (2.44))
$$
\widehat{T} (q) = {\bf 1} \{ 1 +   (N-1-2m) L \epsilon \} + \epsilon H (k') + O(\epsilon^2) 
$$
where the Hamiltonian $H (k')$ is given by
\bea(lll)
H (k') = H_0 + k' H_1, &
H_0 = - 2\sum_{\ell =1}^L  \sum_{j=1}^{N-1}   \frac{ \omega^{m j }Z^j_\ell Z^{-j}_{\ell +1} }{1-\omega^{-j}}& 
H_1 = - 2 \sum_{\ell =1}^L \sum_{j=1}^{N-1}    \frac{ \omega^{m j }X_\ell^j }{1-\omega^{-j}}.
\elea(H01)
with the boundary condition: $Z_{L+1}= \omega^{-r} Z_1, ~ X_{L+1}= X_1$. By (\req(TTform)), one may regard (\req(Ek')) as the $H (k')$-eigenvector decomposition of ${\cal E}_{{\tt F}, P_a, P_b}$ with the $H (k')$-eigenvalues:
\bea(l)
E(s_1, \ldots, s_{m_E}; k') (= E(s; k')) =  \alpha + k' \beta  + N \sum_{i=1}^{m_E} s_i \varepsilon (\theta_i) 
\elea(Esk')
where $\alpha = 2 P_\mu + Nm_E - (N-1-2m) L,  \beta =2(P_b - P_a) - \alpha $, and $\varepsilon (\theta_i) = (1-k')(\overline{w}_i)$, expressed by
\bea(l)
\varepsilon (\theta_i) ~ ( = \varepsilon (\theta_i ; k') ) =  \sqrt{1+k'^2 - 2k' \cos \theta_i} \geq 0 ~ ~ {\rm for} ~ k' \in \RZ . 
\elea(epsilon)
(When $s_E=0$, $\sum_{i=1}^{m_E} $ in (\req(Esk')) and $\prod_{i=1}^{m_E} $ in (\req(cG)) are defined to be $0, 1$ respectively.) Note that for real $k'$, $H(k')$ is an Hermitian operator of the quantum space $\bigotimes^L \CZ^N$ with the Hermitian form inherited from local spin basis. Hence $H(k')$ is indeed an Hermitian operator of ${\cal E}_{{\tt F}, P_a, P_b}$ with the induced Hermitian metric. Therefore (\req(Ek')) can be regarded as an orthonormal eigenvector decomposition of ${\cal E}_{{\tt F}, P_a, P_b}$ with the real eigenvalues (\req(Esk')):
\bea(ll)
 \overline{\vec{v}(s ; k')} \vec{v}(s'; k') = \delta_{s, s'}, ~ ~ s =(s_1, \ldots, s_{m_E}) , s'=(s'_1, \ldots, s'_{m_E}) ~ ~ {\rm for} ~ k' \in \RZ.
\elea(vort)
Then $\{ \vec{v}(s ; k') \}_{k' \in \RZ}$ forms a continuous family of orthonormal basis of ${\cal E}_{{\tt F}, P_a, P_b}$ with the $\pm \infty$-monodromy relation $\vec{v}(s ; + \infty) = \lambda (s) \vec{v}(-s ; - \infty)$ for some $\lambda (s) \in \CZ^*$ with $|\lambda (s)|=1$.  For the description of duality symmetry, we also need the following $k'$-state basis of ${\cal E}_{{\tt F}, P_a, P_b}$ continuous at $k'=\infty$:
\bea(ll)
\vec{w}(s_1, \ldots, s_{m_E} ; k')(=\vec{w}(s ; k')) :=& \left \{ \begin{array}{ll} \vec{v}(s_1, \ldots, s_{m_E} ; k') &{\rm if} ~ k' > 0 \\
                      \lambda (s)  \vec{v}(-s_1, \ldots, -s_{m_E} ; k') &{\rm if} ~ k' < 0 .
                  \end{array} \right.
\elea(wect)
The $T, \widehat{T}$-eigenvalues (\req(TTform))  of $\vec{w}(s ; k')$ are given by setting $w_i=\frac{k'}{|k'|}s_i \overline{w}_i $ in (\req(cG)), with the $H (k')$-eigenvalue 
\bea(l)
\widetilde{E}(s_1, \ldots, s_{m_E}; k') ( = \widetilde{E}(s; k')) =  \alpha + k' \beta  + N \sum_{i=1}^{m_E} \frac{k'}{|k'|} s_i \varepsilon (\theta_i). 
\elea(Etsk)
Note that ${\rm Re} ((1-k') \frac{k'}{|k'|} \overline{w}_i) \longrightarrow \pm 1$ as $k' \longrightarrow \pm 0$, and ${\rm Re} (\frac{1-k'}{|k'|} \overline{w}_i) \longrightarrow 1 $ as $k' \longrightarrow \pm \infty$, corresponding to the continuity of $\vec{w}(s; k')$ at $\infty$ with the $0^\pm$-monodromy relation $\vec{w}(s; 0^+)= \lambda(-s)^{-1} \vec{w}(-s; 0^-)$.
Hence we have CPM eigenvector-decomposition of the $\tau^{(2)}$-eigenspace for large $k'$:
\be
{\cal E}_{{\tt F}, P_a, P_b}= \bigoplus_s \CZ ~ \vec{w}(s ; k') ~ {\rm with} ~ ~ s= (s_1, \ldots, s_{m_E}) ,  ~ 0 \neq k' \in \RZ \cup \{ \infty \} .
\ele(Ewk')
Note that $\vec{v}(s ; k')$'s and $\vec{w}(s ; k')$'s are uniquely determined by their eigenvalue (\req(Esk')) for a general $k'$ up to a norm-one scale factor. In the rest of this paper, we shall also write $\vec{v}(s ; k')$ or $\vec{w}(s ; k')$ as the norm-one vector up to a factor if no ambiguity could arise.

By (\req(pp*)), the dual rapidity $p^*$ of the superintegrable $p$ in (\req(pcood)) is again defined by (\req(pcood)) for $\eta^* = \eta (1/k')$. Hence the dual $\tau^{(2)}$-model, $\tau^{(2) \dagger }({\tt t}^*)$, and CPM, $T^\dagger (q^*), T^\dagger (q^*)$, (with the boundary condition $r^* = Q$, and charge $Q^*=r$) in (\req(Dual)) are again defined by (\req(hsupL)) and (\req(TTform)), but using the coordinates $t^*, x^*, y^*, \mu^*$, quantum numbers $P_a^*, P^*_b, \cdots$ and the Bethe polynomial ${\tt F}^*({\tt t}^*)$. The relation (\req(Dual)) in turn yields the following identification of normalized variables and quantum numbers (\cite{R09} (3.24))\footnote{In $m=0$ case, the duality was found by Baxter in \cite{B89} where the variables $x, y, \mu$ and $x_d, y_d, \mu_d$ in (2.1),(5.4) there are correspond to ${\tt x}, {\tt y}^{-1}, {\tt y} \mu^{-1}$ and ${\tt x}^*, {\tt y}^{* -1}, {\tt y}^* \mu^{* -1}$ respectively in this paper.} :
\bea(llll)
{\tt x}^* = {\tt x} \mu, & {\tt y}^* = {\tt y} \mu^{-1}, & \mu^* = \mu^{-1} , & {\tt t}^* = {\tt t}, \\
P_a^*= P_a , & P^*_b = P_b, & P_\mu \equiv r , & P_\mu^* \equiv Q , \\ 
 J^*= J, & m^*_E = m_E, &  \alpha^*_1 = \alpha_1,   &  S^*_R = S_R 
\elea(duqn)
with ${\tt F}^*({\tt t}^*)= {\tt F}({\tt t})$, ${\tt P}^*({\tt t}^*) = {\tt P}({\tt t}) $, ${\cal G}^*(\lambda^*) = \lambda^{m_E} {\cal G} (\lambda)$, and $(1-k') \overline{w}_i = k' (1-k'^{-1}) \overline{w^*}_i$. In particular, both $\tau^{(2)}({\tt t})$ and $\tau^{(2) \dagger }({\tt t})$ are  defined by the same $L$-operator (\req(hsupL)), with the identical eigenvalue (\req(stauev)), but $\Psi$-related as operators in (\req(Dual)).  Note that the monodromy matrices (\req(Mont2)) of $\tau^{(2)}$ and $\tau^{(2) \dagger}$-model are not related by $\Psi$. The Hamiltonian $H^\dagger (k'^{-1}) (= H_0^\dagger  + k'^{-1} H_1^\dagger)$ of $T\dagger (q^*)$ is $\Psi$-similar to $H(k')$ in (\req(H01)):
\be
H(k') = k' \Psi^{-1} H^\dagger (k'^{-1}) \Psi  ~ ~ (\Leftrightarrow ~ H_0 = \Psi^{-1} H^\dagger_1 \Psi , ~ ~ H_1 = \Psi^{-1} H^\dagger_0 \Psi ).
\ele(Hinf) 
Hence $\Psi$ in (\req(Psi)) induces an isomorphism between the $\tau^{(2)}$-eigenspaces ${\cal E}_{{\tt F}, P_a, P_b}$ in $V_{r, Q}$  and ${\cal E}^\dagger_{{\tt F}, P_a, P_b}$ in $V_{r*, Q^*}$ which sends a $k'$-eigenvector to a $k'^{-1}$-eigenvector with the same eigenvalue with the linear terms related by $\alpha = \beta^\dagger , \beta = \alpha^\dagger$. 
Since $\widetilde{E}(s; k')= k' E^\dagger (s; k'^{-1})$, $\Psi (\vec{w}(s ; k')$ is equal to $\vec{v}^\dagger (s; k'^{-1})$ up to a norm-one factor.  Indeed, we have 
\bea(lll)
\Psi: &\vec{w}(s_1, \ldots, s_{m_E}; k') \mapsto  \lambda(-s) \vec{v}^\dagger (s_1, \ldots, s_{m_E} ; k'^{-1}), &\lambda(-s) = \prod_{i=1}^{m_E}(-s_i ) , \\
 & \vec{v}(s_1, \ldots, s_{m_E}; k')  \mapsto  \lambda(-s) \vec{w}^\dagger (s_1, \ldots, s_{m_E} ; k'^{-1}) .
\elea(vec*)
The factor $\pm 1$ in above are  derived from the Onsager-algebra structure of ${\cal E}_{{\tt F}, P_a, P_b} = {\cal E}^\dagger_{{\tt F}, P_a, P_b}$  with the identification of Onsager-algebra generators, $H_0= H^\dagger_1$ and $H_1= H^\dagger_0$ in (\req(Hinf)), which we shall discuss in Subsection \ref{ssec.OA}.

\subsection{Quantum numbers and Bethe equation in superintegrable $\tau^{(2)}$-model \label{ssec.QN} }
In this subsection, we describe the complete constraints of quantum numbers in the superintegrable CPM, which consist of fours types of sectors, then discuss their relationship under the duality relation of CPM.  
Furthermore we describe another reflective symmetry between sectors with conjugate total momentum through the relation among solutions of Bethe equation

The integers $P_a, P_b$ and $P_\mu$ in (\req(TTform)) are indeed quantum numbers of the superintegrable $\tau^{(2)}$-model, depending only on the $\tau^{(2)}$-eigenvalue. By the finiteness of Boltzmann weights as $x_q$ or $y_q$ tending to zero in (\req(TTform)), together with the behavior of the leading and constant terms of  $\tau^{(2)}(t)$ (\cite{R0805} (4.26) (4.27)),  one finds $P_a, P_b$ are non-negative integers satisfying 
\be
0 \leq P_a + P_b \leq N-1, ~ ~ P_b - P_a \equiv Q +r +(1+2m)L \pmod{N}. 
\ele(Pab1)
Furthermore, there are further constraints involving the number $J$ in (\req(Bethesup))\footnote{The number $J$ here is denoted by $m_p$ in \cite{AMP}.}, which are classified into four types (\cite{R0805} (4.36) (4.37), \cite{R09} (2.25)):
\bea(llll)
I_+: &P_a =0, P_b  \equiv mL + r - J, & 
I_-: & P_b = 0, m L  +Q -J \equiv 0   \\
i_+ :& P_a =0, P_b  \equiv m L  +Q -J , & 
i_-: & P_b = 0,  m L+ r -J \equiv 0  .
\elea(Iipm)
The finiteness of Boltzmann weights when $\mu_q$ or $\mu_q^{-1}$ tends to zero yields the integer $P_\mu (\equiv r \pmod{N})$ satisfy the relation:
\bea(l)
P_b - mL + J  \leq P_\mu \leq (N-1-m)L -P_a - Nm_E-J ~ ~( = P_b  - mL + J + d_E) , 
\elea(Pmu)
where $d_E$ is defined in (\req(ME)) (for $m=0$ case, see \cite{B94} (22)-(24)). Since $d_E$ is an non-negative integer less that $N$, $P_\mu$ is uniquely defined by (\req(Pmu)), hence determined by $P_a, P_b$ and $J$. Using (\req(Pab1)) and (\req(Pmu)), one finds the quantum numbers of sectors in (\req(Iipm)) are given as follows:
\bea(lll)
I_+: &P_a =0, ~ ~ P_b  \equiv mL + r - J, & Q \equiv - (1+m)L - J, \\
& P_\mu = P_b- mL +J ;\\
I_-: & P_a  \equiv -(1+m)L - r - J , ~ ~ P_b = 0, &  Q \equiv - mL + J, \\& P_\mu = (N-1-m)L -P_a - Nm_E-J;  \\
i_+ :& P_a =0 , ~ ~ P_b  \equiv m L  +Q -J , & r \equiv - (1+m)L - J,\\ 
& P_\mu = (N-1-m)L - Nm_E-J;  \\
i_-: &  P_a  \equiv -(1+m)L - Q - J , ~ P_b = 0, & r \equiv - m L + J,\\
& P_\mu = - mL + J ,
\elea(qTn)
by which, the linear term in (\req(Esk')) can be written as 
\be
\alpha + k' \beta = \left\{ \begin{array}{ll} (P_b - P_a - d_E) + k' (P_b - P_a + d_E) & {\rm for ~ sectors ~ in} ~ I^+ \cup i_- ; \\
(P_b - P_a + d_E) + k' (P_b - P_a - d_E) & {\rm for ~ sectors ~ in} ~ I^- \cup i_+ .
 \end{array} \right.
\ele(linH)
Note that there are common sectors in (\req(qTn)), indeed 
\bea(cl)
I_+ \cap I_- = i_+ \cap i_- :& P_a =P_b= 0 \equiv (1+2m)L + 2r ~ (\equiv (1+2m)L + 2Q ); \\
I_\pm \cap i_\pm = &\left\{ \begin{array}{ll} I_\pm = i_\pm  &    {\rm if} ~ r \equiv Q , \\            
                                              \emptyset & {\rm otherwise} , \end{array} \right. \\
 I_+ \cap i_- : & P_a =P_b= 0 \equiv  m L+ r -J \equiv (1+m)L + Q +J, \\
I_- \cap i_+ : & P_a =P_b= 0 \equiv  (1+m)L + r + J \equiv  m L+ Q - J  . 
\elea(Pab0)
By (\req(Pab1)) (\req(qTn)) and (\req(linH)), the above first case  is characterized by either one of the following equivalent conditions:
\bea(l)
I_+ \cap I_- = i_+ \cap i_- :  P_a =P_b=0, Q=r  \Longleftrightarrow \alpha + k' \beta= 0 .
\elea(Ipm)
Since the quantum space $V_{r, Q}$ in (\req(VrQ)) is the union of ${\cal E}_{{\tt F}, P_a, P_b}$ with $P_a, P_b$ satisfying (\req(Pab1)):
\be
V_{r, Q} = \bigcup\{ {\cal E}_{{\tt F}, P_a, P_b}| P_b-P_a \equiv (1+2m)L+Q+r, {\tt F}: {\rm Bethe ~ polynomial} \}, 
\ele(VEdec)
one may determine those sectors in (\req(Iipm)) which appear in the above relation.  Indeed, for ${\cal E}_{{\tt F}, P_a, P_b} \subset V_{r, Q}$, by (\req(qTn)) and (\req(Pab0)), in the case of $(1+2m)L+Q+r \equiv 0$ we find
$$
P_a=P_b=0 , ~ ({\tt F}, P_a, P_b) \in \left\{\begin{array}{ll} 
I_+ \cap I_- = i_+ \cap i_-, & {\rm if } ~ r=Q , \\
(I_+ \cap i_- \setminus (I_- \cup i_+) ) \bigcup ( I_- \cap i_+ \setminus (I_+ \cup i_-)),  & {\rm if } ~ r\neq Q ,; \end{array} \right.\\
$$
and when $(1+2m)L+Q+r \not\equiv 0$, $(P_a, P_b) \equiv (0, (1+2m)L+Q+r)$ or $(-(1+2m)L-Q-r, 0)$ with 
$$
\begin{array}{l}
({\tt F}, P_a=0, P_b) \in \left\{\begin{array}{ll} 
I_+  \setminus (I_- \cup i_-)= i_+  \setminus (I_- \cup i_-), & {\rm if } ~ r=Q ,\\
(I_+  \setminus (I_- \cup i_- \cup i_+)) \bigcup (i_+  \setminus (I_- \cup i_- \cup I_+)),  & {\rm if } ~ r\neq Q ;\end{array} \right.\\
({\tt F}, P_a, P_b=0) \in \left\{\begin{array}{ll} 
I_-  \setminus (I_+ \cup i_+)= i_-  \setminus (I_+ \cup i_+), & {\rm if } ~ r=Q , \\
(I_-  \setminus (I_+ \cup i_+ \cup i_-)) \bigcup (i_-  \setminus (I_+ \cup i_+ \cup I_-)),  & {\rm if } ~ r\neq Q . \end{array} \right.
\end{array}
$$
The dual model $\tau^{(2) \dagger}({\tt t})$ in (\req(Dual)) also consists of four types of sectors, $I_\pm^\dagger, i_\pm^\dagger$ as in (\req(qTn)). By (\req(duqn)) and the identification of their quantum spaces by $\Psi$ (or $\Psi^\dagger$)  in (\req(Psi)), the sectors of $\tau^{(2)}$- and $\tau^{(2) \dagger}$-model with equal $J$ are identified under duality according to
\bea(ll)
i_\pm \longleftrightarrow I^\dagger_\pm , & I_\pm \longleftrightarrow i^\dagger_\pm ~ ~ {\rm respectively}, 
\elea(IiD)
which are consistent with the decomposition in (\req(VEdec)). Note that $\Psi^\dagger \Psi = \omega^{QQ^*} S_R$ (\cite{R09} Section 3.1), and the eigenvectors of $\tau^{(2)}({\tt t})$ and $\tau^{(2) \dagger}({\tt t})$ correspond to each other under $\Psi$ or $\Psi^\dagger$. Using (\req(duqn)) and (\req(qTn)), one finds the exact relation between quantum numbers $P_\mu$ and $P^*_\mu$ of the dual sectors in (\req(IiD)): $P_\mu \pm d_E = P^*_\mu $. In particular, all sectors $({\tt F},P_a, P_b) \in I_+ \cap I_- = i_+ \cap i_-$  in (\req(Ipm)) are fixed by the dual correspondence (\req(IiD)), hence $\Psi$ in (\req(Psi)) induces an automorphism of ${\cal E}_{({\tt F},P_a, P_b)}$: $\Psi: {\cal E}_{({\tt F},P_a, P_b)} \simeq {\cal E}_{({\tt F},P_a, P_b)}$.

Next we establish an one-to-one correspondence between sectors in $I_+$ and $I_-$, also between $i_+$ and $i_-$. For non-zero complex numbers ${\tt v}_j$'s with ${\tt F} ({\tt t}) = \prod_{j=1}^J (1 + \omega {\tt v}_j {\tt t})$, we define 
\be
{\tt F}' ({\tt t}) = \prod_{j=1}^J (1 + \omega {\tt v}'_j {\tt t}), ~ ~ {\tt v}'_j := {\tt v}^{-1}_j \omega^{-3-2m} ~ ~ (1 \leq j \leq J),
\ele(F')
which is related to  ${\tt F}({\tt t})$ by
\be
{\tt F}' (\omega^m {\tt t})  = \frac{\omega^{-(2+m)J}}{\prod_{j=1}^J {\tt v}_j} {\tt t}^J {\tt F} ( \omega^{1+m} {\tt t}^{-1}) .
\ele(F'F)
Note that the relation between ${\tt v}_j$ and ${\tt v}'_j$ is "reciprocal", i.e., ${\tt v}_j'' = {\tt v}_j$  and ${\tt F}'' ({\tt t}) = {\tt F} ({\tt t})$.
\begin{lem}\label{lem:FF'} 
Assume $P_a, P_b$ and $P'_a, P'_b$ are integers satisfying the relation  
\be
P_a+ P_b+ P'_a+P'_b \equiv - (L+2J) \pmod{N}. 
\ele(Pab')

$(i)$. $\{{\tt v}_j \}_{j=1}^J$ is a solution of Bethe equation (\req(Bethesup)) for $(P_a, P_b)$ if and only if 
$\{{\tt v}'_j \}_{j=1}^J$ is a solution of Bethe equation (\req(Bethesup)) for $(P'_a, P'_b)$. 

$(ii)$. Let $\tau^{(2)}({\tt t}), \tau'^{(2)}({\tt t})$ be the $\tau^{(2)}$-eigenvalues in (\req(stauev)) for $(P_a, P_b, {\tt F})$ and $(P'_a, P'_b, {\tt F}')$ respectively, and ${\tt P}({\tt t}), {\tt P}'({\tt t})$ be their corresponding ${\tt t}^N$-polynomials of degree $m_E, m_E'$ in (\req(Pt)). Then 
$$
\begin{array}{ll}
&\tau'^{(2)}(\omega^m {\tt t}) = \omega^{P_b'+P_a+L+J} (-{\tt t})^L \tau^{(2)}(\omega^{m-1} {\tt t}^{-1}) \\
(\Leftrightarrow & 
\tau^{(2)}(\omega^m {\tt t}) = \omega^{P_b+P'_a+L+J} (-{\tt t})^L \tau'^{(2)}(\omega^{m-1} {\tt t}^{-1})),
\end{array}
$$
and ${\tt P}({\tt t}), {\tt P}'({\tt t})$ satisfy the following reciprocal relation:
\be
 \omega^{P'_b+ m(P_a'+P_b')} (\prod_{j=1}^J {\tt v}_j' ) {\tt P}'({\tt t}) 
 = \omega^{P_b+ m(P_a+P_b)} (\prod_{j=1}^J {\tt v}_j)  {\tt P}({\tt t}^{-1}) {\tt t}^{Nm_E} , 
\ele(PPrec)
with
\be
m_E = m'_E , ~ ~ P_a + P_b = d_E' , ~ ~ P_a'+P_b' = d_E
\ele(mm'dd')
where $d_E, d_E'$ are defined in (\req(ME)) for ${\tt P}({\tt t}), {\tt P}'({\tt t})$ respectively.
\end{lem}
{\it Proof.} It is easy to verify $(i)$. The relation of $\tau^{(2)}$-eigenvalues in $(ii)$ follows from (\req(F'F)), by which the term $(1- \omega^{-m+k} {\tt t})^L{\tt F}' (\omega^k {\tt t}) {\tt F}'( \omega^{k+1} {\tt t})$ in the expression of ${\tt P}'({\tt t})$ in (\req(Pt)) is equal to
$$ 
(-1)^L \omega^{-m(L+2J)+k(L+2J)} (\prod_{j=1}^J {\tt v}'_j{\tt v}^{-1}_j) {\tt t}^{2J+L} (1-  \omega^{m-k} {\tt t}^{-1} )^L {\tt F} ( \omega^{-k+2m} {\tt t}^{-1}) {\tt F} ( \omega^{-k+1+2m} {\tt t}^{-1}).
$$
Under the condition (\req(Pab')), one can write ${\tt P}'({\tt t})$ in the form
$$
\omega^{P'_b+m(P_a'+P_b')} (\prod_{j=1}^J {\tt v}_j' ) {\tt P}'({\tt t}) 
 = {\tt t}^{(N-1)L-2J-Nm_E-(P_a+P_b+P'_a+P'_b)} \omega^{P_b+ m(P_a+P_b)} (\prod_{j=1}^J {\tt v}_j)  {\tt t}^{Nm_E} {\tt P}( \omega^{2m} {\tt t}^{-1}). 
$$
Since ${\tt P}'(0) \neq 0$ and $m_E$ is the ${\tt t}^N$-degree of ${\tt P}({\tt t})$, the above expression yields
$$
(N-1)L-2J-Nm_E = P_a+P_b+ P'_a+P'_b, 
$$
hence we obtain (\req(mm'dd')) by (\req(ME)). Then follows (\req(PPrec)) by ${\tt P}( \omega^{2m} {\tt t}^{-1})= {\tt P}({\tt t}^{-1})$.
$\Box$ \par \vspace{.1in} \noindent
{\bf Remark}. By (\req(PPrec)), the roots of ${\tt P}'({\tt t})$ and ${\tt P}({\tt t})$ in the above lemma are reciprocal. Hence the $\theta_i', \theta_i$ for ${\tt P}'({\tt t}), {\tt P}({\tt t})$ in (\req(ci)) are related to $\theta_i' = \pi - \theta_i$ for $1 \leq i \leq m_E$, and the evaluation polynomial $({\tt P}')_{\rm ev}(\xi)$ of ${\tt P}'({\tt t})$ in (\req(Ptrt)) is equal to the polynomial in (\req(P'ev)): $({\tt P}')_{\rm ev}(\xi)= {\tt P}'_{\rm ev}(\xi)$.

With $(P_a, P_b) \in I_\pm , i_\pm$ and $(P'_a, P'_b) \in I_\mp , i_\mp$ respectively with the same $J$ in Lemma \ref{lem:FF'},  the condition (\req(Pab')) holds by (\req(qTn)). Therefore, one finds the one-to-one correspondence of sectors in (\req(Iipm)):
\be
({\tt F}, P_a, P_b) \in I_\pm, i_\pm  \longleftrightarrow ({\tt F}', P'_a, P'_b) \in I_\mp , i_\mp , ~ ~ {\rm respectively}
\ele(conj)
where ${\tt F}'$ is in (\req(F')), with their $\tau^{(2)}$-eigenvalues and  ${\tt P}$-polynomials related in Lemma \ref{lem:FF'}. Furthermore, by (\req(qTn)) and (\req(mm'dd')), the linear terms  of sectors in (\req(linH)) are related by
\bea(ll)
I_\pm : \alpha =  \alpha^\prime , \beta = - \beta^\prime ; & i_\pm : \alpha =  - \alpha^\prime , \beta = \beta^\prime .
\elea(cjlin)
By (\req(F'F)) and (\req(qTn)), the total momentum $S_R, S_R'$ (\req(tmom)) of two sectors in  (\req(conj)) are related by the following conjugate relation:
\bea(ll)
S_R S_R' =  \omega^{(2m+1)r} ~ ~ {\rm or} ~ ~ \omega^{(2m+1)Q} &{\rm for} ~ ({\tt F}, P_a, P_b) \in I_\pm  ~ {\rm or}  ~ i_\pm ~ ~ {\rm respectively}.
\elea(SRcj)
Denote the boundary condition and $\ZZ_N$-charge of ${\cal E}_{{\tt F}, P_a, P_b}$ and ${\cal E}_{{\tt F}', P'_a, P'_b}$ in (\req(conj)) by $(r, Q), (r', Q')$ respectively. Then $(r, Q)=(r', Q')$ if and only if $L+2J \equiv 0 \pmod{N}$; and when $L+2J  \not\equiv 0$,  $r=r' , Q \neq Q'$ in the case $I_\pm$, and $r \neq r' , Q = Q'$ in the case $i_\pm$. 
In particular, the condition (\req(Ipm)) is preserved under the conjugate correspondence (\req(conj)) with $r= Q = r'=Q'$, in which case $S_R S_R'= \omega^{(2m+1)r}$. The sectors fixed by (\req(conj)), i.e.  $({\tt F}, P_a, P_b)= ({\tt F}', P'_a, P'_b)$, are characterized by the following condition by Lemma \ref{lem:FF'}:
\bea(ll)
P_a=P_b=0, Q \equiv r, & {\tt F}(\omega^m {\tt t}) = \frac{\omega^{-(2+m)J}}{\prod_{j=1}^J {\tt v}_j} {\tt t}^J {\tt F} ( \omega^{1+m} {\tt t}^{-1}) \\
&(\Leftrightarrow \tau^{(2)}(\omega^m {\tt t}) = \omega^{L+J} (-{\tt t})^L \tau^{(2)}(\omega^{m-1} {\tt t}^{-1}) ), 
\elea(Prec)
with $S_R = \pm \omega^\frac{(2m+1)r}{2}$. In this situation, ${\tt P}({\tt t})$ is a ${\tt t}^N$-polynomial with ${\tt t}$-degree $Nm_E=(N-1)L-2J$, satisfying the reciprocal condition 
$$
{\tt P}({\tt t}) = \omega^{3J-mL} (\prod_{j=1}^J {\tt v}_j^2) {\tt P}({\tt t}^{-1}){\tt t}^{Nm_E}, ~ ~ {\tt P}(0)= N. 
$$
Note that the condition for ${\tt F}=1$ (i.e. $J=0$) satisfying the relation (\req(Prec)) is $P_a=P_b=Q=r=0, L \equiv 0$, where $S_R = 1$.

In Subsection \ref{ssec.OA}, using the Onsager-algebra symmetry of the eigenspaces, we indicate the existence of a canonical isomorphism between ${\cal E}_{{\tt F}, P_a, P_b}$ and ${\cal E}_{{\tt F}', P'_a, P'_b}$ with the basis correspondence:
\bea(ll)
I_\pm \ni {\cal E}_{{\tt F}, P_a, P_b} \ni \vec{w}(s_1, \ldots , s_{m_E} ;k')   \longleftrightarrow  \vec{w}^{\prime} (-s_1, \ldots , -s_{m_E}  ; -k') \in {\cal E}_{{\tt F}', P'_a, P'_b} \in I^\mp , \\
i_\pm \ni {\cal E}_{{\tt F}, P_a, P_b} \ni \vec{v}(s_1, \ldots , s_{m_E} ;k')   \longleftrightarrow  (-1)^{m_E} \vec{v}^{\prime} (-s_1, \ldots , -s_{m_E}  ; -k') \in {\cal E}_{{\tt F}', P'_a, P'_b} \in i^\mp .
\elea(E'Eb)
Indeed, later in Proposition \ref{prop:inversion}, we shall show the above correspondence is induced by the spin-inversion operator of the quantum space.  
Note that the duality of $\tau^{(2)}$-model interchanges $I_+$ ($i_+$) and $I_-$ ($i_-$)-sectors in (\req(IiD)) with the same Bethe polynomial by (\req(duqn)), hence the inverse correspondence (\req(E'Eb)) commutes with the duality relation.

\section{Onsager-algebra Symmetry and $sl_2$-loop-algebra Symmetry in Superintegrable $\tau^{(2)}$-model \label{sec.OAlp}}
\setcounter{equation}{0}
In this section, we study the degeneracy symmetries of an Onsager sectors in a superintegrable $\tau^{(2)}$-model, and examine the relationship of CPM $k'$-eigenvectors  under the duality and inversion relations. We also describe a procedure of constructing the $k'$-state vector of superintegrable CPM through these symmetries.

\subsection{The Onsager algebra structure of superintegrable $\tau^{(2)}$-eigenspaces \label{ssec.OA}}
The CPM $k'$-vectors in (\req(Ek')) or (\req(Ewk')) can be studied through the superintegrable chiral Potts quantum chain $H(k')$ in (\req(H01)) with the structure known since decades ago \cite{B89}. For our purpose,  an explicit and precise relationship between CPM $k'$-eigenvectors for different $k'$s is needed. We re-examine the Onsager-algebra symmetry of superintegrable chiral Potts quantum chain through the theory of Onsager-algebra representation \cite{DR, Dav, R91}. Since the pair of operators, $\frac{2H_0}{N}$ and $ \frac{-2H_1}{N}$, satisfy the Dolan-Grady relation \cite{GR}, the Hamiltonian $H(k')$ gives rise to an Onsager-algebra representation on ${\cal E}_{{\tt F}, P_a, P_b}$ \cite{Dav, Per}. 
It is known that the Onsager-algebra can be realized as the Lie-subalgebra of the loop-algebra $sl_2[z, z^{-1}]$ fixed by the standard 
involution $(e^\pm, h, z) \leftrightarrow (e^\mp, -h, z^{-1})$  \cite{R91}, and any irreducible representation of the Onsager algebra is always induced from an irreducible $sl_2[z, z^{-1}]$-representation by evaluating $z$ on a finite number of non-zero values not equal to $\pm 1$ \cite{DR, Dav}. For the Onsager-algebra representation ${\cal E}_{{\tt F}, P_a, P_b}$, the  $H(k')$-eigenvalue expression (\req(Esk')) for the basis in (\req(Ek')) yields that the $sl_2[z, z^{-1}]$-representation is obtained by evaluating $z$ at ${\rm e}^{{\rm i} \theta_i}$'s  related to zeros of ${\tt P}({\tt t})$ by (\req(ci)), equivalently ${\rm e}^{{\rm i} \theta_i}= \frac{{\tt t}_i^{N/2} +1}{{\tt t}_i^{N/2} -1}$, then applying the spin-$\frac{1}{2} ~ sl_2$-representation on each evaluated factor.

In this subsection, we discuss the structure of $H(k')$-eigenvectors of ${\cal E}_{{\tt F}, P_a, P_b}$ in (\req(vort)
) for $k' \in \RZ$. First we note that $\theta_i$'s in (\req(Esk')) are real with $ 0 < \theta_i < \pi$, equivalently the roots ${\tt t}^N_i$ of ${\tt P}_{\rm ev}(\xi)$ in (\req(Ptrt)) are all negative real, $ - \infty < {\tt t}^N_i < 0$ with ${\rm Im} ({\tt t}_i^{N/2}) < 0$. With the Hermitian form induced from the quantum space $\bigotimes^L \CZ^N$, ${\cal E}_{{\tt F}, P_a, P_b} $ carries a $\CZ^2$-product structure:
\be
{\cal E}_{{\tt F}, P_a, P_b} \cong \otimes_{i=1}^{m_E} V_i , ~ ~ V_i = \CZ {\tt b}_i^+ + \CZ {\tt b}_i^- ~ ( \cong \CZ^2 ),
\ele(EV)
so that the basis in (\req(vort)) and (\req(wect)) can be expressed as a product form as follows. 
Indeed as operators of ${\cal E}_{{\tt F}, P_a, P_b}$, $\frac{2H_0}{N}$ and $ \frac{-2H_1}{N}$ form an irreducible representation of Onsager algebra, considered as a subalgebra of $sl_2[z, z^{-1}]$ with the identification $\frac{2H_0}{N} = e_i^+ + e_i^-$ and $ \frac{-2H_1}{N} = z e_i^+ + z^{-1} e_i^-$. The representation space ${\cal E}_{{\tt F}, P_a, P_b}$ is factored into product of $m_E$-copies of spin-$\frac{1}{2} ~ sl_2$-representation by evaluating $z$ at ${\rm e}^{{\rm i} \theta_i}$ at the $i$th factor $V_i$ in (\req(EV)), where ${\tt b}_i^\pm$ are the basis for spin-$\frac{1}{2}$ representation. Let $e^\pm_i, h_i$ be the standard operators of the special Lie-algebra $sl (V_i)$ with respect to the basis ${\tt b}_i^\pm$, and $J^x_i := \frac{e^+_i+e^-_i}{2}, J^y_i := \frac{-{\rm i}(e^+_i-e^-_i)}{2}, J^z_i := \frac{h_i}{2}$ are  the unitary operators. Then the Hamiltonian $H(k')$ on ${\cal E}_{{\tt F}, P_a, P_b}$ can be expressed by \footnote{The second relation in (\req(HJ)) for $i_-$-sector with $J=0$, when changing $P_a, k', \theta_i, J^x_i, J^y_i$ to $Q, -k', \pi- \theta_i, -J^z_i, - J^x_i$, is the same as \cite{B89} (2.20).}: 
\bea(ll)
H(k') &= \alpha + k' \beta + N \sum_{i=1}^{m_E} \bigg( e_i^+ + e_i^- - k' (e^{{\rm i} \theta_i}e_i^+ + e^{-{\rm i} \theta_i}e_i^-) \bigg)\\
&= \alpha + k' \beta + 2N \sum_{i=1}^{m_E} \bigg( (1- k' \cos \theta_i) J^x_i + k' \sin \theta_i J^y_i \bigg). 
\elea(HJ)
One may use the above formulas to find the expression of $H(k')$-eigenvectors, i.e.  $\vec{v}(s; k')$ in (\req(Ek')) or $\vec{w}(s; k')$ in (\req(wect)) in terms of ${\tt b}_i^\pm$'s. Corresponding to the $i$th component of the $H(k')$-eigenvalues (\req(Esk')), (\req(Etsk)) of $\vec{v}(s; k')$ and $\vec{w}(s; k')$ for $1 \leq i \leq m_E$, we define the angle-functions, depending on $k'$ and $\theta_i$ in (\req(ci)), by the following analytic relations:
\bea(ll)
{\rm e}^{{\rm i} \vartheta_{i, k'}} := \frac{1- k' \cos \theta_i - {\rm i} k' \sin \theta_i}{\varepsilon (\theta_i;k')}, & 
{\rm e}^{{\rm i} \varphi_{i, k'}} :=  \frac{|k'| (1- k' \cos \theta_i - {\rm i} k' \sin \theta_i)}{k' \varepsilon (\theta_i;k')} ,
\elea(angle)
where $\varepsilon (\theta_i; k') =  \sqrt{1+k'^2 - 2k' \cos \theta_i}$ in (\req(epsilon)), and $\vartheta_{i, k'}, \varphi_{i, k'} $ satisfy the relations:
$$
\begin{array}{ll}
\vartheta_{i, k'} ~ (k' \in \RZ): \vartheta_{i, - \infty} = 2\pi + \theta_i , \varphi_{i, -1}= 2\pi + \frac{\theta_i }{2}, & \vartheta_{i, 0} = 2 \pi , \vartheta_{i, 1}= \frac{3 \pi  + \theta_i }{2}, ~ \vartheta_{i, + \infty} = \pi + \theta_i ;    \\
\varphi_{i, k'}  ~ (k' \in \RZ^* \cup \{\infty \}):  \varphi_{i, 0^-} =  \pi, ~  \varphi_{i, -1}= \pi + \frac{\theta_i }{2}, & \varphi_{i, \infty} = \pi + \theta_i, ~ \varphi_{i, 1}= \frac{3 \pi  + \theta_i }{2}, ~ \varphi_{i, 0^+} = 2 \pi ,
\end{array}
$$
Note that $\vartheta_{i, k'} =\varphi_{i, k'}$ for $k' >0$ and $\vartheta_{i, k'} =\varphi_{i, k'} + \pi$ for $k'<0$. Using the relation $\varepsilon (\theta ; k') = |k'| \varepsilon (\theta ; 1/k')$, one finds ${\rm e}^{{\rm i}(\varphi_{i, k'} + \vartheta_{i, 1/k'} - \theta_i)} = - 1$, hence $\varphi_{i, k'} + \vartheta_{i, 1/k'} =  3\pi + \theta_i $. 
Since the $H(k')$-eigenvectors  are obtained by diagonalizing the $V_i$-operators in (\req(HJ)), we consider two $k'$-basis
${\tt w}_{i,k'}^\pm, {\tt v}_{i,k'}^\pm$ of $V_i$:
\bea(lll)
 V_i & = \CZ {\tt w}_{i,k'}^+ + \CZ {\tt w}_{i,k'}^- &
({\tt w}_{i,k'}^+,  {\tt w}_{i, k'}^- ) =  ({\tt b}_i^+, {\tt b}_i^-)  \left( \begin{array}{cc}
        {\rm e}^{{\rm i} \varphi_{i, k'}/2}  &   - {\rm i} {\rm e}^{{\rm i} \varphi_{i, k'}/2} 
\\ {\rm e}^{-{\rm i} \varphi_{i, k'}/2} & {\rm i} {\rm e}^{-{\rm i} \varphi_{i, k'}/2}
\end{array} \right) \frac{1}{\sqrt{2{\rm i}}} , \\
& = \CZ {\tt v}_{i,k'}^+ + \CZ {\tt v}_{i,k'}^-, &  ({\tt v}_{i,k'}^+,  {\tt v}_{i, k'}^- ) =  ({\tt b}_i^+, {\tt b}_i^-)  \left( \begin{array}{cc}
        {\rm e}^{{\rm i} \vartheta_{i, k'}/2}  &   - {\rm i} {\rm e}^{{\rm i} \vartheta_{i, k'}/2} 
\\ {\rm e}^{-{\rm i} \vartheta_{i, k'}/2} & {\rm i} {\rm e}^{-{\rm i} \vartheta_{i, k'}/2}
\end{array} \right)  \frac{1}{\sqrt{2{\rm i}}} .
\elea(vi)
Then 
\bea(l)
({\tt w}_{i,k'}^+,  {\tt w}_{i, k'}^- ) = \left \{ \begin{array}{ll} ({\tt v}_{i,k'}^+,  {\tt v}_{i, k'}^- ) & {\rm if} ~ k' >0 , \\
({\tt v}_{i,k'}^-,  -{\tt v}_{i, k'}^+ ) & {\rm if} ~ k' <0 , \end{array}
\right.  
\elea(wvi)
with $(-{\tt w}_{i,0^-}^-,  {\tt w}_{i, 0^-}^+ )= ({\tt v}_{i, 0}^+,  {\tt v}_{i, 0}^- )  = ({\tt w}_{i,0^+}^+, 
{\tt w}_{i, 0^+}^- )$ , $
({\tt v}_{i,-\infty}^-,  -{\tt v}_{i, -\infty}^+ ) = ({\tt w}_{i,\infty}^+,  {\tt w}_{i, \infty}^- ) = ({\tt v}_{i,+\infty}^+,  {\tt v}_{i, +\infty}^- ) $. Using 
\bea(lll)
({\tt b}_i^+, {\tt b}_i^-)  &= ({\tt w}_{i, \infty}^+, {\tt w}_{i, \infty}^-) \left( \begin{array}{cc}
       {\rm e}^{-{\rm i} \theta_i/2}  &   - {\rm e}^{{\rm i} \theta_i/2}   \\
{\rm i} {\rm e}^{-{\rm i} \theta_i /2}& {\rm i} {\rm e}^{{\rm i} \theta_i/2} 
\end{array} \right)  \frac{1}{\sqrt{2{\rm i}}}  & = ({\tt v}_{i, 0}^+, {\tt v}_{i, 0}^-) \left( \begin{array}{cc}
      -{\rm i}   &  - {\rm i}   \\ 1 & - 1
\end{array} \right) \frac{1}{\sqrt{2{\rm i}}}  , 
\elea(eoinf)
the basis ${\tt w}_{i,k'}^\pm, {\tt v}_{i,k'}^\pm, $ in (\req(vi)) can be expressed by ${\tt w}_{i, \infty}^\pm$ or ${\tt v}_{i,0}^\pm$ respectively:
\bea(ll)
({\tt w}_{i, k'}^+, {\tt w}_{i, k'}^-)&= ({\tt w}_{i, \infty}^+, {\tt w}_{i, \infty}^-)\left( \begin{array}{cc}
         \sin \frac{\varphi_{i, k'} - \theta_i}{2}&  -  \cos \frac{\varphi_{i, k'} - \theta_i}{2}  \\ \cos \frac{\varphi_{i, k'} - \theta_i}{2} & \sin \frac{\varphi_{i, k'} - \theta_i}{2} \end{array}  \right) ; \\
({\tt v}_{i, k'}^+, {\tt v}_{i, k'}^-)&= ({\tt v}_{i, 0}^+, {\tt v}_{i, 0}^-)\left( \begin{array}{cc}
       - \cos  \frac{\vartheta_{i, k'}}{2}  &    - \sin  \frac{\vartheta_{i, k'}}{2}   \\ \sin  \frac{\vartheta_{i, k'}}{2} &  - \cos  \frac{\vartheta_{i, k'}}{2}
\end{array} \right) .
\elea(vk0)
With respective to the basis ${\tt w}_{i, k'}^\pm$ of $V_i$, one has $e^\pm_{i,k'}, h_{i, k'}$ as the standard generators of the special Lie-algebra $sl (V_i)$ with  the unitary basis $J^{x, y, z}_{i, k'}= \frac{e^+_{i,k'}+e^-_{i,k'}}{2}, \frac{-{\rm i}(e^+_{i' k'}-e^-_{i, k'})}{2}, \frac{h_{i, k'}}{2}$ respectively. Similarly, for the basis ${\tt v}_{i, k'}^\pm$, we have $sl (V_i)$-generators $e^\pm_{i,k'}, ~ h_{i, k'}$ and the unitary basis ${\tt j}^{x, y, z}_{i, k'}$ with the relations ${\tt j}^{x, z}_{i, k'} = \frac{k'}{|k'|} J^{x, z}_{i, k'}$, ${\tt j}^y_{i, k'} = J^y_{i, k'}$. Using (\req(HJ)) (\req(vi)), one finds 
$$
\begin{array}{ll}
H(k') 
&= \alpha + k' \beta + 2N \sum_{i=1}^{m_E} \frac{k'}{|k'|} \varepsilon (\theta_i) J^z_{i, k'} = \alpha + k' \beta + 2N \sum_{i=1}^{m_E}  \varepsilon (\theta_i) {\tt j}^z_{i, k'}.
\end{array}
$$
The above formulas provide the following expression of $H(k')$-eigenvectors $\vec{w}(s ; k')$ or $\vec{v}(s ; k')$ with eigenvalue $\widetilde{E}(s; k')$ or $E(s; k')$ in (\req(Etsk)), (\req(Esk'))  respectively:
\bea(llll)
\vec{w}(s ; k') = \vec{w}(s_1, \ldots, s_{m_E} ; k') & = \otimes_{i=1}^{m_E} {\tt w}_{i,k'}^{s_i}, &
\vec{v}(s ; k') = \vec{v}(s_1, \ldots, s_{m_E} ; k') & = \otimes_{i=1}^{m_E} {\tt v}_{i,k'}^{s_i}. 
\elea(vsv)
By (\req(wvi)), the relation (\req(wect)) holds for $\lambda (s) = \prod_{i=1}^{m_E} s_i$. We shall consider $V_i$ is  a vector space with a Hermitian form induced from ${\tt b}_i^\pm$, and ${\tt w}_{i, k'}^\pm$'s or ${\tt v}_{i, k'}^\pm$'s form a continuous family of $V_i$. Then $\vec{w}(s ; k')$ or $\vec{v}(s ; k')$ in  (\req(vsv)) form an orthonormal basis of ${\cal E}_{{\tt F}, P_a, P_b}$ as in (\req(wect)) or (\req(Ek')).  
By (\req(eoinf)), $H(k')$ can also be expressed by the $sl_2$-operators for $k'=0, \infty$:
\bea(ll)
H(k')& = \alpha + k' \beta + 2N \sum_{i=1}^{m_E} \bigg( (k'-\cos \theta_i)  J^z_{i, \infty} - \sin  \theta_i  J^x_{i, \infty}  \bigg) \\
 &= \alpha + k' \beta + 2N \sum_{i=1}^{m_E} \bigg( (1- k' \cos \theta_i) {\tt j}^z_{i, 0} + k' \sin \theta_i {\tt j}^x_{i, 0} \bigg)  \bigg)  .
\elea(Hk'0)
The second expression in (\req(Hk'0)) is the usual matrix expression of $H(k')$ in literature, e.g. \cite{B89, B08, IG}\footnote{ The matrix form of $H(k')$ in \cite{B89, B08, IG} is with respect to the basis  ${\tt v}^\mp_{i, 0}$, hence differs from the expression here by the $J_x$-similar relation.}. From (\req(HJ)) and (\req(Hk'0)), one then can express the eigenvectors of $H(k')$ for an arbitrary real $k'$ in terms of $H(0)$-eigenvectors or $H(\infty)$-eigenvectors. Indeed, by (\req(vk0)), one finds
$$
\begin{array}{ll}
{\tt w}_{i, k'}^{s_i} &= \sum_{s_i' = \pm 1} {\tt w}_{i, \infty}^{s_i'} \sin (\frac{\varphi_{i, k'} - \theta_i}{2}+  \frac{(s_i - s_i') \pi}{4}),\\
{\tt v}_{i, k'}^{s_i} &= \sum_{s_i' = \pm 1} {\tt v}_{i, 0}^{s_i'} \sin  (\frac{\vartheta_{i, k'}- \pi }{2} +  \frac{(s_i - s_i') \pi}{4}) \\

\end{array}
$$
hence the basis elements at $k'$ in  (\req(vsv)) can be expressed by $\vec{v}(s' ; 0)$ or $\vec{w}(s' ; \infty)$ by\footnote{By (\req(vk0)), $H(k')$-eigenvectors indeed induce a $\RZ$-structure of ${\cal E}_{{\tt F}, P_a, P_b}$ for $k' \in \RZ \cup \{\infty \} (:= {\tt RP}^1)$. The eigenvectors $\vec{w}(s; k')$'s or $\vec{v}(s; k')$'s form a continuous family of orthonormal basis for $k' \neq 0 , \infty$ respectively, but not for all $k' \in {\tt RP}^1$. In the terminology of algebraic geometry,  $H(k')$-eigenvectors form a torsion vector bundle over ${\tt RP}^1$ with a $\RZ$-structure. The bundle is trivial over ${\tt RP}^1 \setminus \{ 0 \}$ and ${\tt RP}^1 \setminus \{ \infty \}$ with local trivial structure using the central bundle $\vec{w}(s' ; \infty)$, $\vec{v}(s' ; 0)$ respectively. The relation (\req(k'0if)) is the expression of $\vec{w}(s; k')$ or $\vec{v}(s; k')$ in the local trivial-frame coordinates.}
\bea(ll)
\vec{w}(s_1, \ldots, s_{m_E} ; k')  & = \sum_{s_1', \ldots s_{m_E}'} \vec{w}(s_1', \ldots, s_{m_E}' ; \infty ) \prod_{i=1}^{m_E} \sin (\frac{\varphi_{i, k'} - \theta_i}{2}+  \frac{(s_i - s_i') \pi}{4}) ,\\
\vec{v}(s_1, \ldots, s_{m_E} ; k') & = \sum_{s_1', \ldots s_{m_E}'} \vec{v}(s_1', \ldots, s_{m_E}' ;0)\prod_{i=1}^{m_E} \sin  (\frac{\vartheta_{i, k'}- \pi }{2} +  \frac{(s_i - s_i') \pi}{4}). \\
\elea(k'0if)
The basis $\vec{w}(s' ;\infty)$'s or $\vec{v}(s' ;0)$'s will be served as the basic $\tau^{(2)}$-eigenvectors of a sector. 
In Subsections \ref{ssec:sl2t2}, \ref{ssec.XXZhw}, and Section \ref{sec.loopt2}, we shall derive a scheme of producing a local-spin-vector form of the basic $\tau^{(2)}$-eigenvectors. The formula (\req(k'0if)) then provides the $k'$-dependent state vectors of CPM in terms of the local spin basis.

Next we consider the relation between the chiral Potts quantum chain $H(k')$  and $H^\dagger(k'^{-1})= H_0^\dagger + k^{-1} H_1^\dagger$ of (\req(HJ)) at $k'^{-1}$. By (\req(Dual)), (\req(duqn)) and (\req(Hinf)), 
the $\tau^{(2) \dagger}$-eigenspace  ${\cal E}^\dagger_{{\tt F}, P_a, P_b}$ of boundary condition $r^* (=Q)$ and charge $Q^* (= r)$ sector can be identified with $\tau^{(2)}$-eigenspace ${\cal E}_{{\tt F}, P_a, P_b}$ through $\psi$.  As the same Onsager-algebra representation by identifying Onsager-generators $H_0= H_1^\dagger, H_1 = H_0^\dagger$, ${\cal E}^\dagger_{{\tt F}, P_a, P_b} = \otimes_{i=1}^{m_E} V_i $ with $V_i$ in (\req(EV)) expressed by the basis as those in (\req(EV)) and (\req(vi)):
\bea(ll)
{\cal E}^\dagger_{{\tt F}, P_a, P_b} = \otimes_{i=1}^{m_E} V_i , ~ V_i = \CZ {\tt b}_i^{\dagger +} + \CZ {\tt b}_i^{\dagger -} = \CZ {\tt w}_{i,k'}^{\dagger +}  + \CZ {\tt w}_{i,k'}^{\dagger -} =\CZ {\tt v}_{i,k'}^{\dagger +}  + \CZ {\tt v}_{i,k'}^{\dagger -} ,
\elea(EVdag)
where the basis ${\tt w}_{i,k'}^{\dagger \pm}$, ${\tt v}_{i,k'}^{\dagger \pm}$ are related to ${\tt b}_i^{\dagger \pm}$  by the same relation in (\req(vi)). The equality $H (k') = k'H^\dagger(k'^{-1})$ yields the equality of linear terms $\alpha = \beta^\dagger, \beta = \alpha^\dagger$  and the Onsager-algebra forms of $H (k')$ and $H^\dagger(k'^{-1})$:
$$
J_i^x = -\cos \theta_i J_i^{x \dagger} + \sin \theta_i J_i^{y \dagger}, ~ -\cos \theta_i J_i^x + \sin \theta_i J_i^y = J_i^{x \dagger} ~ ~ (i=1, \ldots, m_E),
$$
by which the basis ${\tt b}_i^{\dagger \pm}, {\tt b}_i^{\pm }$ of $V_i$ satisfy the uniquely unitary relation (up to $\pm$ sign):
$$
({\tt b}_i^{\dagger +}, {\tt b}_i^{\dagger -}) = ({\tt b}_i^+, {\tt b}_i^-)
\left( \begin{array}{cc}
       0    &    {\rm i} {\rm e}^{{\rm i} \theta_i/2}   \\
-{\rm i} {\rm e}^{-{\rm i} \theta_i /2}&  0
\end{array} \right).
$$
By (\req(vi)) and the relation $\varphi_{i, k'}+\vartheta_{i, k'}=  \theta_i + 3\pi $, one finds 
$$
\begin{array}{ll}
({\tt w}_{i, k'}^+, {\tt w}_{i, k'}^-) =  (-{\tt v}_{i, 1/k'}^{\dagger +}, {\tt v}_{i, 1/k'}^{ \dagger -}) , &
({\tt v}_{i, k'}^+, {\tt v}_{i, k'}^-) =  (-{\tt w}_{i, 1/k'}^{\dagger +}, {\tt w}_{i, 1/k'}^{ \dagger -}) ,
\end{array}
$$
then follows the relation (\req(vec*)) by (\req(vsv)).

We now examine the Onsager algebra structures between the inverse sectors in (\req(conj)). First we consider the case when ${\cal E}_{{\tt F}, P_a, P_b} \in I_\pm$ and ${\cal E}'_{{\tt F}', P'_a, P'_b} \in I_\mp$ in (\req(conj)). By Remark of Lemma \ref{lem:FF'} and (\req(cjlin)), the linear terms of $H(k')$ of ${\cal E}_{{\tt F}, P_a, P_b}$ in (\req(HJ)) and $H^\prime(-k')$ of ${\cal E}'_{{\tt F}', P'_a, P'_b}$ are related by $\alpha + k' \beta = \alpha^\prime - k' \beta^\prime$, and the angles $\theta_i$'s of $H(k')$ in (\req(HJ)) are related to $\theta'_i$'s of $H^\prime(-k')$ by $\theta'_i + \theta_i = \pi$ for $1 \leq i \leq m_E$. Identify ${\cal E}'_{{\tt F}', P'_a, P'_b}$ with ${\cal E}_{{\tt F}, P_a, P_b}$ in (\req(EV)):
$$
{\cal E}_{{\tt F}', P'_a, P'_b} = \otimes_{i=1}^{m_E} V_i , ~ ~ V_i = \CZ {\tt b}_i^{\prime +} + \CZ {\tt b}_i^{\prime -} 
$$
with the factor $V_i$ expressed by
$$
V_i = \CZ {\tt b}_i^{\prime +} + \CZ {\tt b}_i^{\prime -} = \CZ {\tt w}_{i,k'}^{\prime +}  + \CZ {\tt w}_{i,k'}^{\prime -},~ ({\tt w}_{i,k'}^{\prime  +},  {\tt w}_{i, k'}^{\prime -} ) =  ({\tt b}_i^{+ \prime }, {\tt b}_i^{\prime  }-)  \left( \begin{array}{cc}
        {\rm e}^{{\rm i} \varphi^\prime _{i, k'}/2}  &   - {\rm i} {\rm e}^{{\rm i} \varphi^\prime _{i, k'}/2} 
\\ {\rm e}^{-{\rm i} \varphi^\prime _{i, k'}/2} & {\rm i} {\rm e}^{-{\rm i} \varphi^\prime _{i, k'}/2}
\end{array} \right) \frac{1}{\sqrt{2{\rm i}}}
$$
where ${\rm e}^{{\rm i} \varphi^\prime_{i, k'}} :=  \frac{|k'|}{k'} (1- k' \cos \theta^\prime_i - {\rm i} k' \sin \theta^\prime_i)/ \varepsilon (\theta^\prime_i; k')$ as in (\req(vi)). Since $\varepsilon (\theta^\prime_i; -k')= \varepsilon (\theta_i; k')$, we find ${\rm e}^{{\rm i} \varphi^\prime_{i, - k'}+{\rm i} \varphi_{i, k'}} = -1 $, hence $\varphi^\prime_{i, - k'}+ \varphi_{i, k'} = 3 \pi$. With respect to the basis ${\tt b}_i^{\prime \pm}$ of $V_i$, the Hamiltonian $H'(k')$ on ${\cal E}'_{{\tt F}', P'_a, P'_b}$ can be expressed by 
$$
H'(k') =  \alpha' + k' \beta' + 2N  \sum_{i=1}^{m_E} \bigg( (1- \cos \theta'_i ) J^{\prime x}_i + k' \sin \theta'_i J^{\prime y}_i \bigg). 
$$
The relation $H(k') = H'(-k')$, i.e. $H_0 = H'_0, H_1 = -H'_1$, is equivalent to $\alpha = \alpha^\prime,  \beta = - \beta^\prime$, and $J^x_i = J^{\prime x}_i$, $ J^y_i = - J^{\prime y}_i$ for all $i$. This yields the unique (up to $\pm$-sign) unitary relation  between ${\tt b}_i^\pm$ and ${\tt b}_i^{\prime \pm}$:
$$
  ({\tt b}_i^+, {\tt b}_i^-) = (- {\tt b}^{\prime -}_i , - {\tt b}_i^{\prime +} ),
$$
which implies 
$$
({\tt w}_{i, k'}^+,  {\tt w}_{i, k'}^- ) = ({\tt w}_{i,-k'}^{\prime -},  {\tt w}_{i, -k'}^{\prime +} ).
$$
Then follows the isomorphism in (\req(E'Eb)) for $I_\pm$-sectors. For $i$-sector case, the duality isomorphism (\req(vec*)) sends $i_\pm \ni {\cal E}_{{\tt F}, P_a, P_b} ( \subset V_{r, Q})$ and $i_\mp \ni {\cal E}'_{{\tt F}', P'_a, P'_b} (\subset V_{r', Q})$ to $I_\pm \ni {\cal E}^\dagger_{{\tt F}, P_a, P_b} (\subset V_{r^*, Q^*})$ and $I_\mp \ni {\cal E}^{\prime \dagger }_{{\tt F}', P'_a, P'_b} (\subset V_{r^*, Q^{\prime *}})$ respectively with $ Q^*=r, Q^{\prime *} = r'$and $r^*=Q$. Then the inverse relation of ${\cal E}_{{\tt F}, P_a, P_b}$ and ${\cal E}'_{{\tt F}', P'_a, P'_b}$ in (\req(E'Eb)) follows from the inverse relation of ${\cal E}^\dagger_{{\tt F}, P_a, P_b}$ and ${\cal E}^{\prime \dagger }_{{\tt F}', P'_a, P'_b}$.

\subsection{The $sl_2$-loop algebra structure of superintegrable $\tau^{(2)}$-eigenspaces \label{ssec:sl2t2}}
In this subsection, we define a loop-algebra $sl_2[z, z^{-1}]$-structure on ${\cal E}_{{\tt F}, P_a,P_b}$ based on the $(\stackrel{m_E}{\oplus} sl_2)$-structure (\req(vsv)) of ${\cal E}_{{\tt F}, P_a,P_b}$ at $k'=0$ or $\infty$. The choice of $k'$ is suggested by the structure of Bethe state of ${\cal E}_{{\tt F}, P_a,P_b}$ later appeared in (\req(Betv)) and (\req(BetvF)) of Subsection \ref{ssec.XXZhw}. For convenience, we introduce the following notions for the basic $\tau^{2)}$-eigenvectors:
\bea(l)
\vec{\bf u}(s) (= \vec{\bf u}(s_1, \ldots, s_{m_E}) = \left\{ 
\begin{array}{ll} \vec{w}(s ; \infty ) & {\rm if} ~ ({\tt F}, P_a,P_b) \in I_\pm , \\
                  \vec{v}(s ; 0 ) & {\rm if} ~ ({\tt F}, P_a,P_b) \in i_\pm . \end{array} \right. 
\elea(usl)
Using the above basis, the $(\stackrel{m_E}{\oplus} sl_2)$-structure of ${\cal E}_{{\tt F}, P_a,P_b}$ consists of operators ${\bf h}_i, {\bf e}^\pm_i$ for $1 \leq i \leq m_E$ defined by
\bea(l)
{\bf h}_i: \vec{\bf u}(s_1, \ldots, s_i, \ldots, s_{m_E}) \mapsto s_i \vec{\bf u}(s_1, \ldots, s_i, \ldots, s_{m_E}) , \\
{\bf e}^\pm_i: \vec{\bf u}(s_1, \ldots, s_i, \ldots, s_{m_E})\mapsto \frac{(1\mp s_i)}{2} \vec{\bf u}(s_1, \ldots, -s_i,\ldots, s_{m_E}). 
\elea(hei)
We use the bold-face character ${\bf h}_i, {\bf e}^\pm_i$ as the generators of a $sl_2$-product structure of ${\cal E}_{{\tt F}, P_a,P_b}$ to distinguish from $h_i, e^\pm_i$ of Subsection \ref{ssec.OA} in the discussion of Onsager-algebra structure of ${\cal E}_{{\tt F}, P_a,P_b}$. The $sl_2$-loop-algebra structure of ${\cal E}_{{\tt F}, P_a,P_b}$ will be defined through the operators ${\bf h}_i, {\bf e}^\pm_i$ in (\req(hei)).
Indeed, using the product structure in (\req(EV)), $\vec{\bf u}(s)= \otimes_{i=1}^{m_E} {\bf u}_i^{s_i}$ with ${\bf u}_i^{s_i}= {\tt w}_{i, \infty'}^{s_i}$ or ${\tt v}_{i, 0}^{s_i}$, then $({\bf h}_i, {\bf e}^\pm_i) = (h_{i, \infty} , e^\pm_{i, \infty}$) or $({\tt h}_{i, 0} ,  {\tt e}^\pm_{i, 0})$ according $({\tt F}, P_a,P_b) \in I_\pm$ or $i_\pm$ respectively.
For $({\tt F}, P_a,P_b) \in I_\pm \cap i_\pm$ or $I_\pm \cap i_\mp$, there are two $sl_2$-product structures of ${\cal E}_{{\tt F}, P_a,P_b}$, one induced from $\vec{w}(s ; \infty )$, denoted by $\vec{\bf u}_\infty(s), {\bf h}_{i, \infty}, {\bf e}^\pm_{i, \infty}$, and the other, $\vec{\bf u}_0(s), {\bf h}_{i, 0}, {\bf e}^\pm_{i, 0}$, induced from $\vec{v}(s ; 0 )$. By (\req(vec*)), one has $\Psi (\vec{\bf u}_\infty(s)) = \prod_{i=1}^{m_E}(-s_i ) \vec{\bf u}_0(s)$.
For a  $sl_2$-product structure of ${\cal E}_{{\tt F}, P_a,P_b}$ in (\req(usl)), we use the evaluation polynomials ${\tt P}_{\rm ev} (\xi )$ in (\req(Ptrt)) to  define a $sl_2[z, z^{-1}]$-structure of ${\cal E}_{{\tt F}, P_a,P_b}$ as follows. For ${\bf g}= {\bf e}^\pm, {\bf h} \in sl_2$, the element ${\bf g} z^n$ in $sl_2[z, z^{-1}]$ will be denoted by ${\bf g}(n)= {\bf g} z^n$ for $n \in \ZZ$. It is well known that the loop algebra  is generated by the Chevalley basis, $
{\sc E}_1 = {\bf e}^+ (0),  {\sc F}_1 = {\bf e}^- (0), {\sc E}_0 = {\bf e}^- (-1),  {\sc F}_0 = {\bf e}^+ (1),  {\sc H}_1 = -{\sc H}_0 = {\bf h} (0)$.
Indeed, using ${\bf h}(1) = [{\bf e}^+ (1) , {\bf e}^- (0) ], {\bf h}(-1) = [{\bf e}^+ (0) , {\bf e}^- (-1) ]$, the rest loop-algebra operators are given by 
${\rm ad}_{{\bf h}(1)}^n ({\bf e}^\pm (0))= (\pm 2)^n {\bf e}^\pm (n)$ , ${\rm ad}_{{\bf h}(-1)}^n ({\bf e}^\pm (0))= (\pm 2)^n {\bf e}^\pm (-n)$ for $n \geq 0$. The loop-algebra structure on ${\cal E}_{{\tt F}, P_a,P_b}$ is obtained by  evaluating the loop-variable $z$ on the inverse roots ${\tt a}_i$'s of ${\tt P}_{\rm ev} (\xi )$ in (\req(Ptrt)):
\be
sl_2[z, z^{-1}] \longrightarrow sl ({\cal E}_{{\tt F}, P_a, P_b}), ~ ~ ~ {\bf g}(n) \mapsto \sum_{i=1}^{\rm m} {\bf g}_i {\tt a}^n_i , ~ ~ ({\bf g} = {\bf e}^\pm, {\bf h} , ~ ~ n \in \ZZ), ~  ~ {\rm m} := m_E ,
\ele(lk') 
where ${\bf g}_i$'s are the operators in (\req(hei)). We shall also denote the image of ${\bf g}(n)$ in above by the same letter ${\bf g}(n)$ if no confusion could arise. Then the ${\cal E}_{{\tt F}, P_a, P_b}$-operators ${\bf g}(n)$'s satisfy the loop-algebra condition. In particular, the Serre relation holds: $[ {\bf e}^\pm (j), {\bf e}^\pm (j), [ {\bf e}^\pm (j), {\bf e}^\mp (k)]]]=0$ for $j \neq k$ and $j, k =0,1$. 
One may express the product operators in (\req(hei)) in terms of loop-operators in (\req(lk')). Indeed, we define the polynomials of degree $({\rm m}-1)$ associated with ${\tt P}_{\rm ev} (\xi ), {\tt P}'_{\rm ev} (\xi )$ in (\req(Ptrt)) (\req(P'ev)):
$$
\begin{array}{ll}
 {\tt P}_{{\rm ev}, j}(\xi ) = \frac{{\tt P}_{\rm ev} (\xi )}{ 1- {\tt a}_j \xi  }  , & {\tt P}'_{{\rm ev}, j}(\xi ) = \frac{{\tt P}'_{\rm ev} (\xi )}{ 1- {\tt a}_j^{-1} \xi} , ~ ~ (1 \leq j \leq {\rm m}),
\end{array}
$$
which are related by ${\tt P}'_{{\rm ev}, j}(\xi ) = {\tt P}_{{\rm ev}, j}(\xi^{-1} ) (- \xi)^{{\rm m}-1} \prod_{i \neq j}{\tt a}_i^{-1} $. The inverse of relations (\req(lk')) for $n_0 \leq n \leq n_0+{\rm m}-1$ with a fixed $n_0 \in \ZZ$, 
$$
\begin{array}{l}
({\bf g}(n_0), \ldots, {\bf g}(n), \ldots, {\bf g}(n_0+{\rm m}-1)) = 
({\bf g}_1, {\bf g}_2, \ldots, {\bf g}_{\rm m})  \left( \begin{array}{ccccc}
     {\tt a}^{n_0}_1    & \ldots &{\tt a}^{n}_1 &\ldots & {\tt a}^{n_0+{\rm m}-1}_1   \\  {\tt a}^{n_0}_2 & \ldots &{\tt a}^{n}_2 &\ldots &  {\tt a}^{n_0+{\rm m}-1}_2 \\ \vdots & \ldots & \vdots  &\ldots&\vdots  \\ {\tt a}^{n_0}_{\rm m} & \ldots &{\tt a}^{n}_{\rm m} &\ldots &  {\tt a}^{n_0+{\rm m}-1}_{\rm m}
\end{array} \right) , 
\end{array}
$$
yields  
\be
({\bf g}_1, {\bf g}_2, \ldots, {\bf g}_{\rm m}) = ({\bf g}(n_0), \ldots, {\bf g}(n), \ldots, {\bf g}(n_0+{\rm m}-1)) \left( \begin{array}{cccc}
    \frac{c'_{0, 1}}{{\tt a}^{n_0}_1}  &\frac{c'_{0, 2}}{{\tt a}^{n_0}_2} & \ldots & \frac{c'_{0, {\rm m}}}{{\tt a}^{n_0}_{\rm m}}   \\  
  \vdots & \vdots& \vdots &  \vdots  \\ 
   \frac{c'_{ k , 1}}{{\tt a}^{n_0}_1} & \frac{c'_{ k , 2}}{{\tt a}^{n_0}_2} &\vdots &\frac{c'_{k , {\rm m}}}{{\tt a}^{n_0}_{\rm m}}  \\ 
 \vdots & \vdots  & \vdots &  \vdots \\ 
  \frac{c'_{{\rm m}-1, 1}}{{\tt a}^{n_0}_1}&\frac{c'_{{\rm m}-1, 2}}{{\tt a}^{n_0}_2}   & \ldots &  \frac{c'_{{\rm m}-1, {\rm m}}}{{\tt a}^{n_0}_{\rm m}} 
\end{array} \right)  ,
\ele(gign')
where $k := n - n_0$, and $c'_{i, j}$'s are defined by $\frac{{\tt P}'_{{\rm ev}, j}(\xi ) }{{\tt P}'_{{\rm ev}, j}({\tt a}_j )} = \sum_{i=0}^{{\rm m}-1} c'_{i, j} \xi^i$.  Equivalently, the inverse of $(\req(lk'))_{n_0-{\rm m}+1 \leq n \leq n_0}$ yields
\be
({\bf g}_1, {\bf g}_2, \ldots, {\bf g}_{\rm m}) = ({\bf g}(n_0), \ldots, {\bf g}(n), \ldots, {\bf g}(n_0-{\rm m}+1)) \left( \begin{array}{cccc}
     \frac{c_{0, 1}}{{\tt a}^{n_0}_1}  &\frac{c_{0, 2}}{{\tt a}^{n_0}_2} & \ldots & \frac{c_{0, {\rm m}}}{{\tt a}^{n_0}_{\rm m}}   \\  
  \vdots & \vdots& \vdots &  \vdots  \\ 
   \frac{c_{ k , 1}}{{\tt a}^{n_0}_1} & \frac{c_{ k , 2}}{{\tt a}^{n_0}_2} &\vdots &\frac{c_{k , {\rm m}}}{{\tt a}^{n_0}_{\rm m}}    \\ 
 \vdots & \vdots  & \vdots &  \vdots \\ 
  \frac{c_{{\rm m}-1, 1}}{{\tt a}^{n_0}_1}&\frac{c_{{\rm m}-1, 2}}{{\tt a}^{n_0}_2}   & \ldots &  \frac{c_{{\rm m}-1, {\rm m}}}{{\tt a}^{n_0}_{\rm m}}   
\end{array} \right)  ,
\ele(gign)
where $k := n_0-n$ , and  $\sum_{i=0}^{{\rm m}-1} c_{i, j} \xi^i = \frac{{\tt P}_{{\rm ev}, j}(\xi ) }{{\tt P}_{{\rm ev}, j}({\tt a}^{-1}_j )}$.  Note that by $c_{k, j}' = c_{{\rm m}-1-k, j} {\tt a}^{-{\rm m}+1}_j$, the relation $(\req(gign))_{n_0'}$ is the same as $(\req(gign'))_{n_0}$ with $n_0' = n_0+{\rm m}-1$. The relation (\req(gign')) or (\req(gign)) enables one to construct a local spin operator form for $sl_2$-operators ${\bf g}_i$'s from the loop-algebra operators ${\bf g}(n)$'s  through the polynomial ${\tt P}_{\rm ev}(\xi )$ if the local-spin-operator form of ${\bf g}(n)$'s are found. This method has been employed in the study of ground state sector in \cite{AuP, AuP7}.   For this purpose, we consider the following loop-algebra currents  on ${\cal E}_{{\tt F}, P_a,P_b}$ (\cite{FM01} (1.20), \cite{R06F} (4.35)):
\be
{\bf E}^- (\xi) = \sum_{n=0}^\infty {\bf e}^- (n) \xi^n , ~ {\bf E}^+ (\xi) = \sum_{n=1}^\infty {\bf e}^+ (n) \xi^{n-1} ,
\ele(Epm)
whose poles both coincide with zeros of ${\tt P}_{\rm ev}(\xi )$: 
\bea(l)
{\tt P}_{\rm ev}(\xi ){\bf E}^- (\xi)  = \sum_{i=1}^m e^-_i \prod_{j \neq i} (1- {\tt a}_j \xi) =\sum_{k=0}^{m-1} (-1)^k \rho^-_k \xi^k ,   \\
{\tt P}_{\rm ev}(\xi ){\bf E}^+ (\xi)  = \sum_{i=1}^m  e^+_i {\tt a}_i \prod_{j \neq i} (1- {\tt a}_j \xi) =\sum_{k=0}^{m-1} (-1)^k \rho^+_k \xi^k ,
\elea(Eex)
with $\rho^-_0 = {\bf e}^-(0), \rho^-_{m-1} = (\prod_{i=1}^m {\tt a}_i){\bf e}^- ( - 1)$, and 
$\rho^+_0 = {\bf e}^+(1), \rho^+_{m-1} = (\prod_{i=1}^m {\tt a}_i){\bf e}^+ (0)$. The $sl_2$-operators ${\bf e}^\pm_i ~ (1 \leq i \leq {\rm m})$ are related to ${\bf E}^\pm (\xi)$  by (\req(gign)) respectively. 
For the  problem of CPM-eigenvectors in ${\cal E}_{{\tt F}, P_a,P_b}$, it suffices to find an expression of the following states and operators in terms of local spin basis:
\bea(lll)
{\rm Sector} ~ I_+, i_+ :& (\vec{\bf u}(+, \ldots, +) , ~ {\bf E}^- (\xi))  \\
{\rm Sector} ~ I_-, i_- : & (\vec{\bf u}(-, \ldots, -) , ~ {\bf E}^+ (\xi)) 
\elea(vvE)
where $\vec{\bf u}(s)$'s are defined in (\req(usl)). 
Using (\req(gign)), one can derive the local-spin-operator form $sl_2$-operators ${\bf e}^\pm_i ~ (1 \leq i \leq m_E)$ from ${\bf E}^\pm (\xi)$ respectively. Then the local-spin-vector form of $\vec{\bf u}(s_1, \ldots, s_{m_E})$ are given by 
\bea(l)
\vec{\bf u}(s_1, \ldots, s_{m_E}) = \left \{ \begin{array}{ll}(\prod_{s_i = -1}{\bf e}^-_i) \vec{\bf u}(+, \ldots, +) & {\rm for ~ sector} ~ I_+, i_+ , \\
(\prod_{s_i =1}{\bf e}^+_i) \vec{\bf u}(-, \ldots, - ) & {\rm for ~ sector} ~ I_-, i_- . \end{array} \right.
\elea(uspin)
Using (\req(k'0if)), one then express the $k'$-dependent state vectors of CPM in terms of the local spin basis. We are going to construct the local spin form of states and operators in (\req(vvE)) by the algebraic-Bethe-ansatz method through the equivalent XXZ-chain of the superintegrable $\tau^{(2)}$-model as in \cite{NiD, ND08, R06F}. The vector in (\req(vvE)) has been identified with the Bethe state in ${\cal E}_{{\tt F}, P_a,P_b}$ in \cite{R09}, which will be recalled in Subsection  \ref{ssec.XXZhw}. The current in (\req(vvE)), up to a scale polynomial,  is identified with the Fabricius-McCoy current with the local spin form, which will be constructed in  Section \ref{sec.loopt2}. \par \vspace{.1in} \noindent
{\bf Remark}. There are two-parameter isomorphisms of $sl_2[z, z^{-1}]$, $\widetilde{\bf e}^\pm (0) = a^{\pm 1} {\bf e}^\pm (0), \widetilde{\bf e}^\pm (\pm 1)= b^{\pm 1} {\bf e}^\pm (\pm 1)$ for $a, b \in \CZ^*$, and the identification of mode basis: $
\widetilde{\bf e}^\pm (n) = a^{-n \pm 1}b^n {\bf e}^+(n)$, $\widetilde{\bf h} (n) = a^{-n}b^n {\bf h} (n)$. 
The corresponding $(\otimes^{\rm m} sl_2)$-structures on ${\cal E}_{{\tt F}, P_a,P_b}$  are related by
$$
\begin{array}{lll}
(\widetilde{\bf e}^\pm_1,  \ldots, \widetilde{\bf e}^\pm_m) & = a^{\pm 1} ({\bf e}^\pm_1,  \ldots, {\bf e}^\pm_m)  \bigg({\tt a}^k_j \bigg) {\rm dia}[a^{-k}b^k]  \bigg( c'_{k,l}\bigg) & = a^{\pm 1}({\bf e}^\pm_1,  \ldots, {\bf e}^\pm_m) \bigg({\tt a'}^k_j\bigg) {\rm dia}[a^k b^{-k}] \bigg(c_{k , j}\bigg) , \\
(\widetilde{\bf h}_1,  \ldots, \widetilde{\bf h}_m) & = a^{\pm 1} ({\bf h}_1,  \ldots, {\bf h}_m)  \bigg({\tt a}^k_j \bigg) {\rm dia}[a^{-k}b^k]  \bigg( c'_{k,l}\bigg) &= ({\bf h}_1,  \ldots, {\bf h}_m) \bigg({\tt a'}^k_j\bigg) {\rm dia}[a^k b^{-k}] \bigg(c_{k , j}\bigg) , 
\end{array} 
$$
where ${\tt a'}^k_j, c'_{k,l}, {\tt a}^k_j, c_{k , j}$ are in (\req(gign')), (\req(gign)) with indices $0 \leq k \leq {\rm m}-1$ and $1 \leq j \leq {\rm m}$. The currents in (\req(Epm)) are connected by $\widetilde{\bf E}^- (\xi) = a^{-1} {\bf E}^- (a^{-1} b \xi), \widetilde{\bf E}^+ (\xi) = a {\bf E}^+ (ab^{-1} \xi)$.

\section{Degeneracy of XXZ chains and Superintegrable $\tau^{(2)}$-models \label{sec.Degt2}}
\setcounter{equation}{0}
In this section, we investigate the degeneracy of a superintegrable $\tau^{(2)}$-model through its equivalent spin-$\frac{N-1}{2}$ XXZ chain as in \cite{NiD, ND08, R06F, R09}. We shall recall the result in \cite{R09} Section 4.2 about the realization of the Bethe state as a basic $\tau^{(2)}$-eigenvector of an Onsager sector. The inversion correspondence of quantum spaces in superintegrable CPM will be also identified through the local operators in XXZ chains.

\subsection{XXZ-chains with $U_q(sl_2)$-cyclic representation \label{ssec.XXZ}}
Using the generators $K^\frac{\pm 1}{2}, e^{\pm}$ of quantum group $U_q (sl_2)$ for an arbitrary $q$ (with relations $ K^{\frac{1}{2}} e^{\pm } K^\frac{-1}{2}  =  q^{\pm 1} e^{\pm}, [e^+ , e^- ] = \frac{K-K^{-1}}{q - q^{-1}}$), one finds a two-parameter family of  $L$-operators
\be
{\bf L} (s)   =  \left( \begin{array}{cc}
         \rho^{-1} \nu^\frac{1}{2} s K^\frac{-1}{2}   -  \nu^\frac{-1}{2} s^{-1} K^\frac{1}{2}   &  (q- q^{-1}) e^-    \\
        (q - q^{-1}) e^+ &  \nu^\frac{1}{2} s K^\frac{1}{2} -  \rho \nu^\frac{-1}{2} s^{-1} K^\frac{-1}{2} 
\end{array} \right) ~ ~ ,0 \neq \rho , \nu \in \CZ,
\ele(6vL)
of YB solutions:
$$
R_{\rm 6v} (s/s') ({\bf L}(s) \bigotimes_{aux}1) ( 1
\bigotimes_{aux} {\bf L}(s')) = (1
\bigotimes_{aux} {\bf L}(s'))( {\bf L}(s)
\bigotimes_{aux} 1) R_{\rm 6v} (s/s'), 
$$
where $R_{\rm 6v}$ is the symmetric six-vertex $R$-matrix
$$
R_{\rm 6v} (s) = \left( \begin{array}{cccc}
        s^{-1} q - s q^{-1}  & 0 & 0 & 0 \\
        0 &s^{-1} - s & q - q^{-1} &  0 \\ 
        0 & q -q^{-1} &s^{-1} - s & 0 \\
     0 & 0 &0 & s^{-1} q - s q^{-1} 
\end{array} \right) .
$$
The $q^{-2r}$-twisted trace
\be
{\bf t} (s) = {\bf A} (s) + q^{-2r} {\bf D} (s), ~ s \in \CZ,
\ele(t6v)
of the monodromy matrix of size $L$,
\be
\bigotimes_{\ell=1}^L {\bf L}_\ell (s)  =  \left( \begin{array}{cc}
        {\bf A} (s)  & {\bf B} (s) \\
        {\bf C} (s) & {\bf D} (s)
\end{array} \right), ~ ~ {\bf L}_\ell= {\bf L},
\ele(M6V)
form a commuting family for $s \in \CZ$ with coefficients in $\stackrel{L}{\bigotimes} U_q (sl_2)$, which commute with $K^\frac{1}{2}:=\stackrel{L}{\bigotimes}K_\ell^\frac{1}{2}$.
The leading and lowest terms of entries in (\req(M6V)),\footnote{ ${\sf A}_+, {\sf D}_-, {\sf B}_\pm, {\sf C}_\pm$ here differ those in \cite{R09} by some scales. Indeed, those in (\req(pmTerm)) are equal to $\rho^{-L}{\sf A}_+, , \rho^{L} {\sf D}_-, \nu^\frac{\pm 1}{2}{\sf B}_\pm , \nu^\frac{\pm 1}{2}{\sf C}_\pm$ in \cite{R09}.}
$$
\begin{array}{ll}
{}^{{\sf A}_\pm}_{{\sf D}_\pm} = \lim_{s^{\pm 1}  \rightarrow \infty} \nu^\frac{\mp L}{2}  (\pm s)^{\mp L}  {}^{\bf A}_{\sf D}(s), &{}^{{\sf B}_\pm}_{{\sf C}_\pm} = \lim_{s^{\pm 1} \rightarrow \infty}  \frac{\nu^\frac{\mp (L-1)}{2} (\pm s)^{\mp (L-1)}}{ q-q^{-1}}{}^{\bf B}_{\sf C}(s), 
\end{array}
$$
give rise to the quantum affine algebra $U_q (\widehat{sl}_2)$:
\bea(l)
{\sf A}_- = {\sf D}_+ = K^\frac{1}{2}, ~ ~  \rho^L {\sf A}_+ = \rho^{-L} {\sf D}_- =  K^\frac{-1}{2}, \\
{\sf B}_\pm  =\rho^\frac{\mp (L-1)}{2} \sum_{i=1}^L  \rho^{\frac{L+1}{2}-i}  \underbrace{K^{\frac{\mp 1}{2}} \otimes \cdots \otimes K^{\frac{\mp 1}{2}}}_{i-1}\otimes e^- \otimes  \underbrace{K^{ \frac{\pm 1}{2}} \otimes \cdots \otimes K^{ \frac{\pm 1}{2}}}_{L-i}, \\
{\sf C}_\pm = \rho^\frac{\mp (L-1)}{2} \sum_{i=1}^L  \rho^{\frac{-L-1}{2}+i}  \underbrace{K^{\frac{\pm 1}{2}} \otimes \cdots \otimes K^{\frac{\pm 1}{2}}}_{i-1}\otimes e^+ \otimes  \underbrace{K^{ \frac{\mp 1}{2}} \otimes \cdots \otimes K^{ \frac{\mp 1}{2}}}_{L-i} ,
\elea(pmTerm)
with the Chevalley generators, $
k_0^{-1} =k_1  = K$, $e_1  = {\sf C}_+ , f_1  = {\sf B}_-$, $e_0 =  {\sf B}_+ , f_0 = {\sf C}_-$, 
and the Hopf-algebra structure: 
$$
\begin{array}{ll}
\bigtriangleup (k_i) = k_i \otimes k_i, ~ ~ ~ i=0, 1 ,& \\
\bigtriangleup (e_1 ) =  k^\frac{1}{2}_1 \otimes e_1    +   \rho^{- 1} e_1 \otimes k^\frac{1}{2}_0 , &
\bigtriangleup (f_1 ) =  k^\frac{1}{2}_1 \otimes f_1    +  \rho f_1 \otimes k^\frac{1}{2}_0 , \\
\bigtriangleup (e_0 ) =  \rho^{-1} k^\frac{1}{2}_0 \otimes e_0    +  e_0 \otimes k^\frac{1}{2}_1 , &
\bigtriangleup (f_0 ) =  \rho k^\frac{1}{2}_0 \otimes f_0    +  f_0 \otimes k^\frac{1}{2}_1 .
\end{array}
$$
In particular, one obtains the well-known homogeneous XXZ chain of spin-$\frac{d-1}{2}$  by setting $\rho =1 , \nu = q^{d-2}$ and the spin-$\frac{d-1}{2}$ (highest weight) representation of $U_q (sl_2)$ on  $\CZ^d = \oplus_{k=0}^{d-1} \CZ {\bf e}^k $:
\be
K^{\frac{1}{2}} ({\bf e}^k) = q^{\frac{d-1-2k}{2}} {\bf e}^k , \ \ e^+ ( {\bf e}^k ) = [ k  ] {\bf e}^{k-1} , \ \ e^-( {\bf e}^k ) = [ d-1-k ] {\bf e}^{k+1},
\ele(spinrp)
where $[n] (=[n]_q) = \frac{q^n - q^{-n}}{q- q^{-1}}$ and $ e^+ ( {\bf e}^{0} ) = e^- ( {\bf e}^{d-1} )= 0$ (see, e.g. \cite{KiR, R06Q, R06F} and references therein).

When the anisotropic parameter $q$ in (\req(6vL)) is a $N$th primitive root of unity, there is a three-parameter family of $U_q ( sl_2)$-cyclic representation on $\CZ^N$, $s_{\phi, \phi^\prime, \varepsilon}$ labeled by non-zero complex numbers $\phi, \phi^\prime, \varepsilon$: 
\bea(ll)
K^\frac{1}{2} \widehat{|k} \rangle =  q^{k+\frac{\phi^\prime-\phi}{2}} \widehat{|k} \rangle , & \\
e^+  \widehat{|k} \rangle =  q^{ \varepsilon} \frac{  q^{\phi - k}-  q^{- \phi + k}  }{q-q^{-1}} \widehat{|k + 1 }\rangle, &
e^-  \widehat{|k} \rangle =  q^{ -\varepsilon} \frac{ q^{\phi^\prime  +k}-  q^{- \phi^\prime  - k}  }{q-q^{-1}} \widehat{|k - 1} \rangle ,  \\
\elea(crep)
(see, e.g. \cite{DJMM, DK}), where $\widehat{|k} \rangle ~ (k \in \ZZ_N)$ are the Fourier basis of $\CZ^N$ in (\req(Fb)). 
With the cyclic representation $s_{\phi, \phi^\prime, \varepsilon}$ on ${\bf L}$ in (\req(M6V)), the monodromy matrix 
\be
\bigotimes_{\ell=1}^L {\cal L}_\ell (s)  =  \left( \begin{array}{cc}
        {\cal A} (s)  & {\cal B} (s) \\
        {\cal C} (s) & {\cal D} (s)
\end{array} \right),  ~ ~ {\cal L}_\ell (s) = s_{\phi , \phi^\prime, \varepsilon} {\bf L} (s),
\ele(6vM)
gives rise to the transfer matrix of XXZ chain for the $U_q (sl_2)$-cyclic representation $s_{\phi , \phi^\prime, \varepsilon}$  with the boundary condition (\req(sBy)):
\be
{\cal T}(s) =   {\cal A} (s) + q^{-2r} {\cal D} (s) =  (\stackrel{L}{\otimes} s_{\phi , \phi^\prime, \varepsilon}) {\bf t}(s).
\ele(TcXZ)
It is known that the above XXZ chains of $U_q (sl_2)$-cyclic representation  (\req(TcXZ))  are equivalent to the $\tau^{(2)}$-models  (\req(tau2)) \cite{R0806}. For simplicity, hereafter in this paper we shall consider only the  odd $N$ case for  the $N$th root-of-unity $q$ :
$$
N = 2M+1, ~ ~ q:= \omega^M  ~ ( = \omega^\frac{-1}{2}).
$$
Now the cyclic representation (\req(crep)) can be expressed by Weyl operators (\req(XZF)):
$$
\begin{array}{lll}
K^\frac{1}{2} = q^\frac{\phi^\prime- \phi}{2}  \widehat{Z}^\frac{-1}{2},& 
e^+  = q^{ \varepsilon} \frac{(  q^{\phi +1} \widehat{Z}^{\frac{ 1}{2}} -  q^{- \phi -1} \widehat{Z}^{  \frac{- 1}{2}})\widehat{X}}{q-q^{-1}} &
e^-  =q^{ -\varepsilon} \frac{(  q^{\phi^\prime +1} \widehat{Z}^{  \frac{- 1}{2}} -  q^{- \phi^\prime -1} \widehat{Z}^{ \frac{ 1}{2}})\widehat{X}^{-1}}{q-q^{-1}}.
\end{array}
$$
By setting $t = s^2$, up to the gauge transform ${\rm dia}[1, -sq]$, the modified $L$-operator $-s \nu^\frac{1}{2} K_\ell^{\frac{-1}{2}} {\cal L}_\ell (s)$ of (\req(6vM)) is equivalent to  $L_\ell (t)$ in (\req(Mont2)) by the identification of parameters:
$$
\begin{array}{l}
{\sf a} = \rho \nu^\frac{-1}{2}  q^{\frac{-\phi - \phi^\prime}{2} - \varepsilon} , ~ \omega {\sf a} {\sf a'} = \rho^2 \nu^{-1}, ~  {\sf b}  = \nu^\frac{-1}{2} q^{\frac{-\phi - \phi^\prime}{2}+ \varepsilon},  ~   {\sf b }{\sf b'} = \nu^{-1},  ~ {\sf c}  = \rho^{-1} q^{\phi - \phi^\prime} , 
\end{array}
$$
in which case, the $\tau^{(2)}(t;p)$ in (\req(tau2)) and $T (s)$ in (\req(TcXZ)) are related by 
$$
\tau^{(2)}(t; p) = (-q^{-1}s)^L \nu^\frac{L}{2}K^{\frac{-1}{2}} {\cal T}(q^{-1}s) , ~ ~ ~ t = s^2.
$$
In particular, the XXZ chain for $(\rho, \nu) = (\omega^{m-M},1)$, $(\phi, \phi^\prime , \varepsilon) \equiv (-1-m,  m, 
M ) \pmod{N}$ in the $L$-operator of (\req(6vM)):
\be 
{\cal L}_\ell (s)   =  \left( \begin{array}{cc}
         q^{1+2m} s K^\frac{-1}{2}   -   s^{-1} K^\frac{1}{2}   &  (q- q^{-1}) e^-    \\
        (q - q^{-1}) e^+ &   s K^\frac{1}{2} -  q^{-1-2m}   s^{-1} K^\frac{-1}{2} 
\end{array} \right) 
\ele(xxzsp) 
with the  $U_q (sl_2)$-representation   
\bea(lll)
K^\frac{1}{2} =  q^{\frac{1}{2} + m}\widehat{Z}^\frac{-1}{2}, & 
e^+  = q^\frac{-1}{2} \frac{(q^{-m}\widehat{Z}^\frac{1}{2}   -q^{m } \widehat{Z}^\frac{-1}{2}  )}{q-q^{-1}}\widehat{X} ,
& e^- =q^\frac{1}{2} \frac{(q^{1+m} \widehat{Z}^\frac{-1}{2}-  q^{-1-m}\widehat{Z}^\frac{1}{2}) }{q-q^{-1}}\widehat{X}^{-1},
\elea(KXZ)
is equivalent to the superintegrable $\tau^{(2)}$-model (\req(hsupL)) by the gauge ${\rm dia}[1, -sq]$ and identification of spectral parameters: ${\tt t}= s^2$ as in \cite{R075}. Denote the local operators in the second Onsager-operator $H_1$ of (\req(H01)) by\footnote{The $S^z$ in Section 4.2 of \cite{R09} differs from $S^z$ here by a minus sign.} 
\bea(ll)
H_1 = 2 \sum_\ell S^z_\ell , ~ ~ ~ 
S^z (=S^z_\ell) := - \sum_{j=1}^{N-1} \frac{ \omega^{m j }\widehat{Z}_\ell^j }{1-\omega^{-j}} .
\elea(H1S)
Using the equality $\sum_{j=1}^{N-1} \omega^{kj}(1-\omega^{-j})^{-1} = (N-1 -2k)/2$  for $ 0\leq k \leq N-1$, one finds
\bea(ll)
S^z ({\bf e}^k) = (k- \frac{N-1}{2}) {\bf e}^k , ~ ~ ~ {\bf e}^k:= \widehat{|k-m} \rangle ~ ~  ~ (k=0, \ldots, N-1).
\elea(ek)
In terms of the above basis ${\bf e}^k$'s, (\req(KXZ)) is expressed by
\bea(lll)
K^\frac{1}{2}({\bf e}^k) =  q^{k- \frac{N-1}{2}} {\bf e}^k, &
e^+  ({\bf e}^k) = q^\frac{-1}{2} [N-1-k] {\bf e}^{k+1} , &
e^-({\bf e}^k)  =q^\frac{1}{2} [k] {\bf e}^{k-1},  
\elea(spN-1)
which is equivalent  to the spin-$\frac{N-1}{2}$ (highest weight) $U_q(sl_2)$-representation (\req(spinrp)) by the  similar isomorphism of $\CZ^N$, ${\bf e}^k \mapsto q^\frac{k}{2} {\bf e}^{N-1-k}$. Note that the operator $K (= \prod_\ell K_\ell)$ is now expressed by $K= q^{H_1}$. It is known that  conjugation of the spin-inversion of local states, $\jmath ({\bf e}^k) = {\bf e}^{N-1-k} ~ ( 0 \leq k \leq N-1 )$, of the spin representation (\req(spN-1)) is given by 
$$
\jmath \cdot K^\frac{1}{2} \cdot \jmath = K^\frac{-1}{2}, ~   \jmath \cdot e^\pm \cdot \jmath = q^{\mp 1} e^\mp .
$$
Hence the local $L$-operator ${\cal L}_\ell (s)$ in (\req(xxzsp)) at the site $\ell$ is gauge-equivalent to the transpose of ${\cal L}_{\ell} (-q^{-1-2m}s^{-1})$ by the diagonal matrix $ {\rm dia}[q, 1 ]$ under the conjugation of the local spin-inversion operator $\jmath_\ell$: 
$$
\jmath_\ell \cdot {\cal L}_\ell (s) \cdot \jmath_\ell = {\rm dia}[q, 1 ] ~ {\cal L}_{\ell} (-q^{-1-2m}s^{-1})^{t} {\rm dia}[q^{-1}, 1 ] . 
$$
Since that the monodromy matrix for the transpose $L$-operator ${\cal L}_{\ell} (s)^{t}$ is conjugate to the transpose of the monodromy matrix (\req(6vM)) under the transformation of quantum space $\bigotimes^L \CZ^N$ sending $\otimes_{\ell=1}^L v_\ell$ to  $\otimes_{\ell=1}^L v'_\ell$ with $v'_{L+1-\ell} = v_\ell$, we find that the monodromy entries in (\req(6vM)) satisfies the inversion property:
\bea(lll)
\jmath \cdot {}^{{\cal A} (s)}_{{\cal D}(s)}  \cdot \jmath = {}^{{\cal A} (-q^{-1-2m}s^{-1})}_{{\cal D} (-q^{-1-2m}s^{-1})} & ~ &\jmath \cdot {}^{{\cal B} (s)}_{{\cal C}(s)}  \cdot \jmath = {}^{{\cal C} (-q^{-1-2m}s^{-1})q}_{ {\cal B} (-q^{-1-2m}s^{-1})q^{-1}} 
\elea(inversion)
where $\jmath$ is the spin-inversion operator of quantum spaces:
\bea(ll)
\jmath  : \bigotimes^L  \CZ^N \longrightarrow \bigotimes^L \CZ^N, & \otimes_{\ell=1}^L {\bf e}^{k_\ell}_\ell \mapsto \otimes_{\ell=1}^L {\bf e}^{k'_\ell}_\ell , ~ ~ k'_{L+1-\ell} = N-1-k_{\ell} 
\elea(jmath)
for $ 0 \leq k_\ell \leq N-1$ and $1 \leq \ell \leq L$. Using (\req(ek)), one can express $\jmath$ in terms of local spin basis in (\req(Fb)) by 
\bea(ll)
\jmath ~ | \widehat{k_1}, \ldots, \widehat{k_L} \rangle  = | \widehat{k_1'}, \ldots, \widehat{k_L'} \rangle , &  k_{L+1-\ell}' \equiv  N-1-2m- k_\ell \pmod{N} ; \\
\jmath ~ | \sigma_1, \ldots, \sigma_L \rangle  = \omega^{-(1+2m) \sum_\ell \sigma_\ell } | \sigma'_1, \ldots, \sigma'_L \rangle , & | \sigma'_{L+1-\ell} \rangle := | - \sigma_{\ell} \rangle ,
\elea(jsig)
which defines an isomorphism between $V_{r, Q}$ and $V_{r, Q'}$ with the charge relation $Q+ Q'\equiv -(1+2m)L$. 
Hence $K^\frac{1}{2}$ and the Hamiltonian $H (k')$ with boundary condition $r$ in (\req(H01)) satisfy the following $\jmath $-conjugation relations:
\bea(ll)
\jmath ~ K^\frac{1}{2} ~ \jmath = K^\frac{1}{2}, & \jmath ~ H (k') ~ \jmath = H (-k').
\elea(Hjth)

\subsection{Degeneracy of XXZ-chain with spin-$\frac{N-1}{2}$ $U_q(sl_2)$-representation  and superintegrable $\tau^{(2)}$-model \label{ssec.XXZhw}}
In this subsection, we study the $sl_2$-loop-algebra symmetry of spin-$\frac{N-1}{2}$ XXZ-chain with $L$-operator in (\req(xxzsp)). The leading and lowest terms ${\cal A}_\pm, {\cal B}_\pm,  {\cal C}_\pm,  {\cal D}_\pm$  of  monodromy entries  (\req(6vM)) are obtained by employing the representation (\req(KXZ)) (or equivalently (\req(spN-1)) ) in (\req(pmTerm)), and they satisfy the condition of ABCD-algebra:  
\bea(l)
[{\cal A}(s), {\cal A}(s')]= [{\cal B}(s), {\cal B}(s')]=[{\cal C}(s), {\cal C}(s')]=[{\cal D}(s), {\cal D}(s')]=0 ; \\ 
{\cal A}(s){\cal B}(s') = f_{s, s'}  {\cal B}(s'){\cal A}(s) -  g_{s, s'} {\cal B}(s){\cal A}(s'), \\
{\cal A}(s){\cal C}(s') = f_{s',s} {\cal C}(s'){\cal A}(s) - g_{s', s} {\cal C}(s){\cal A}(s'), ~ ~ ~ ~
({\cal A} \leftrightarrow {\cal D} ~ {\rm and} ~ {\cal B} \leftrightarrow {\cal C}) , 
\elea(6ABCD)
where $f_{s, s'} :=  \frac{s^2 q^2 - s'^2  }{q (s^2- s'^2 )}$, $g_{s, s'} := \frac{s s'(q^2 - 1)}{q (s^2- s'^2 )}$. Then follow the relations
\bea(ll)
{\cal A}(s) \prod_{i=1}^n {\cal B}(s_i) =
&(\prod_{i=1}^n f_{s, i})  (\prod_{i=1}^n {\cal B}(s_i)) {\cal A}(s) \\
&- \sum_{k=1}^n g_{s, k} (\prod_{i=1, i \neq k }^n f_{k, i})  {\cal B} (s)  \prod_{i=1, i \neq k }^n {\cal B}(s_i) {\cal A}(s_k), \\
{\cal D}(s) \prod_{i=1}^n {\cal B}(s_i) =
&(\prod_{i=1}^n f_{i, s} ) (\prod_{i=1}^n {\cal B}(s_i)) {\cal D}(s) \\
&- \sum_{k=1}^n g_{k, s}  (\prod_{i=1, i\neq k}^n f_{i, k} ){\cal B}(s) \prod_{i=1, i \neq k }^n {\cal B}(s_i) {\cal D}(s_k)  ; \\
&({\cal A} \leftrightarrow {\cal D} ~ {\rm and} ~ {\cal B} \leftrightarrow {\cal C}) , \\
\elea(6ADs)
here we write only $i$ for $s_i$ in the subscripts of $f_{s ,s'}, g_{s, s'}$.
With $s$ or $s'$ tends to $\infty, 0$ in (\req(6ABCD)), one finds
\bea(ll)
{}^{{\cal A}_\pm {\cal B}(s)}_{{\cal D}_\pm {\cal C}(s)} = {}^{{\cal B}(s){\cal A}_\pm}_{{\cal C}(s){\cal D}_\pm} q^{\pm 1}  , &{}^{{\cal D}_\pm {\cal B}(s)}_{{\cal A}_\pm {\cal C}(s)} =  {}^{{\cal B}(s){\cal D}_\pm}_{{\cal C}(s){\cal A}_\pm} q^{\mp 1} .
\elea(ApmB)
Since (\req(spN-1)) is equivalent  to the spin-$\frac{N-1}{2}$ highest weight $U_q(sl_2)$-representation, one may define
the normalized $n$th power of ${\cal B}_\pm,  {\cal C}_\pm$ as in \cite{DFM, NiD, ND08},
${\cal B}_\pm^{(n)} = \frac{{\cal B}_\pm^n}{[n]!}$ , ${\cal C}_\pm^{(n)} = \frac{{\cal C}_\pm^n}{[n]!}, ~ (n \geq 0)$, on a generic $q$ first, then taking the limit on the $N$th root of unity $q$, where $[n]!= \prod_{i=1}^n [i]_q$ and $[0]!:=1$. By induction argument, one finds the following expression of ${\cal B}_\pm^{(n)}, {\cal C}_\pm^{(n)}$:
\bea(ll)
{\cal B}_\pm^{(n)} =& \sum_{ 0 \leq k_i < N, \  k_1+\cdots+ k_L=n } \frac{1}{[k_1]! \cdots [k_L]!} \otimes_{i=1}^L  K_i^{\frac{\pm 1}{2}(\sum_{j (<i)} - \sum_{j (>i)})k_j  }  (e_i^-)^{k_i} \rho^{\mp \sum_{j ({}^{>}_{<} i)} k_j} , \\
{\cal C}_\pm^{(n)} =& \sum_{ 0 \leq k_i < N, \  k_1+\cdots+ k_L=n } \frac{1}{[k_1]! \cdots [k_L]!} \otimes_{i=1}^L  K_i^{\frac{\mp 1}{2}( \sum_{j (<i)} - \sum_{j (>i)})k_j  } (e_i^+)^{k_i}  \rho^{\mp \sum_{j ({}^{<i}_{>i})} k_j},
\elea(BCnpm)
where $\rho =  q^{-2m -1} (=  \omega^{m-M})$.
By setting $s_i = x q^{-i} ~ (i=1, \ldots, n)$ in (\req(6ADs)), with the same argument in \cite{NiD} Section 3, \cite{ND08} Section 4 or \cite{R06F} Section 4.2, one finds the relations: 
\bea(ll)
{\cal A}(s) {\cal B}_\pm^{(n)} = q^{\mp n }  {\cal B}_\pm^{(n)}   {\cal A}(s) +  s^{\pm 1}   {\cal B} (s)  {\cal B}_\pm^{(n-1)} {\cal A}_\pm ,& \\
{\cal D}(s) {\cal B}_\pm^{(n)} = q^{\pm n}  {\cal B}_\pm^{(n)}    {\cal D}(s) -  s^{\pm 1}  {\cal B}(s)  {\cal B}_\pm^{(n-1)} {\cal D}_\pm , &
({\cal A} \leftrightarrow {\cal D} ~ {\rm and} ~ {\cal B} \leftrightarrow {\cal C}) . 
\elea(ADBpm)
For non-negative integers $n, n', n''$ with $n \equiv n' \equiv -n'' \pmod{N}$, by (\req(ApmB)) and (\req(ADBpm)), one finds   
$$
\begin{array}{l}
~ [{}^{{\cal A}(s)}_{{\cal D}(s)}, {\cal C}_+^{( n')}{\cal B}_+^{(n)}] =  \pm s  \bigg( {\cal C}_+^{( n')}{\cal B}(s)  {\cal B}_+^{(n-1)}- {\cal C} (s)  {\cal C}_+^{(n'-1)}{\cal B}_+^{(n)}\bigg){}^{q^n {\cal A}_+}_{q^{-n}{\cal D}_+ } , \\
~ [{}^{{\cal A}(s)}_{{\cal D}(s)}  , {\cal C}_+^{(n'')} {\cal B}_-^{(n)} ]
=   \pm s^{- 1}  {\cal C}_+^{(n'')}    {\cal B} (s)  {\cal B}_-^{(n-1)} {}^{q^{n''}  {\cal A}_-}_{q^{- n'' } {\cal D}_-}  \mp   s  {\cal C}(s)  {\cal C}_+^{(n''-1)} {\cal B}_-^{(n)} {}^{q^n {\cal A}_+}_{q^{-n} {\cal D}_+} ,
\end{array}
$$
hence
\bea(ll)
[{\cal T}(s), {\cal C}_+^{( n')}{\cal B}_+^{(n)}]= & s ( {\cal C}_+^{(n')} {\cal B}(s){\cal B}_+^{(n-1)}  -   {\cal C}(s)  {\cal C}_+^{( n'-1)} {\cal B}_+^{(n)})(q^n {\cal A}_+ -q^{ -2r- n}  {\cal D}_+) , \\

[{\cal T}(s) , {\cal C}_+^{(n'')} {\cal B}_-^{(n)}]
= &s^{- 1}  {\cal C}_+^{(n'')}    {\cal B} (s)  {\cal B}_-^{(n-1)} (q^{n''}  {\cal A}_- - q^{- n''-2r } {\cal D}_-) - 
s {\cal C}(s)  {\cal C}_+^{(n''-1)} {\cal B}_-^{(n)} \times \\
&(q^n {\cal A}_+ - q^{-n-2r} {\cal D}_+ ).
 \elea(TCC+)
Similarly, we have
\bea(l) 
~ [{\cal T}(s), {\cal B}_+^{(n')}{\cal C}_+^{(n)}]
=  -s   \bigg(  {\cal B}_+^{(n')}  {\cal C}(s)  {\cal C}_+^{(n-1)} 
- {\cal B}(s)  {\cal B}_+^{(n'-1)} {\cal C}_+^{(n)}\bigg) (q^{-n }{\cal A}_+ -  q^{-2r+n}  {\cal D}_+), \\
~ [ {\cal T}(s), {\cal C}_-^{(n')}     {\cal B}_-^{(n)}  ] =   s^{- 1} (  {\cal C}_-^{(n')}  {\cal B} (s)  {\cal B}_-^{(n-1)}  -   {\cal C}(s)  {\cal C}_-^{(n'-1)}  {\cal B}_-^{(n)} )(q^{-n}  {\cal A}_- - q^{-2r+ n} {\cal D}_-), \\
~ [{\cal T}(s), {\cal B}_-^{(n')}{\cal C}_-^{(n)}]= -s^{- 1} \bigg(  {\cal B}_-^{(n')} {\cal C} (s)  {\cal C}_-^{(n-1)}   -   {\cal B}(s)  {\cal B}_-^{(n'-1)} {\cal C}_-^{(n)}\bigg)(q^{n} {\cal A}_- - q^{-2r- n}   {\cal D}_-), 
\elea(TBC+)
and 
\bea(ll)
[{\cal T}(s), {\cal C}_-^{(n'')} {\cal B}_+^{(n)} ]
=   & s   {\cal C}_-^{(n'')}  {\cal B} (s)  {\cal B}_+^{(n-1)}(q^{- n''}  {\cal A}_+  -  q^{ n'' -2r} {\cal D}_+  ) - 
s^{- 1}  {\cal C}(s)  {\cal C}_-^{(n''-1)} {\cal B}_+^{(n)} \times \\
&( q^{-n} {\cal A}_- -  q^{n-2r} {\cal D}_-), \\

[{\cal T}(s) , {\cal B}_-^{(n)}{\cal C}_+^{(n'')}] 
=  &-  s  {\cal B}_-^{(n)}  {\cal C}(s)  {\cal C}_+^{(n''-1)}( q^{ n }  {\cal A}_+  - q^{- n-2r}   {\cal D}_+ ) + 
 s^{- 1}   {\cal B} (s)  {\cal B}_-^{(n-1)} {\cal C}_+^{(n'')} \times \\
& (q^{n''}{\cal A}_- - q^{-n''-2r} {\cal D}_- ), \\

[{\cal T}(s), {\cal B}_+^{(n)}{\cal C}_-^{(n'')}] = & -  s^{- 1}   {\cal B}_+^{(n)}{\cal C}(s)  {\cal C}_-^{(n''-1)}
( q^{- n } {\cal A}_- - q^{ n-2r}  {\cal D}_- ) + s   {\cal B} (s)  {\cal B}_+^{(n-1)}  {\cal C}_-^{(n'')} \times \\
& (q^{-n''} {\cal A}_+- q^{n''-2r} {\cal D}_+ ).
\elea(TCBpm)
Using the above formulas, we can generalize the results in \cite{NiD, ND08, R06F, Tar} as follows:
\begin{lem}\label{lem:loop}
The superintegrable $\tau^{(2)}$-model $\tau^{(2)} ({\tt t})$ (\req(stauev)) with the $L$-operator (\req(hsupL)) is equivalent to the XXZ chain ${\cal T}(s)$ (\req(TcXZ)) defined by the $L$-operator (\req(xxzsp)) employing the representation (\req(KXZ)), so that the equality holds\footnote{The relation (\req(tauT)) here is the same as formula (4.9) in \cite{R075}. }:
\be
\tau^{(2)} ({\tt t}) = (-q^{-1}s)^L K^{\frac{-1}{2}} {\cal T}(q^{-1}s) , ~ ~ ~ {\tt t} = s^2.
\ele(tauT)
Indeed, the monodromy entries (\req(Mont2)), (\req(6vM)) of $\tau^{(2)}$-model and  XXZ chain (respectively) are related by
\bea(ll)
{}^{A ({\tt t})}_{D ({\tt t})} = (-s)^L  K^\frac{-1}{2} {}^{{\cal A} (s)}_{{\cal D} (s)}, & {}^{B({\tt t})}_{ C({\tt t})} = (-s)^{L\mp 1}  q^{\mp 1} K^\frac{-1}{2} {}^{{\cal B} (s)}_{{\cal C} (s)} .
\elea(tTMon)
Then the relation (\req(stauev)) in  $\tau^{(2)}$-model is equivalent to the following one in XXZ chain:
\be
(-s)^L K^{\frac{-1}{2}} {\cal T}(s) = q^{2P_a+J } (1- q^{2m+2} s^2 )^L \prod_{i=1}^J f_{s, i}  + q^{-2P_b-J}(1- q^{2m} s^2 )^L \prod_{i=1}^J f_{i,s}
\ele(Tev)
where $f_{s, i}, f_{i, s}$ are in (\req(6ADs)) for the roots $s_i^2 (= - (\omega {\tt v}_i)^{-1})$ of ${\tt F}({\tt t})$ in (\req(stauev)), satisfying the Bethe equation equivalent to (\req(Bethesup)):
\be
\bigg( \frac{1-q^{2m+2} s_j^2 }{1-q^{2m} s_j^2 }\bigg)^L = - q^{-2(P_a+P_b+J)} \prod_{i=1}^J \frac{q^2 s_i^2 - s_j^2}{s_i^2 - q^2 s_j^2},~ ~ j=1, \ldots, J .
\ele(BeXXZ)
Both the operators $\tau^{(2)} ({\tt t})$ and ${\cal T}(s)$ (on the quantum space $V_{r, Q}$) commute with
\bea(llll)
{\cal C}_+^{( n_+')}{\cal B}_+^{(n_+)}, &{\cal B}_+^{(p_+')}{\cal C}_+^{(p_+)},& {\cal C}_-^{(n_-')}{\cal B}_-^{(n_-)},& 
 {\cal B}_-^{(p_-')}{\cal C}_-^{(p_-)}
\elea(nmpm)
where ${\cal B}_\pm^{(n)}, {\cal C}_\pm^{(n)}$ are defined in (\req(BCnpm)) with non-negative integer powers satisfying
\bea(l)
n_+ \equiv  n+' \equiv -p_+ \equiv -p_+' \equiv Q- r , \\
n_- \equiv n'_- \equiv -p_- \equiv -p_-' \equiv (1+2m)L +Q+r (\equiv P_b - P_a) \pmod{N}.
\elea(nn')
When $2r + (2m+1)L \equiv 0$, $\tau^{(2)} ({\tt t})$ and ${\cal T}(s)$ commute with 
\bea(llll)
{\cal C}_+^{(n'')} {\cal B}_-^{(n)}, &{\cal C}_-^{(n'')} {\cal B}_+^{(n)} ,& {\cal B}_-^{(n)}{\cal C}_+^{(n'')},& 
 {\cal B}_+^{(n)}{\cal C}_-^{(n'')}
\elea(CBpm)
where $n, n''$ are non-negative integers satisfying $n \equiv -n'' \equiv Q +L(m-M) (\equiv -H_1) \pmod{N}$.
\end{lem}
{\it Proof.} By the discussion in Subsection \ref{ssec.XXZhw}, (\req(tauT)) and (\req(tTMon)) are easily obtained, then follow the relations (\req(Tev)) and (\req(BeXXZ)). 
By (\req(KXZ)), $K = q^{L(1 + 2 m)}\widehat{Z}^{-1} = \omega^{L(M-m)} X^{-1}$. Since (\req(TCC+)) and (\req(TBC+)) are zeros whenever the last terms in the formulas vanish, the expression of ${\cal A}_\pm, {\cal D}_\pm$ in (\req(pmTerm)), together with (\req(ApmB)), implies that the operators in (\req(nmpm)) commute with ${\cal T}(s)$ and $K$, hence also with $\tau^{(2)} ({\tt t}) $ by (\req(tauT)). Using (\req(TCC+)) and (\req(TCBpm)), ${\cal T}(s)$ commutes with 
operators in (\req(CBpm)) if $K q^{-2n}= \rho^{-L}  q^{2r} = \rho^{L}  q^{-2r}$, hence follows the conclusion by the assumption of odd $N=2M+1$. $\Box$
\par \vspace{.1in} \noindent
By (\req(duqn)),  $\tau^{(2)}({\tt t})$ and ${\cal T}(s)$ in the above lemma can also be regarded as the transfer matrices of the dual $\tau^{(2)}$-model $\tau^{(2) \dagger} ({\tt t})$ in (\req(Dual)), and its equivalent XXZ chain ${\cal T}^\dagger(s)$ defined by (\req(xxzsp)) and  (\req(KXZ)): $
\tau^{(2) \dagger} ({\tt t})= \tau^{(2)}({\tt t}) $ and ${\cal T}^\dagger(s) = {\cal T}(s) $, but with the boundary condition $r^* = Q$.
Then $\tau^{(2) \dagger} ({\tt t})$ and ${\cal T}^\dagger(s)$ on $V_{r^*, Q^*}~ (Q^* =r)$ commute with the operators ${\cal C}_\pm^{\dagger (n')}{\cal B}_\pm^{\dagger (n)}$ as in (\req(nmpm)) with $n= n_\pm^\dagger, p_\pm^\dagger$ defined by $r^*, Q^*$ in (\req(nn')). Since $n_+^\dagger = p_+ , n_-^\dagger = n_-$, the $\tau^{(2)}$-duality relation in (\req(Dual)) in turn yields the commutativity of $V_{r, Q}$-operators $\tau^{(2)} ({\tt t}), {\cal T}(s)$  and 
\bea(llll)
\Psi^{-1}{\cal C}_+^{\dagger (p_+')}{\cal B}_+^{\dagger (p_+)}\Psi,& \Psi^{-1}{\cal B}_+^{\dagger (n_+')}{\cal C}_+^{\dagger (n_+)}\Psi, & \Psi^{-1}{\cal C}_-^{\dagger (n_-')}{\cal B}_-^{\dagger (n_-)}\Psi, &
 \Psi^{-1}{\cal B}_-^{\dagger (p_-')}{\cal C}_-^{\dagger (p_-)}\Psi
\elea(nmpmd)
with $\Psi$ in (\req(Psi)), and non-negative-integer powers in (\req(nn')). Furthermore, when $2Q + (2m+1)L \equiv 0$, $\tau^{(2)} ({\tt t})$ and ${\cal T}(s)$ commute with 
\bea(llll)
\Psi^{-1}{\cal C}_+^{({n^*}'')} {\cal B}_-^{(n^*)}\Psi, & \Psi^{-1}{\cal C}_-^{({n^*}'')} {\cal B}_+^{(n^*)}\Psi , & \Psi^{-1}{\cal B}_-^{(n^*)}{\cal C}_+^{{n^*}'')}\Psi, &
 \Psi^{-1}{\cal B}_+^{(n^*)}{\cal C}_-^{({n^*}'')}\Psi
\elea(nmCBpm)
 with $n^* \equiv -{n^*}'' \equiv r +L(m-M) (\equiv -H_0) \pmod{N}$.

By (\req(nn')), the powers of operators in (\req(nmpm)), (\req(nmpmd)) (or (\req(CBpm)), (\req(nmCBpm))) are all $N$-multiples if and only if $Q=r, 
P_a=P_b=0$, which is equivalent to those sectors in $I_+ \cap I_- = i_+ \cap i_-$ by (\req(Pab0)). In this situation, (\req(nmpm)) are generated by ${\cal B}_\pm^{(N)}, {\cal C}_\pm^{(N)}$ as the Chevalley generators ${}^{E_0}_{F_1}, {}^{E_1}_{F_0}$ respectively, which give rise to a $sl_2$-loop-algebra on a sector in $V_{r, Q} ~ (r=Q \equiv \frac{-(1+2m)L}{2})$ with the Bethe state as the highest (or lowest) weight vector (\cite{NiD, ND08, Tar} for the case $r=m=0$, \cite{R06F}\footnote{The $N$th root of unity and spectral parameters in \cite{R06F} (for the case $d=N$) are slightly different with those used in this paper. Indeed with the same $N$th root of unity $q$, the $\omega, s, t$, and $t'$ (in Section 4.3),  of \cite{R06F} are respectively equal to $\omega^{-1}, sq^\frac{-(N-2)}{2}, {\tt t}q^3, {\tt t}q$  here. The Bethe equation $(4.22)_{d=N}$ for the eigenvalue $(4.21)_{d=N}$ in \cite{R06F} can be identified with (\req(BeXXZ)) and $K \cdot (\req(Tev))$ in the $I_+$-sectors  for the case $m=M, r=0$ here.} for the case $r=0, m=M$, and \cite{R09}). Furthermore, the Bethe state of each sector in (\req(qTn)) was obtained in  \cite{R09} Section 4.2 by the algebraic-Bethe-ansatz method as follows. With the local basis ${\bf e}_\ell^k$'s in (\req(ek)), we define the pseudo-vacuum of $\tau^{(2)}({\tt t})$ by
\bea(lll)
\Omega^+ ~ (=\Omega^+_r) :=  \bigotimes^L_{\ell=1} {\bf e}_\ell^{N-1}, & \Omega^- ~ (=\Omega^-_r) :=  \bigotimes_{\ell =1}^L {\bf e}_\ell^0 ,  & ({\bf e}^k_{L+1} = \omega^{r(m-k)}{\bf e}^k_1),
\elea(vac)
which can also expressed by $\Omega_r^\pm = | \widehat{k_1^\pm}, \ldots, \widehat{k_L^\pm} \rangle ~ (| \widehat{k_{L+1}^\pm}  \rangle = \omega^{-r k_1} | \widehat{k_1^\pm}\rangle$) with $k_\ell^+ = N-1-m$ and $k_\ell^- = -m$ for all $\ell$. Then 
$$
\begin{array}{ll}
C({\tt t}) \Omega^+ = {\cal C}(s) \Omega^+ = 0, & B({\tt t}) \Omega^- = {\cal B}(s) \Omega^- = 0 \\
{}^{A({\tt t})}_{D({\tt t})}\Omega^+  = {}^{(1- \omega^{-m-1}{\tt t})^L}_{(\omega^m - {\tt t})^L} \Omega^+ & (\Leftrightarrow ~ {}^{{\cal A} (s)}_{{\cal D} (s)}\Omega^+ =  {}^{( s q^{2m+3/2}- s^{-1}q^{-1/2})^L}_{(sq^{-1/2} -s^{-1}q^{-2m-1/2} )^L} \Omega^+ ), \\
{}^{A({\tt t})}_{D({\tt t})}\Omega^-  = {}^{(1- \omega^{-m} {\tt t})^L}_{(\omega^{1+m}- {\tt t})^L}\Omega^- & (\Leftrightarrow ~ 
 {}^{{\cal A} (s)}_{{\cal D} (s)} \Omega^-  = {}^{(s q^{2m+1/2}-s^{-1} q^{1/2})^L}_{(s q^{1/2}-s^{-1} q^{-2m-3/2} )^L} \Omega^- ),
\end{array}
$$
where $A, B, ..$ and ${\cal A}, {\cal B}, ..$ are the monodromy-entries in (\req(Mont2)) and (\req(6vM)) respectively. Define
\bea(l)
\psi^+ (= \psi^+ ({\tt v}^+_1, \ldots,{\tt v}^+_J)) = \prod_{j=1}^J B(-(\omega {\tt v}^+_j)^{-1}) \Omega_r^+,  \\
\psi^- (= \psi^-({\tt v}^-_1, \ldots,{\tt v}^-_J)) = \prod_{j=1}^J C(-(\omega {\tt v}^-_j)^{-1}) \Omega_r^- , 
\elea(Betv)
where ${\tt v}^\pm_j$'s are Bethe roots (\req(Bethesup)) of ${\tt F}({\tt t})$ in (\req(stauev)). Then $\psi^\pm$ is the Bethe state of $I_\pm$-sectors respectively with $J$ and quantum numbers in (\req(Pab0)). By the commutation relation between $B({\tt t}), C({\tt t})$ and $H_1$ in (\req(H01)), $[H_1, B({\tt t})] = -2 B({\tt t})$, $[H_1, C({\tt t})] = 2 C({\tt t})$, and using (\req(qTn)), one finds $(H_1 - \beta ) \psi^\pm = \pm N m_E \psi^\pm$ (respectively) where $\beta$ is the linear term  in (\req(Esk')). Hence $\psi^\pm$ can be identified with the basis element $\vec{w} (\pm ; k' )$ of ${\cal E}_{{\tt F}, P_a , P_b }$ at $k'= \infty$ in (\req(Ewk'))\footnote{The $\vec{v} (\pm, \ldots, \pm ; \infty )$ in \cite{R09} Section 4.2 are equal to $\vec{w} (\pm, \ldots, \pm ; \infty )$ in this paper.} : 
$$
\begin{array}{ll}
\frac{\psi^+}{\parallel \psi^+ \parallel} = \vec{\bf u} (+, \ldots, + ), & \frac{\psi^-}{\parallel \psi^- \parallel} = \vec{\bf u} (-, \ldots, -).
\end{array}
$$ 
With the same Bethe-state argument for the dual model $\tau^{(2) \dagger}({\tt t})$, through the duality correspondence (\req(Dual)), $\tau^{(2) \dagger}$-Bethe states are converted to Bethe states  of $\tau^{(2)}({\tt t})$ via the isomorphism $\Psi$ in (\req(Psi)):
\bea(ll)
\phi^+ (= \phi^+ ({\tt v}^+_1, \ldots, {\tt v}^+_J)) & = \prod_{j=1}^{J} B (-(\omega {\tt v}^+_j)^{-1})  \Psi^{-1} \Omega_Q^{\dagger +}  ( = \Psi^{-1} \prod_{j=1}^{J} B^\dagger (-(\omega {\tt v}^+_j)^{-1}) \Omega_Q^{\dagger +} ),  \\
\phi^- (= \phi^- ({\tt v}^-_1, \ldots, {\tt v}^-_J)) & = \prod_{j=1}^{J} C (-(\omega {\tt v}^-_j)^{-1}) \Psi^{-1} \Omega_Q^{\dagger -} (= \Psi^{-1}\prod_{j=1}^{J} C^\dagger (-(\omega {\tt v}^-_j)^{-1}) \Omega_Q^{\dagger -}),  
\elea(BetvF)
where  $\Psi^{-1} \Omega_Q^\pm = |Q; n_1^\pm, \ldots, n_L^\pm \rangle ~ ~ (|(n_{L+1}^\pm \rangle\rangle = \omega^{-Qn_1^\pm} |(n_1^\pm \rangle\rangle  $ with $n_\ell^+ = N-1-m$ and $n_\ell^- = -m$ for all $\ell$.  Then $\phi^\pm$ are the Bethe states of $i_\pm$-sectors (respectively) with quantum numbers in (\req(Pab0)). 
By (\req(vec*)), $\phi^\pm$ can be identified with the basis element $\vec{v} (\pm ; k' )$ of  of ${\cal E}_{{\tt F}, P_a , P_b }$ in (\req(Ek')) at $k'=0$: 
$$
\begin{array}{ll}
\frac{\phi^+}{\parallel \phi^+ \parallel} = \vec{\bf u} (+, \ldots, + ),  \frac{\phi^-}{\parallel \phi^- \parallel}  = \vec{\bf u} (-, \ldots, -).
\end{array}
$$
Under the dual correspondence (\req(IiD)), $\Psi: {\cal E}_{({\tt F},P_a, P_b)} \simeq {\cal E}^\dagger_{({\tt F},P_a, P_b)}$, the Bethe states $ \psi^\pm, \phi^\pm$  are identified with $ \phi^{\pm \dagger}, \psi^{\pm \dagger}$ respectively.
\par \vspace{.1in} \noindent

By (\req(inversion)) and (\req(tTMon)), the $\jmath$-conjugate relation of monodromy entries (\req(Mont2)) of $\tau^{(2)}$-model 
$\tau^{(2)}( {\tt t}) (=\tau^{(2) \dagger}({\tt t}))$ for $\jmath$ in (\req(jmath)) are given by  
\bea(ll)
\jmath \cdot {}^{A ({\tt t})}_{D({\tt t})}  \cdot \jmath = (-{\tt t})^L q^{(1+2m)L} K  {}^{A (\omega^{1+2m}{\tt t}^{-1})}_{D (\omega^{1+2m}{\tt t}^{-1})} ; &\jmath \cdot {}^{B ({\tt t})}_{C ({\tt t})}  \cdot \jmath = - (-{\tt t})^L q^{(1+2m)L} K {}^{C (\omega^{1+2m}{\tt t}^{-1}) q^{2m}}_{ B (\omega^{1+2m}{\tt t}^{-1}) q^{-2m} } .
\elea(parit)
By (\req(tau2)), the $\tau^{(2)}$-matrix satisfy the following inversion relation:
\bea(l) 
\jmath \cdot \tau^{(2)}( \omega^m {\tt t}) \cdot \jmath = (-{\tt t})^L q^{-L} K \tau^{(2)}( \omega^{m-1} {\tt t}^{-1}) , \\
\jmath^* \cdot \tau^{(2)}( \omega^m {\tt t}) \cdot \jmath^{*} = (-{\tt t})^L q^{-L} K^* \tau^{(2)}( \omega^{m-1} {\tt t}^{-1}) 
\elea(taupa)
where $\jmath^* = \Psi^{-1} \jmath \Psi $, $K^* = \Psi^{-1} K \Psi$ with $\Psi$ defined in (\req(Psi)). Note that in general  $\jmath^* \neq \jmath$. The  first relation in (\req(taupa)) holds only for sectors with the same boundary condition $r$, and the second relation, induced from $\tau^{(2) \dagger}({\tt t})$-matrix via the identification (\req(Dual)) and (\req(duqn)), holds for sectors with the same charge $Q$. Then $\jmath$ or $\jmath^*$ permute the $\tau^{(2)}$-eigenspaces ${\cal E}_{{\tt F}, P_a, P_b}$'s. Indeed, we have the following result: 
\begin{prop}\label{prop:inversion} 
Under the spin-inversion operator $\jmath$ or $\jmath^*$,  the $\tau^{(2)}$-eigenspaces and Bethe states (\req(Betv)) or (\req(BetvF)) are in one-to-correspondence:
\bea(l)
\jmath : ({\cal E}_{{\tt F}, P_a, P_b}, \psi^\pm) \in I_\pm \longleftrightarrow ({\cal E}_{{\tt F}', P'_a, P'_b}, \psi^\mp)  \in I_\mp, \\
\jmath^* : ({\cal E}_{{\tt F}, P_a, P_b}, \phi^\pm ) \in i_\pm \longleftrightarrow ({\cal E}_{{\tt F}', P'_a, P'_b}, \phi^\mp ) \in i_\mp , \\
\elea(jcor)
such that $({\tt F}, P_a, P_b)$ and $({\tt F}', P'_a, P'_b)$ are related by (\req(F'F)) with the results in Lemma \ref{lem:FF'} valid for their $\tau^{(2)}$-eigenvalues and evaluation polynomials. Furthermore, the isomorphism in (\req(jcor)) induces the inversion relation between the correspondence of $k'$ and $(-k')$-eigenvectors in (\req(E'Eb)). 
\end{prop}
{\it Proof.} By (\req(taupa)), $\jmath ({\cal E}_{{\tt F}, P_a, P_b})$ or $\jmath^* ({\cal E}_{{\tt F}', P'_a, P'_b})$ are $\tau^{(2)}$-eigenspaces. Let ${\tt v}_j = {\tt v}_j^\pm ~ ( 1 \leq j \leq J)$ be the Bethe roots  in the definition of Bethe state $\psi^\pm , \phi^\pm$ of ${\cal E}_{{\tt F}, P_a, P_b}$ in (\req(Betv)) or (\req(BetvF)). Since $\jmath (\Omega_r^\pm ) = \Omega_r^\mp $ and $\jmath^* (\Omega_Q^\pm ) = \Omega_Q^\mp $, using the $\jmath$-conjugate relations of $B ({\tt t}), C ({\tt t})$ in (\req(parit)),  one finds $\jmath (\psi^\pm) = \psi^\mp$ , $\jmath^*(\phi^\pm) = \phi^\mp$, and  the Bethe roots, ${\tt v}_j = {\tt v}_j^\pm$ and ${\tt v}'_j = {\tt v}_j^\mp$, are related  by (\req(F')). Hence $({\tt F}, P_a, P_b)$ and $({\tt F}', P'_a, P'_b)$ satisfy conditions in Lemma \ref{lem:FF'}, where the eigenvalue relations in Lemma \ref{lem:FF'} $(ii)$ are equivalent to (\req(taupa)) with $q^{-L} K =\omega^{-Lm} X^{-1}$  and $q^{-L} K^* = \omega^{-Lm-r}$. Then we obtain the isomorphism of sectors in (\req(jcor)). In order to show (\req(E'Eb)), we first consider the $I_\pm$-sector case. By (\req(Hjth)), $\jmath (\vec{w}(s; k'))$ is a $H(-k')$-eigenvector in ${\cal E}_{{\tt F}', P'_a, P'_b}$ with the eigenvalue $\widetilde{E}(s; k')$ in (\req(Etsk)). As $\jmath$ is an isometric isomorphism, $\jmath (\vec{w}(s; k')) = \vec{w}^\prime(s^\prime; -k')$ for some $s^\prime= (s_1^\prime, \ldots, s^\prime_{m_E})$, and $\widetilde{E}(s; k') = \widetilde{E}^\prime(s^\prime; -k') ( = \alpha^\prime - \beta^\prime k' + N \sum_{i=1}^{m_E}\frac{k'}{|k'|} s^\prime_i \varepsilon (\theta^\prime_i ; -k'))$. Hence $\alpha + k' \beta = \alpha^\prime - \beta^\prime k'$ as in (\req(cjlin)). Using (\req(ci)) and (\req(PPrec)), one finds $\theta_i + \theta^\prime_i = \pi$, $\varepsilon (\theta_i ; k') =  \varepsilon (\theta^\prime_i ; -k')$, hence $s_i^\prime = - s_i$ for all $i$. Therefore $\jmath$ induces the correspondence in (\req(E'Eb)) for $I_\pm$-sectors.  Using the $\jmath$-correspondence (\req(E'Eb)) about $I_\pm$-sectors for the dual $\tau^{(2) \dagger}$-model, together with the duality correspondence (\req(vec*)), one obtains $\jmath^*$ in (\req(jcor)) gives rise to the inverse correspondence (\req(E'Eb)) for $i_\pm$-sectors.
$\Box$ \par  \noindent
{\bf Remark}. (I) The isomorphism (\req(E'Eb)) in Proposition \ref{prop:inversion} for $k'= \infty$ or $0$ with $s_i = \pm 1$ for all $i$  is the same as the correspondence of Bethe states in (\req(jcor)) by (\req(usl)). Indeed, (\req(E'Eb)) induces the correspondence of $\tau^{(2)}$-eigenvectors in (\req(usl)):
$$
\begin{array}{ll}
I_\pm \ni {\cal E}_{{\tt F}, P_a, P_b} \ni \vec{\bf u}(s_1, \ldots, s_{m_E}) \stackrel{\jmath}{\longleftrightarrow} 
\vec{\bf u}^{\prime} (-s_1, \ldots , -s_{m_E} ) \in {\cal E}_{{\tt F}', P'_a, P'_b} \in I^\mp , \\
i_\pm \ni {\cal E}_{{\tt F}, P_a, P_b} \ni \vec{\bf u}(s_1, \ldots , s_{m_E}) \stackrel{\jmath^*}{\longleftrightarrow} (-1)^{m_E} \vec{\bf u}^{\prime} (-s_1, \ldots , -s_{m_E}) \in {\cal E}_{{\tt F}', P'_a, P'_b} \in i^\mp .
\end{array}
$$
By (\req(PPrec)), (\req(hei)) and  (\req(lk')), the correspondence (\req(jcor)) induces the equivalence of the $sl_2$-product and loop-algebra structure of $\tau^{(2)}$-eigenspaces: 
$$
\begin{array}{lllll}
I_\pm: &  \jmath \cdot {\bf e}_i^{\prime \pm} \cdot \jmath = {\bf e}_i^\mp , &\jmath \cdot {\bf h}_i^{\prime} \cdot \jmath = - {\bf h}_i , &  \jmath \cdot {\bf e}^{\prime \pm}(n) \cdot \jmath = {\bf e}^\mp (-n), & \jmath \cdot {\bf h}^{\prime}(n) \cdot \jmath = - {\bf h}(-n) ;  \\
i_\pm : & \jmath^* \cdot {\bf e}_i^{\prime \pm} \cdot \jmath^* = {\bf e}_i^\mp , &\jmath^* \cdot {\bf h}_i^{\prime} \cdot \jmath^* = - {\bf h}_i , &  \jmath^* \cdot {\bf e}^{\prime \pm}(n) \cdot \jmath^* = {\bf e}^\mp (-n), & \jmath^* \cdot {\bf h}^{\prime}(n) \cdot \jmath^* = - {\bf h}(-n) , 
\end{array}
$$
for $i=1, \ldots, m_E$, and $n \in \ZZ$. In particular, the currents ${\bf E}^\mp (\xi)$ of ${\cal E}_{{\tt F}, P_a, P_b}$ and ${\bf E}^{\prime \pm} (\xi)$ of ${\cal E}_{{\tt F}', P'_a, P'_b}$ in (\req(Epm)) are related by
\bea(lll)
\jmath \cdot {\bf E}^\mp (\xi^{-1} ) \cdot  \jmath = - \xi {\bf E}^{\prime \pm} (\xi) & {\rm or} & \jmath^* \cdot {\bf E}^\mp (\xi^{-1} ) \cdot  \jmath^* = - \xi {\bf E}^{\prime \pm} (\xi).
\elea(jEj)
\par \vspace{.05in} \noindent
(II) By (\req(jsig)), $\jmath^*$ is an isomorphism between $V_{r, Q}$ and $V_{r', Q}$ with $r+r' \equiv -(1+2m)L$ which sends $|Q; n_1, \ldots, n_L \rangle$ to $|Q; n'_1, \ldots, n'_L \rangle$ with $n'_\ell = N-1-2m - n_{L+1-\ell}$.
\par \vspace{.05in} \noindent
(III) For a sector ${\cal E}_{{\tt F}, P_a , P_b }$ in $V_{r, Q}$, operators in (\req(nmpm)), (\req(CBpm)), (\req(nmpmd)) and (\req(nmCBpm)) can be used to describe the $\tau^{(2)}$-degeneracy of ${\cal E}_{{\tt F}, P_a , P_b }$, where these operators, when applied to the Bethe state, are expected to generate all vectors as in $(I_+ \cap I_-)$-sectors. 
In the case $m=r=0, L \equiv 0$,  the loop-algebra structure of the ground-state sector (i.e. $i_\mp$-sector with $J=0$ ) was investigated in \cite{AuP7, AuP9}\footnote{In the paper we use different conventions than those in \cite{AuP7, AuP9}, where $Q, m_Q, |\Omega \rangle,  |\overline{\Omega} \rangle,$, ${\bf x}^\pm_{n, Q}$, ${\bf E}^\pm_{k, Q}, {\bf H}_k^Q$ are corresponding to $N-Q, m_E, \Psi^{-1} |\Omega^-_Q \rangle, \Psi^{-1} |\Omega^+_Q \rangle $, ${\bf e}^\pm (n)$, ${\bf e}^\pm_k , {\bf h}_k$  here respectively.} through these operators with partial success. The operators ${\bf e}^- (n), {\bf e}^+ (n')$ for $n \geq 1, n' \geq 0$ are constructed from (\req(nmpmd)), (\req(nmCBpm)) for $(i_-)$-ground-state sector in \cite{AuP9} Section 3.5, but the relationship between the complete $sl_2$-loop algebra structure of $\tau^{(2)}$-eigenspace in Subsection \ref{ssec:sl2t2} and operators in (\req(nmpmd)) and (\req(nmCBpm)) is still unknown. However, using the local-operator form of ${\bf e}^\pm (n')$ for $n' \geq 1$ and the relation (\req(gign')) or (\req(gign)), one can derive a local-operator expression of the $sl_2$-product structure ${\bf e}_i^\pm$, hence  ${\bf h}_i$ of ${\cal E}_{{\tt F}=1, P_a , P_b=0 }$. By using the Bethe state $|\Omega \rangle (=\Psi^{-1} |\Omega^-_Q \rangle)$ and the relation (\req(uspin)), one then obtains the local-spin form of  $\tau^{(2)}$-eigenvectors. In \cite{AuP7, AuP9}, the eigenvector $\vec{v}(s_1, \ldots, s_{m_E} ; k')$ of superintegrable CPM transfer matrix $T$ or $\widehat{T}$ in (\req(ThatT)) and (\req(TTform)) are constructed from $\vec{v}(s'_1, \ldots, s'_{m_E} ; 0)$'s through linear transformations of $V_i$ in (\req(EV)), which can be identified with the $\CZ$-algebra spanned by ${\bf 1}_i, {\bf h}_i, {\bf e}_i^\pm$. Write $T, \widehat{T}$ as a product of $V_i$-transformations, expressed in a rotation-operator form (\cite{AuP9} (130) and (137)). By the expression of $\langle \Omega |T |\Omega \rangle,  \langle \Omega |{\bf e}_i^- T |\Omega \rangle$ and $\langle \Omega |\widehat{T} |\Omega \rangle,  \langle \Omega | \widehat{T} {\bf e}_i^+ |\Omega \rangle$, the local-operator form of rotation matrices are derived in \cite{AuP9} ((150)-(152) , (161-(162)) so that one obtains the $k'$-dependent local-spin form of $\vec{v}(s ; k')$ up to a scaling factor (\cite{AuP9} (159)). On the other hand, through the Onsager-algebra representation generated by the quantum spin chain $H(k')$ (\req(H01)) and using the form of (\req(HJ)) and (\req(Hk'0)), one may also find the expression of $\vec{v}(s ; k')$ or $\vec{w}(s ; k')$ in terms of $\vec{v}(s ; 0)$ or $\vec{w}(s ; 0)$ by (\req(k'0if)). Equivalently, the formula (\req(k'0if)) is the expression of $k'$-eigenvectors  in terms of $\infty$-eigenvectors or $0$-eigenvectors in (\req(Hk'0)). This approach can be applied to the general case for an arbitrary sector as well. By this method, the effort of non-trivial calculation about rotation matrices in  \cite{AuP7, AuP9} can be avoided, even for the ground state sector. Hence the expression of $k'$-dependence of the CPM state vectors in the local spin basis can be obtained  by (\req(k'0if)) if the local-spin-vector form of $\vec{\bf u}(s)$ in (\req(usl)) is known. By (\req(uspin)), the CPM-eigenvector problem is then reduced to  the local-spin form of the states and currents in  (\req(vvE)). As the states in (\req(vvE)) are identified with the Bethe states in (\req(Betv)) or (\req(BetvF)) expressed in the local spin basis, it suffices to find an equivalent form for the current ${\bf E}^\pm (\xi)$ in  (\req(vvE)) in terms of monodromy entries in (\req(6vM)), given by the Fabricius-McCoy current discussed later in Section \ref{sec.loopt2}. Note that ${\bf E}^- (\xi)$ or ${\bf E}^+ (\xi)$ is defined in (\req(Epm)), whose operator-coefficients constitute only one-half of the nilpotent part of the loop-algebra $sl_2[z, z^{-1}]$, in particular no Serre-relation among the generators. The structure of one-half nilpotent subalgebra is simpler than the subalgebra constructed in \cite{AuP9} for the ground state $i_-$-sector, but it is sufficient to produce the $sl_2$-product algebra elements in the $\tau^{(2)}$-eigenvector formula (\req(uspin)).

\section{The $sl_2$-loop-algebra Generators in Superintegrable $\tau^{(2)}$-model \label{sec.loopt2}}
\setcounter{equation}{0}
In this section, we derive the currents in (\req(vvE)) of a superintegrable $\tau^{(2)}({\tt t})$ through its equivalent the XXZ chain ${\cal T}(s)$ in Lemma \ref{lem:loop} with the identification of spectral variables ${\tt t}= s^2$. By extending the method in \cite{FM01, R06F}, the local-operator-representation of those currents will be obtained by 
the Fabricius-McCoy current which is a series in terms of the monodromy matrix (\req(6vM)) of ${\cal T}(s)$. 
As in \cite{R06F} Section 4.2, we construct the Fabricius-McCoy current for a sector in (\req(qTn)) by differentiating the relations in (\req(6ADs)). For convenience, we use the subscript of variables $s, q $ to indicate the logarithmic partial-derivative $s \partial_s , q \partial_q, \ldots$ of an operator or function, e.g, 
${\cal B}_s = s (\partial_s {\cal B})$,  ${\cal B}_q  = q (\partial_q {\cal B})$, $(f_{s, s'})_s = s (\partial_s f_{s, s'})$, $(f_{s, s'})_{s'} = s' (\partial_{s'} f_{s, s'})$ etc. By the same argument in \cite{R06F} Proposition 4.1, the highest weight representation (\req(spN-1)) of $U_q(sl_2)$-operators in the monodromy matrix (\req(6vM)) yields the vanishing of averages of the off-diagonal entries, $\langle {\cal B} \rangle = \langle {\cal C} \rangle  = 0$, where $\langle {\sf O} \rangle = \prod_{i=0}^{N-1} {\sf O} (q^i s)$. The vanishing of $\langle {\cal B} \rangle$  and $\langle {\cal C} \rangle$ mirrors the automatic vanishing of certain constraint equations in Bethe's ansatz, hence should create the complete $N$-strings in the degeneracy. As in \cite{FM01} (1.36)-(1.40), we replace the vanishing creation operators by derivatives of $\langle {\cal B} \rangle, \langle {\cal C} \rangle$ to form the Fabricius-McCoy current of XXZ chain ${\cal T}(s)$. Note that the $s$-derivative of $\langle {\cal B} \rangle, \langle {\cal C} \rangle$ vanishes: 
\be
  \langle {\cal B} \rangle_s (= \sum_{n=0}^{N-1}{\cal B}_s (sq^n) \prod_{i=0, i \neq n}^{N-1} {\cal B}(sq^i ))= 0,  \ ~ \  ~ \  ~ \  {\cal B} \leftrightarrow {\cal C} = 0 ,
\ele(Bs0)
and the leading and lowest terms of the $q$-derivative of $\langle {\cal B} \rangle , \langle {\cal C} \rangle$  are 
\bea(l) 
{}^{{\cal B}^{(N)}_\pm}_{{\cal C}^{(N)}_\pm} = (2N^2)^{-1} \lim_{s^{\pm 1}\rightarrow \infty}   (\pm s)^{\mp N(L-1)}  {}^{\langle {\cal B} \rangle_q}_{\langle {\cal C} \rangle_q} (s),   \\
\elea(SBq)
by $\prod_{j=1}^{N-1} (1-\omega^j) = N$ (\cite{R06F} (4.34)). 

\subsection{The Fabricius-McCoy current of $I_\pm$-sectors  \label{ssec.Ipm}}
For a sector in $I_+$ of (\req(qTn)) with the Bethe state $\psi^+ = \psi^+ ({\tt v}^+_1, \ldots,{\tt v}^+_J)$ in (\req(Betv)), the Fabricius-McCoy current is a current in the form 
\bea(ll)
{\cal B}^{(N)} (s) = \langle {\cal B} (s) \rangle_q + {\cal B}^{(N)}_{\varphi} (s) , & {\cal B}^{(N)}_{\varphi} (s)= \sum_{n=0}^{N-1} \varphi(sq^n) {\cal B}_s (sq^n) \prod_{i=0, i \neq n }^{N-1} {\cal B}(sq^i) ,
\elea(BNc)
for some suitable $\varphi(s)$ such that the following properties hold:
\bea(l)
{\cal T} (s) (\prod_{i = 1}^{n} {\cal B}^{(N)} (x_i )) \psi^+  = (\prod_{i = 1}^n {\cal B}^{(N)} (x_i )) {\cal T} (s)  \psi^+ , ~ ~ (n \geq 1) ,
\elea(Bcur) 
where $s, s', x_i \in \CZ$ (\cite{FM01} (2.1)-(2.14), \cite{R06F} (4.39) (4.40)). Note that ${\cal B}^{(N)} (s) = {\cal B}^{(N)} (qs)$ by the construction of ${\cal B}^{(N)} (s)$. Since $\langle {\cal B} (s) \rangle_q, {\cal B}_s (s) $ commute with ${\cal B} (s')$ for all $s, s'$, $[{\cal B}^{(N)} (s), {\cal B} (s')] = [{\cal B}^{(N)} (s), {\cal B}^{(N)} (s')]\footnote{Here we assume $(k^\frac{1}{2})_q = 0 = (e^\pm)_q $ so that $[\langle {\cal B} (s) \rangle_q, \langle {\cal B} (s') \rangle_q ] =0$.}=0$. By differentiating relations in (\req(6ADs)) for $n=N$, then using the identities   
$$
\begin{array}{ll}
f_{k, k+1} = 0 , \ \ \prod_{i=1}^{N} f_{s, i} = \prod_{i=1}^N f_{i, s} = 1 , & ( \prod_{i=1, i \neq k }^N f_{k, i})_q =   ( \prod_{i=1, i \neq k }^N f_{i, k})_q =   \frac{2 }{q-q^{-1}} ,   \\
( \prod_{i=1, i \neq k}^N f_{k, i})_{s_n} = ( \delta_{k, n}- \delta_{k, n-1}) \frac{2}{q- q^{-1}}  , & ( \prod_{i=1, i \neq k}^N f_{i, k})_{s_n} = ( \delta_{k, n}- \delta_{k, n+1}) \frac{-2}{q- q^{-1}} ,
\end{array}
$$
for $s_i = x q^i (1 \leq i \leq N)$, one finds the following relations between ${\cal A}(s), {\cal D}(s)$ and ${\cal B}^{(N)} (x)$ in (\req(BNc)):
\bea(l)
~[{\cal A}(s), {\cal B}^{(N)} (x)] 
=  \frac{2{\cal B} (s) }{q - q^{-1}}  \sum_{k=0}^{N-1} g_{s, k} (\varphi(xq^{k+1})- \varphi(xq^k)-1 ) ( \prod_{i=0, i \neq k}^{N-1} {\cal B}(xq^i) ){\cal A}(xq^k) ; \\
~ [ {\cal D}(s), {\cal B}^{(N)} (x)] = \frac{2{\cal B} (s) }{q- q^{-1}}\sum_{k=0}^{N-1} g_{s,k} (\varphi(xq^{k-1})- \varphi(xq^k)+1)   (\prod_{i=0, i \neq k }^{N-1} {\cal B}(xq^i)) {\cal D}(xq^k), 
\elea(ADBN)
(for the details, see \cite{R06F} Section 4.2). Using $\langle {\cal B} \rangle =0$, one can show the relations 
\bea(l)
\bigg( \prod_{i=1}^{N-1} {\cal B}(sq^i) \bigg) {}^{{\cal A}(s)}_{{\cal D}(s)} \prod_{i = 1}^{n} {\cal B}^{(N)} (x_i ) \psi^+   = 
{}^{a (s)^L \prod_{j=1}^J f_{s, j}}_{d (s)^L \prod_{j=1}^J f_{j, s} }  \bigg( \prod_{i=1}^{N-1} {\cal B}(sq^i) \bigg) \prod_{i = 1}^{n} {\cal B}^{(N)} (x_i )\psi^+ ,  
\elea(Bpsi)
where $s, x_i \in \CZ$, $s_j^2 = - (\omega {\tt v}_j^+)^{-1}$ as in (\req(BeXXZ)), $a (s):= s q^{2m+3/2}- s^{-1}q^{-1/2}, d(s):= sq^{-1/2} -s^{-1}q^{-2m-1/2}$, by induction  first on $n=0$ by (\req(6ADs)) with the Bethe roots $s_i ~ (i=1, \ldots, J)$, then applying $\prod_{i=1}^{N-1} {\cal B}(sq^i)$ on relations in (\req(ADBN)) for the general $n$.
Evaluating (\req(ADBN)) on the Bethe state $\psi^+$, then using $(\req(Bpsi))_{n=0}$ with $s= xq^k ~ ( 0 \leq k \leq N-1)$, the identity (\req(Bcur)) for $n=1$ is equivalent to the following constraint of $\varphi(s)$:
\be
(\varphi(sq)- \varphi(s)-1 ) a^L(s) \prod_{i=1}^J f_{s, s_i} + (\varphi(sq^{-1})- \varphi(s)+ 1 ) q^{-2r} d^L(s) \prod_{i=1}^J f_{s_i, s} = 0, ~ ~  ~ (s \in \CZ)
\ele(varpc)
where $s_i^2 = - (\omega {\tt v}_i^+)^{-1}$ as in (\req(Bpsi)). Applying $(-s)^L K^\frac{-1}{2}$ on (\req(varpc)), then changing $s$ to $q^{-1}s$, one finds the equivalent form of (\req(varpc)) by (\req(qTn)): 
\be
(\varphi(s)- \varphi(q^{-1}s)-1 ) (1- \omega^{-m} {\tt t})^L {\tt F}({\tt t})  = (\varphi(q^{-1}s)-\varphi(q^{-2}s)- 1 )(1- \omega^{1-m} {\tt t})^L {\tt F}(\omega^2 {\tt t})  \omega^{P_b+P_a} . 
\ele(vart)
By (\req(Pt)) and (\req(Ptrt)), ${\tt P}_{\rm ev}({\tt t}^N )$ can be written in the form, 
$$
 {\tt P}_{\rm ev}({\tt t}^N ) \omega^{P_b} {\tt P}(0) ~ (=\omega^{P_b} {\tt P}({\tt t})) = \sum_{k=0}^{N-1} p(\omega^k {\tt t}) , ~ ~ p({\tt t}) := \frac{(1-  {\tt t}^N )^L {\tt t}^{-(P_a+P_b)}}{(1- \omega^{-m} {\tt t})^L {\tt F} ({\tt t}) {\tt F}( \omega {\tt t}) } .
$$
Since $p({\tt t})$ satisfies the relation (\req(vart)), $\varphi(s)- \varphi(q^{-1}s)-1 = \alpha ({\tt t}^N) 
p({\tt t})$ for some ${\tt t}^N$-function $\alpha$. Then $\alpha ({\tt t}^N)= \frac{-N}{\omega^{P_b} {\tt P}({\tt t})}$ by 
$\varphi(s) = \varphi(q^Ns)$, and the solution of $\varphi$-function in (\req(vart)) is uniquely, up to additive $s^N$-functions (which define the same current in (\req(BNc)) by $\langle {\cal B} \rangle = 0$),  given by 
\be
\varphi (s)  = \frac{\sum_{k=1}^N k p(\omega^{k-1} {\tt t})}{{\tt P}_{\rm ev}({\tt t}^N ) \omega^{P_b} {\tt P}(0) } ~ \bigg( = \frac{\sum_{k=1}^N k p(\omega^{k-1} {\tt t})}{\omega^{P_b} {\tt P}({\tt t}) } \bigg) .
\ele(varpsl)
With the  above $\varphi(s)$ in (\req(BNc)), we defines the Fabricius-McCoy current ${\cal B}^{(N)} (s)$  satisfying (\req(Bcur)) for $n=1$. By replacing $\psi^+$ by $\prod_{i = 1}^{n} {\cal B}^{(N)} (x_i ) \psi^+$, and using (\req(ADBN)) and (\req(ADBN))  in the above argument, the relation (\req(Bcur)) follows by induction for the general $n$.  Furthermore, using the method in \cite{FM01} ((3.1)-(3.5)), one can show  the poles of ${\cal B}^{(N)} (s)$  equal to zeros of ${\tt P}_{\rm ev}({\tt t}^N )$, first by the Bethe equation (\req(Bethesup)), to determine the behavior of $\varphi(s){\tt P}_{\rm ev}({\tt t}^N )$  near a pole ${\tt t} = -( \omega^{1+n} {\tt v}^+_i)^{-1} ~ ( 1 \leq n \leq N) $ as 
$$
{\varphi}(s){\tt P}_{\rm ev}({\tt t}^N ) \longrightarrow \bigg(\frac{1-N \delta_{n, N}}{1+ \omega^{1+n} {\tt v}^+_i {\tt t}}\bigg) \frac{(-\omega)^{P_a+P_b} ({\tt v}_i^+)^{(-N+1)L+P_a+P_b+2J-2}({\tt v}^{+ N}_i- 1)^L  }{\omega^{P_b} {\tt P}(0)(1-\omega)({\tt v}^+_i + \omega^{-1-m}) \prod_{j \neq i} ({\tt v}_i^+ - {\tt v}^+_j)({\tt v}^+_i- \omega {\tt v}^+_j)}, 
$$
then by (\req(Bs0)), the current ${\tt P}_{\rm ev}({\tt t}^N ) {\cal B}^{(N)}_{\varphi} (s)$ in (\req(BNc)) near ${\tt t} = -( \omega^{1+n} {\tt v}^+_i)^{-1} = (q^{-n}s_i)^2 $ with behavior as 
$$
\bigg(\frac{-N}{1+ \omega^{1+n} {\tt v}^+_i {\tt t}}\bigg) \frac{(-\omega)^{P_a+P_b} ({\tt v}_i^+)^{(-N+1)L+P_a+P_b+2J-2}({\tt v}^{+ N}_i- 1)^L  }{\omega^{P_b} {\tt P}(0)(1-\omega)({\tt v}^+_i + \omega^{-1-m}) \prod_{j \neq i} ({\tt v}_i^+ - {\tt v}^+_j)({\tt v}^+_i- \omega {\tt v}^+_j)} {\cal B}_s (s_i) \prod_{k=1 }^{N-1} {\cal B}(s_i q^{-k}). 
$$
By (\req(tTMon)) and $-( \omega {\tt v}^+_i)^{-1} = s_i^2$, the above operator vanishes when applied to the Bethe state $\psi^+$, and the same vanishing property for  $\prod_i {\cal B}^{(N)} (x_i )\psi^+$ holds, even though $\varphi (s)$ has poles at the Bethe roots $s_i^2 = -( \omega {\tt v}^+_i)^{-1}$.  Therefore, ${\tt P}_{\rm ev}({\tt t}^N ) (-s)^{N(L-1)} {\cal B}^{(N)}_{\varphi} (s) $ is a regular current of ${\cal E}_{{\tt F}, P_a, P_b}$ with ${\tt t}$-degree $\leq N(L-1)+(N-1)L-P_a-P_b-2J$, invariant under ${\tt t} \mapsto \omega {\tt t}$, hence a ${\tt t}^N$-current with  ${\tt t}^N$-degree $ \leq L +m_E-1$ where $m_E$ is in (\req(ME)):
\be
{\tt P}_{\rm ev}({\tt t}^N ) (-s)^{N(L-1)} {\cal B}^{(N)}_{\varphi} (s) = \sum_{k=0}^{L+m_E-1} (-1)^{kN} {\cal R}^-_k {\tt t}^{kN}. 
\ele(Bvph)
Hence the current ${\tt P}_{\rm ev}({\tt t}^N ) (-s)^{N(L-1)} {\cal B}^{(N)} (s)$ has the ${\tt t}^N$-degree $\leq (L-1)+m_E$. 
On the other hand, ${\bf E}^-({\tt t}^N)$ in (\req(Epm)) also satisfies the relation (\req(Bcur)) with poles equal to ${\tt P}_{\rm ev}({\tt t}^N )$'s zeros  by (\req(Eex)). Since a ${\cal E}_{{\tt F}, P_a,P_b}$-current with the property (\req(Bcur)) and poles the same as ${\tt P}_{\rm ev}({\tt t}^N )$'s zeros is unique  up to scale functions, $(-s)^{N(L-1)}  {\cal B}^{(N)}(s)$ differs from  ${\bf E}^- ({\tt t}^N)$ by some ${\tt t}^N$-polynomial $\gamma ({\tt t}^N)$ with degree $\leq L$ :
\be
(-s)^{N(L-1)}  {\cal B}^{(N)}(s) = \gamma^- ({\tt t}^N) {\bf E}^- ({\tt t}^N), ~ ~ \gamma^- (\xi) = \sum_{n=0}^L \gamma^-_n (-\xi)^n . 
\ele(BE)
Furthermore, by (\req(Eex)) and (\req(SBq)), the constant and ${\tt t}^{N(L-1)+Nm_E}$-coefficient of ${\tt P}_{\rm ev}({\tt t}^N )$-multiple of currents in (\req(BE)) yield 
$$
\begin{array}{ll}
{\cal B}^{(N)}_- + \frac{{\cal R}^-_0}{2N^2} = \frac{\gamma^-_0}{2N^2} {\bf e}^-(0), &{\cal B}^{(N)}_+ + \frac{{\cal R}^-_{L+m_E-1}}{2N^2 \prod_{i=1}^{m_E} {\tt a}_i } =  \frac{\gamma^-_L   }{2N^2} {\bf e}^- (-1).  
\end{array}
$$
By the Remark in Subsection \ref{ssec:sl2t2}, the above ${\bf e}^-(0), {\bf e}^- (-1)$ can be normalized so that the local-operator expression of the loop-operator ${\bf e}^-(0), {\bf e}^- (-1)$ is given by
\bea(lll)
\gamma^-_{0} = \gamma^-_{L} = 2N^2, &
{\cal B}^{(N)}_- + \frac{{\cal R}^-_0}{2N^2} =  {\bf e}^-(0), &{\cal B}^{(N)}_+ + \frac{{\cal R}^-_{L+m_E-1}}{2N^2 \prod_{i=1}^{m_E} {\tt a}_i  } = {\bf e}^- (-1).   
\elea(modid)

For an $I_-$-sector ${\cal E}_{{\tt F}, P_a,P_b}$ with Bethe state $\psi^-$ in (\req(vvE)), we consider its $\jmath$-inverse $I_+$-sector and Bethe state $({\cal E}_{{\tt F}^\prime, P_a^\prime,P_b}, \psi^+)$ in (\req(taupa)). Let 
${\cal B}^{\prime (N)} (s)$ be the Fabricius-McCoy current (\req(BNc)) of ${\cal E}_{{\tt F}^\prime, P_a^\prime,P_b^\prime}$ defined by $\varphi^\prime (s)$ in (\req(varpsl)) using the evaluation polynomial ${\tt P}({\tt t})$  of  ${\cal E}_{{\tt F}^\prime, P_a^\prime,P_b^\prime}$:
$$
\varphi^\prime (s) = \frac{\sum_{k=1}^N k p^\prime(\omega^{k-1} {\tt t})}{\omega^{P_b} {\tt P}^\prime({\tt t}) }, ~  
p^\prime({\tt t}) = \frac{(1-  {\tt t}^N )^L {\tt t}^{-(P^\prime_a+P^\prime_b)}}{(1- \omega^{-m} {\tt t})^L {\tt F}^\prime ({\tt t}) {\tt F}^\prime( \omega {\tt t}) } .
$$
The Fabricius-McCoy current of ${\cal E}_{{\tt F}, P_a,P_b}$ is defined by 
\be 
{\cal C}^{(N)} (s) = \jmath \cdot {\cal B}^{\prime (N)} (-q^{-1-2m}s^{-1}) \cdot \jmath
\ele(CBN)
for the consistency of the inversion relation of monodromy entries in (\req(inversion)). Indeed, one can express the above current in terms of the monodromy ${\cal C}$-entry  of ${\cal E}_{{\tt F}, P_a,P_b}$ as follows:
\begin{lem}\label{lem:CNf}
\bea(ll)
{\cal C}^{(N)} (s) = \langle {\cal C} (s) \rangle_q + {\cal C}^{(N)}_\varphi (s) , & {\cal C}^{(N)}_\varphi (s)=  \sum_{n=0}^{N-1} \varphi(sq^n) {\cal C}_s (sq^n) \prod_{i=0, i \neq n }^{N-1} {\cal C}(sq^i) ,
\elea(CNc)
where $\varphi(s)$ is defined in (\req(varpsl)).
\end{lem}
{\it Proof.} First, we find the relation between $\varphi(s)$ and  $\varphi^\prime (s)$. By (\req(ME) and (\req(mm'dd')), the quantum numbers of the inverse sectors are related by $P_a + P_b + P_a'+P_b' = (N-1)L-2J - N m_E$. By (\req(F')) , (\req(F'F)) and (\req(PPrec)), $p^\prime ({\tt t}), {\tt P}^\prime({\tt t})$ are related to $p( {\tt t}), {\tt P}({\tt t})$ in (\req(varpsl)) by
$$
\begin{array}{ll}
p^\prime(\omega^{1+2m} {\tt t}^{-1}) &= p(\omega^{-1} {\tt t}) {\tt t}^{-Nm_E}  \omega^{m(P_a+P_b-P^\prime_a-P^\prime_b)+(3+2m)J} (\prod_{j=1}^J {\tt v}_j)^2 , \\
\omega^{P'_b} {\tt P}'({\tt t}^{-1}) 
 &= \omega^{P_b}{\tt P}({\tt t}) {\tt t}^{-Nm_E}  \omega^{ m(P_a+P_b- P_a'-P_b')+(3+2m)J} (\prod_{j=1}^J {\tt v}_j)^2 ,
\end{array}
$$
which yield the relation
\be
\varphi^\prime (-q^{-1-2m}s^{-1}) = \frac{\sum_{k=1}^N k p(\omega^{-k} {\tt t})  }{\omega^{P_b}{\tt P}({\tt t})} = N+1- \varphi(s). 
\ele(phipp)
By (\req(inversion)),  the chain rule of differentiation implies $\jmath {\cal B}^\prime_s \jmath (-q^{-1-2m}s^{-1}q^n) = - {\cal C} (sq^{-n})_s$. Using  (\req(Bs0)) and (\req(phipp)),   one finds
\bea(lll)
\jmath \cdot \langle {\cal B}^\prime \rangle_q (-q^{-1-2m}s^{-1}) \cdot \jmath &= \langle {\cal C} (s) \rangle_q + (1+2m)  \jmath  \langle {\cal B}^\prime \rangle_s (-q^{-1-2m}s^{-1})  \jmath  &= \langle {\cal C} (s) \rangle_q , \\
\jmath \cdot {\cal B}^{\prime (N)}_{\varphi^\prime} (-q^{-1-2m}s^{-1}) \cdot \jmath 
&=  {\cal C}^{(N)}_\varphi (s) - (N+1)\langle {\cal C} \rangle_s &=  {\cal C}^{(N)}_\varphi (s).
\elea(jBC)
Then follows (\req(CNc)).  
$\Box$ \par \noindent
{\bf Remark.} By (\req(phipp)), the relation (\req(varpc)) of $\varphi^\prime$ in sector $I_+$ is equivalent to the following relation of $\varphi$ in sector $I_-$:
$$
(\varphi(sq^{-1})- \varphi(s)+1 ) a^L(sq^{-1}) \prod_{i=1}^J f_{s_i, s} + (\varphi(sq)- \varphi(s)- 1 ) q^{-2r} d^L(sq) \prod_{i=1}^J f_{s, s_i} = 0, 
$$
which is again equivalent to (\req(vart)). One can also derive the Fabricius-McCoy current (\req(CNc)) of ${\cal E}_{{\tt F}, P_a,P_b}$ directly by the same argument as that for (\req(BNc)).
\par \vspace{.1in} \noindent
By (\req(P'ev)), Remark of Lemma \ref{lem:FF'} and (\req(jEj)), the substitution of $s$ by $-q^{-1-2m}s^{-1}$ in the $\jmath$-conjugation of ${\cal B}^{\prime (N)}$ in (\req(Bvph))and (\req(BE)) yields 
\bea(ll)
{\tt P}_{\rm ev}({\tt t}^N ) (-s)^{N(L-1)} {\cal C}^{(N)}_\varphi (s) &= \sum_{k=0}^{L+m_E-1} (-1)^{kN} {\cal R}^+_k {\tt t}^{kN}, \\
(-s)^{N(L-1)}  {\cal C}^{(N)}(s) = \gamma^+ ({\tt t}^N) {\bf E}^+ ({\tt t}^N), &\gamma^+ (\xi) = \sum_{n=0}^L \gamma^+_n (-\xi)^n 
\elea(Cvph)
where ${\cal R}^+_k = (\prod_{i=1}^{m_E} {\tt a}_i) \jmath {\cal R}^{\prime -}_{L+m_E-1-k} \jmath $, and $\gamma^+_n = \gamma^{\prime -}_{L-n} $. By (\req(SBq)) and the first relation in (\req(jBC)), one has $\jmath \cdot {\cal B}^{\prime (N)}_\pm \cdot \jmath = {\cal C}^{(N)}_\mp$. Hence the normalization condition for ${\cal B}^{\prime (N)}$ in (\req(modid)) is equivalent to its reciprocal condition for ${\cal C}^{(N)}$:
\bea(lll)
\gamma^-_{0} = \gamma^-_{L} = 2N^2, & {\cal C}^{(N)}_- + \frac{{\cal R}^+_0}{2N^2} =  {\bf e}^+(1), &{\cal C}^{(N)}_+ + \frac{{\cal R}^+_{L+m_E-1}}{2N^2 \prod_{i=1}^{m_E} {\tt a}_i  } = {\bf e}^+ (0).   
\elea(modi)
Hence we have shown the following result. 
\begin{prop}\label{prop:curI}
(i) For a sector in $I_+$, the Fabricius-McCoy current ${\cal B}^{(N)} (s)$ is defined by (\req(BNc)) with $\varphi$ in (\req(varpsl)), which is equal to a multiple of the current ${\bf E}^-({\tt t}^N)$ via the relations (\req(BE)). The $sl_2$-loop-algebra generators ${\bf e}^-(0), {\bf e}^-(-1)$ in Subsection \ref{ssec:sl2t2} are expressed by local operators in (\req(modid)).

(ii) For a sector in $I_-$, the Fabricius-McCoy current is ${\cal C}^{(N)} (s)$ in (\req(CNc))) with $\varphi$ in (\req(varpsl)), which is equal to a multiple of the current ${\bf E}^+({\tt t}^N)$ in (\req(Epm))  via the second relation of (\req(Cvph)). The $sl_2$-loop-algebra generators ${\bf e}^+(1), {\bf e}^+(0)$ in Subsection \ref{ssec:sl2t2} are expressed by local operators in (\req(modi)).

(iii) For a $I_+$-sector $({\cal E}_{{\tt F}^\prime, P_a^\prime,P_b^\prime}, \psi^{ +})$ and $I_-$-sector $({\cal E}_{{\tt F}, P_a, P_b}, \psi^-)$ in the inversion correspondence (\req(jcor)), the Fabricius-McCoy currents are connected by the inverse relation (\req(CBN)).
\end{prop}
$\Box$ \par  \vspace{.05in} \noindent  
{\bf Remark}. (I) For a sector ${\cal E}_{{\tt F}, P_a, P_b}$ in $I_+ \cap I_-$, by (\req(Ipm)) we have $P_a=P_b=d_E=0$, and $H_1 \equiv 0 \pmod{N}$. Then ${\cal E}_{{\tt F}, P_a, P_b}$ contains the Bethe states $\psi^\pm$ so that  they are $\jmath$-conjugate under the inversion correspondence (\req(jcor)). In this case, by (\req(varpsl)), ${\cal R}^\mp_0$ and ${\cal R}^\mp_{L+m_E-1}$ are zeros as they are equal to some scale-multiples of the lowest and leading term of $\langle {\cal B} \rangle_s, \langle {\cal C} \rangle_s$ in (\req(Bs0)). By (\req(modid)) and (\req(modi)), ${\cal B}^{(N)}_-  =  {\bf e}^-_\infty(0), {\cal B}^{(N)}_+ = {\bf e}^-_\infty (-1)$ and ${\cal C}^{(N)}_-  =  {\bf e}^+_\infty(1), {\cal C}^{(N)}_+ = {\bf e}^+_\infty (0)$ so that they form the $sl_2$-loop-algebra generators of ${\cal E}_{{\tt F}, P_a, P_b}$  as in \cite{R06F}\footnote{There are misprints in \cite{R06F} about the $sl_2$-loop algebra mode basis associated with (4.34), ( $e(-1)= T^{+ N}, f(1) =  T^{- N}$ should be $e(1)= T^{+ N}, f(-1) =  T^{- N}$), the first term  in (4.43) ($-\sum_{k=0}^N k p(\omega t)$ should be $-\sum_{k=1}^N k p(\omega^{k-1} t)$)), and the equality of currents in Theorem 4.1 where only the first term in (\req(modid)) here was stated. Note that $\varphi(s)$ in \cite{R06F} (4.43) differs from the $\varphi(s)$ in (\req(varpsl)) here (for the case $m=M, r=0$) by the additive function $-(N+1) {\tt P}_{\rm ev} ({\tt t}^N)$.} with $\jmath \cdot {\bf e}^+_\infty(n) \cdot \jmath = {\bf e}^-_\infty (-n)$ for $n \in \ZZ$. Using (\req(spN-1)), one can  verify $[{\cal C}^{(N)}_\pm {\cal B}^{(N)}_\mp ] = \frac{H_1}{N} =h(0 )$  as in \cite{DFM}.
For other sectors in $I_\pm$, the operators ${\cal R}^\mp_0, {\cal R}^\mp_{L+m_E-1}$ in (\req(modid)) or (\req(modi)) could be non-zero in the one-half-algebra-generator expression of ${\bf e}^-(0), {\bf e}^-(-1)$ or ${\bf e}^+(1), {\bf e}^+(0)$, with $h(0) = \frac{H_1- \beta }{N}$, where $\beta$ is linear term in (\req(Esk')).
\par \vspace{.05in} \noindent 
(II) In Proposition \ref{prop:curI} $(i), (ii)$, the ${\cal R}^\mp_k$'s in (\req(Bvph)), (\req(Cvph)) are operators in the local spin basis, depending on ${\tt F}(\tt t)$ and $P_a, P_b$. Hence the Fabricius-McCoy current is expressed by local-spin operators. As ${\tt P}_{\rm ev}(\xi) {\bf E}^\mp (\xi)$ is a degree-$(m_E-1)$ $\xi$-polynomial  with operator-coefficients, one may solve the polynomial $\gamma^\mp (\xi)$ and the local-operator coefficients of ${\tt P}_{\rm ev}(\xi) {\bf E}^\mp (\xi)$ by using the ${\tt P}_{\rm ev}({\tt t}^N)$-multiple of Fabricius-McCoy current in (\req(BE)) or (\req(Cvph)), under the normalization condition  (\req(modid)), (\req(modi)) respectively. Therefore we obtain the local-operator expression of currents in (\req(vvE)), hence follows the-local-spin vectors in (\req(uspin)) for $I_\pm$-sectors.

\subsection{The Fabricius-McCoy current of $i_\pm$-sectors  \label{ssec.ipm}}
The eigenvectors of $\tau^{(2)}$-model in (\req(uspin)) for $i_\pm$-sectors can be obtained from the Fabricius-McCoy current of the dual model $\tau^{(2) \dagger}({\tt t})$ in Proposition \ref{prop:curI}. Indeed, through the duality correspondence (\req(IiD)), we define the Fabricius-McCoy current for $i_\pm$-sectors as follows:
\begin{prop}\label{prop:curipm}
For an $i_\pm$-sector of $\tau^{(2)}({\tt t})$ , the Fabricius-McCoy current is defined by
\bea(lll)
{\cal B}^{(N)} (s) := \Psi^{-1} {\cal B}^{(N) \dagger} (s) \Psi , & {\cal C}^{(N)} (s) := \Psi^{-1} {\cal C}^{(N) \dagger} (s) \Psi &{\rm for} ~ i_+, i_-{\rm -sector ~ respectively}, 
\elea(BCPsi)
where ${\cal B}^{(N) \dagger} (s), {\cal C}^{(N) \dagger} (s)$ are the Fabricius-McCoy currents of $\tau^{(2) \dagger}({\tt t})$ in Proposition \ref{prop:curI} for the $I_\pm^\dagger$-sector dual to the $i_\pm$-sector of $\tau^{(2)}({\tt t})$ in (\req(IiD)), and $\Psi$ is the duality correspondence  in (\req(Psi)).
The Fabricius-McCoy current is a polynomial-multiple of ${\bf E}^\mp ({\tt t}^N)$ in (\req(Epm))  so that the relations (\req(BE)) and 
(\req(Cvph)) hold. The $sl_2$-loop-algebra generators, ${\bf e}^-(0), {\bf e}^-(-1)$ for $i_+$-sectors, and ${\bf e}^+(1), {\bf e}^+(0)$ for $i_-$-sectors, are expressed by the local operators by (\req(modid)) and (\req(modi)) respectively.  Furthermore, for a $i_+$-sector $({\cal E}_{{\tt F}^\prime, P_a^\prime,P_b^\prime}, \phi^{ +})$ and $i_-$-sector $({\cal E}_{{\tt F}, P_a, P_b}, \phi^-)$ in the inversion correspondence (\req(jcor)), the Fabricius-McCoy currents are connected by the inverse relation:
\be
{\cal C}^{(N)} (s) = \jmath^* \cdot {\cal B}^{\prime (N)} (-q^{-1-2m}s^{-1}) \cdot \jmath^* .
\ele(CBNw)
\end{prop}
$\Box$ \vspace{.05in} \par  \noindent
Note that $\tau^{(2)}({\tt t})$ and $\tau^{(2) \dagger}({\tt t})$ share the same polynomial ${\tt P}_{\rm ev}({\tt t}^N)$ in (\req(Ptrt)). By (\req(Dual)), the Fabricius-McCoy current ${\cal B}^{(N)} (s)$ or ${\cal C}^{(N)} (s)$ for a $i_\pm$-sector is again defined by (\req(BNc)) or (\req(CNc)) with $\varphi$ in (\req(varpsl)). As in Remark (II) of Proposition \ref{prop:curI}, one can obtain the local spin form of $\tau^{(2)}$-eigenvectors in (\req(uspin)) for $i_\pm$-sectors. For a $(i_+ \cap i_-)$-sector ${\cal E}_{{\tt F}, P_a, P_b}$, by (\req(qTn)), ${\cal E}_{{\tt F}, P_a, P_b}$ is invariant under the conjugations of $\Psi, \jmath$ and $\jmath^*$. The Fabricius-McCoy currents ${\cal B}_\infty^{(N)}, {\cal C}_\infty^{(N)}$ in  Proposition \ref{prop:curI} and ${\cal B}_0^{(N)}, {\cal C}_0^{(N)}$ in  Proposition \ref{prop:curipm} are related by (\req(BCPsi)). The $sl_2$-loop algebra of ${\cal E}_{{\tt F}, P_a, P_b}$ generated by ${\cal B}^{(N)}_{0, -}  =  {\bf e}^-_0(0), {\cal B}^{(N)}_{0,+} = {\bf e}^-_0 (-1)$ , ${\cal C}^{(N)}_{0, -}  =  {\bf e}^+_0(1)$ and ${\cal C}^{(N)}_{0,+} = {\bf e}^+_0 (0)$ is $\Psi$-conjugate to the loop algebra in Proposition \ref{prop:curI} Remark (I), with the $\tau^{(2)}$-eigenvectors in ${\cal E}_{{\tt F}, P_a, P_b}$ related by $\Psi (\vec{\bf u}_\infty(s)) = \vec{\bf u}_0(s) \prod_{i=1}^{m_E}(-s_i ) $. \par  \vspace{.1in} \noindent
{\bf Remark.} (I) The definition of Fabricius-McCoy current of a superintegrable $\tau^{(2)}$-model defined in this section strongly relies on its equivalent XXZ-chain in Lemma \ref{lem:loop}. In (\req(BNc)) or (\req(CNc)), by (\req(tTMon)), the second term ${\cal B}^{(N)}_{\varphi} (s)$, ${\cal C}^{(N)}_{\varphi} (s)$ can be phrased in terms of the $\tau^{(2)}$ monodromy entry $B({\tt t})$ , $C({\tt t})$ in (\req(Mont2)) respectively. The first term  $\langle {\cal B} (s) \rangle_q$ (or $\langle {\cal C} (s) \rangle_q$) is constructed through a process of deforming the $N$th root of unity $q$ to a generic $\tilde{q}$ in (\req(spN-1)), using the $L$-operator (\req(xxzsp)) as the solution of YB equation for the symmetry $R_{\rm 6v}$-matrix. However, there exist no finite-dimensional Weyl operators $\widehat{Z}, \widehat{X}$ for a generic $\tilde{q}$ such that the relations (\req(XZF)) and (\req(KXZ)) hold. It is not clear how to deform the $L$-operator (\req(hsupL)) to a YB solution for the asymmetry $R$-matrix so that the relation (\req(tauT)) is still valid for a generic $\tilde{q}$.
Since the operator-coefficients of the Fabricius-McCoy current are expressed by products of local operators in (\req(spN-1)), we can use (\req(KXZ)) to express the Fabricius-McCoy current in terms of local Weyl operators in (\req(XZF)).
Nevertheless the approach of local-spin form of ${\bf E}^\mp(\xi)$ through the Fabricius-McCoy current is in essence based on the general theory of XXZ-chain.
\par \vspace{.05in} \noindent
(II) The Fabricius-McCoy current ${\cal B}^{(N)} (s)$ (\req(BNc)) in Propositions \ref{prop:curI} and \ref{prop:curipm}, hence ${\bf E}^- (\xi)$, is defined by the monodromy  entry ${\cal B} (s)$ of XXZ-chain only. Similarly, 
${\cal C}^{(N)} (s)$  and ${\bf E}^- (\xi)$ in (\req(CNc)) depends only on the monodromy  entry ${\cal C} (s)$. But in general, the equation (\req(BE)) or (\req(Cvph)) is related to many terms of coefficients in the monodromy entry, not only evolved with the leading or lowest term, even for the relation (\req(modid)) or (\req(modi)). For example, in the ground-state $i_-$-sector with $m=r=0, L \equiv 0$ case in \cite{AuP9} where $P_a \neq 0$ and $P_b=0$,  ${\cal R}^+_0$ in (\req(modi)) is expressed by products of ${\cal C}_0 , ..., {\cal C}_{P_a}$, and ${\cal R}^+_{L+m_E-1}$ by products of 
${\cal C}_{L-1+N-P_a} , ..., {\cal C}_{L-1}$, where ${\cal C}_j$'s are defined by $(-s)^{(L-1)}  {\cal C}(s) = \sum_{j=0}^{L-1} {\cal C}_j (-{\tt t})^j $. Hence the expressions of ${\bf e}^+(1), {\bf e}^+(0)$ in (\req(modid)) or (\req(modi)) are different from those in \cite{AuP9}, which are in the form of (\req(nmpmd)),(\req(nmCBpm)). The relationship between these two approaches is not clear at this moment.

\section{Concluding Remarks \label{sec.F}}
In this work, the $k'$-dependent CPM-eigenvectors are constructed by using symmetries in the superintegrable $\tau^{(2)}$-model and CPM. First in Subsection \ref{ssec.QN}, by the constraints of quantum numbers, we classify all Onsager sectors of a superintegrable $\tau^{(2)}$-model into type $I_\pm$ and $i_\pm$,  and discuss their relationship under two reflective symmetries: the duality of superintegrable CPM in \cite{B89, R09}, and the inversion relation in (\req(conj)). The duality interchanges sectors of $I$ and $i$; while the inversion interchanges the $+$ and $-$ sectors. 
The quantum-space correspondence of  (\req(Psi)) in duality relation was found in \cite{R09}, while the correspondence of quantum spaces in inversion symmetry of superintegrable  $\tau^{(2)}$-model was constructed through the local operators in the XXZ chain equivalent to the $\tau^{(2)}$-model (Proposition \ref{prop:inversion}). Furthermore, the correspondences of $k'$-CPM-state vectors under these two reflective symmetries are explicitly found through the theory of Onsager-algebra representation. The Onsager-algebra symmetry also produces the $k'$-dependent expression (\req(k'0if)) of CPM-state vectors 
$$
\begin{array}{lll}
I_\pm: &\vec{w}(s_1, \ldots, s_{m_E} ; k')  & = \sum_{s_1', \ldots s_{m_E}'} {\bf u}(s_1, \ldots, s_{m_E})  \prod_{i=1}^{m_E} \sin (\frac{\varphi_{i, k'} - \theta_i}{2}+  \frac{(s_i - s_i') \pi}{4}) ,\\
i_\pm :& \vec{v}(s_1, \ldots, s_{m_E} ; k') & = \sum_{s_1', \ldots s_{m_E}'} {\bf u}(s_1, \ldots, s_{m_E})  \prod_{i=1}^{m_E} \sin  (\frac{\vartheta_{i, k'}- \pi }{2} +  \frac{(s_i - s_i') \pi}{4}). \\
\end{array}
$$
where ${\rm e}^{\theta_i}$'s are the evaluation values of the Onsager-algebra representation of a sector ${\cal E}_{{\tt F}, P_a, P_b}$, $\vartheta_{i, k'}$ and $\varphi_{i, k'}$ are $k'$-dependent angle-functions defined in (\req(angle)), and $\vec{\bf u}(s)$ are the basic $\tau^{(2)}$-eigenvectors in (\req(usl)). We may express the basic $\tau^{(2)}$-eigenvectors $\vec{\bf u}(s)$ in the local spin basis through the theory of spin-$\frac{N-1}{2}$ XXZ chains. Indeed as in \cite{NiD, ND08, R06F, R09}, the loop-algebra symmetry of an Onsager sector in superintegrable $\tau^{(2)}$-model can be defined through its equivalent spin-$\frac{N-1}{2}$ XXZ chain in Sections \ref{sec.Degt2} and \ref{sec.loopt2}. First, the Bethe state (\req(Betv)) or (\req(BetvF)) in algebraic-Bethe-ansatz can be realized as the basic $\tau^{(2)}$-vector with the highest or lowest weight among basic $\tau^{(2)}$-eigenvectors in a sector. Then using the algebraic-Bethe-ansatz techniques of XXZ chains, we are able to construct the Fabricius-McCoy currents of all sectors of a superintegrable  $\tau^{(2)}$-model compatible with duality and inversion. As the Fabricius-McCoy current is expressed by local operators proportional to ${\bf E}^\mp (\xi)$ in (\req(vvE)), together with the basic Bethe-$\tau^{(2)}$-state, one then obtains the local-vector form of basic $\tau^{(2)}$-eigenvectors $\vec{\bf u}(s)$ in an Onsager sector by using the relations in (\req(uspin)).

\section*{Acknowledgements}
The author is pleased to thank Professor B. M. McCoy for the invitation and hospitality in Simons Center Workshop (January 2010) "Correlation Functions for Integrable Models" at Stony Brook University, where part of the results in this work was reported.


\begin{thebibliography}{99}
\bibitem{AMP} G. Albertini, B. M. McCoy, and 
J. H. H. Perk, Eigenvalue spectrum of the
superintegrable chiral Potts model, In {\it Integrable system in quantum field theory and statistical mechanics,}  Adv. Stud. Pure Math., 19, Kinokuniya Academic, Academic Press, Boston, MA (1989) 1--55.
%
\bibitem{AMPT} H. Au-Yang, B. M. McCoy,  
J. H. H. Perk and S. Tang, Solvable models in statistical mechanics and Riemann surfaces of genus greater than one, {\it Algebraic Analysis}, Vol. 1 , eds. M. Kashiwara and T. Kawai, Academic Press, San Diego (1988), 29--40.
%
\bibitem{AuP} H. Au-Yang and J. H. H. Perk, Onsager's star-triangle equation: Master key to integrability, In {\it Integrable system in quantum field theory and statistical mechanics,}  Adv. Stud. Pure Math., 19, Kinokuniya Academic, Academic Press, Boston, MA (1989) 57--94.
%
\bibitem{AuP7} H. Au-Yang and J.H.H. Perk, Eigenvectors the superintegrable model I: $\goth{sl}_2$ generators, J. Phys. A: Math. Theor. 41 (2008) 275201; arXiv: 0710.5257; Eigenvectors in the superintegrable model II: ground state sector,  J. Phys. A: Math. Theor. 42 (2009) 375208; arXiv: 0803.3029.
%
\bibitem{AuP9} H. Au-Yang and J.H.H. Perk, Quantum loop subalgebra and eigenvectors of the superintegrable chiral Potts transfer matrices, J. Phys. A: Math. Theor. 44 (2011) 025205; arXiv: 0907.0362.
%
\bibitem{Bax} R. J. Baxter, Exactly solved models in statistical mechanics, Academic Press (1982).
%
\bibitem{B88} R. J. Baxter, Free energy of the solvable chiral Potts model, J. Stat. Phys. 52 (1988) 639--667.
%
\bibitem{B89} R. J. Baxter, Superintegrable chiral Potts model: Thermodynamic properties, an "Inverse" model, and a simple associated Hamiltonian, J. Stat. Phys. 57 (1989) 1--39.
%
\bibitem{B90} R. J. Baxter, Chiral Potts model: eigenvalues of the transfer matrix, Phys. Lett. A 146 (1990) 110--114.
%
\bibitem{B91} R. J. Baxter, Calculation of the eigenvalues of the transfer matrix of the chiral Potts model, {\it Proc. Fourth Asia-Pacific Physics Conference} (Seoul, Korea, 1990) Vol 1, World-Scientific, Singapore (1991) 42--58.
%
\bibitem{B93} R. J. Baxter, Chiral Potts model with skewed boundary conditions, J. Stat. Phys. 73 (1993) 461--495.
%
\bibitem{B94} R. J. Baxter, Interfacial tension of the chiral Potts model, J. Phys. A: Math. Gen. 27 (1994) 1837--1849.
%
\bibitem{B05a} R. J. Baxter, The order parameter of the chiral Potts model, J. Stat. Phys. 120 (2005) 1-36; cond-mat/0501226.
%
\bibitem{B05b} R. J. Baxter, Derivation of the order parameter of the chiral Potts model, Phys. Rev. Lett. 94 (2005) 130602; cond-mat/0501227.
%
\bibitem{B08} R. J. Baxter, A conjecture for the superintegrable chiral Potts model, J. Stat. Phys. 132 (2008) 983-1000, arXiv. 0803.4037.
%
\bibitem{B09a} R. J. Baxter, Some remarks on a generalization of the superintegrable chiral Potts model, arXiv:0906.3551.
%
\bibitem{B09b} R. J. Baxter, Spontaneous magnetization of the superintegrable chiral Potts model: calculation of the determinant $D_PQ$, arXiv:0912.4549. 
%
\bibitem{B10} R. J. Baxter, Proof of the determinantal form of the spontaneous magnetization of the superintegrable chiral Potts model, arXiv:1001.0281.
%
\bibitem{BBP} R. J. Baxter, V.V. Bazhanov and
J.H.H. Perk,  Functional relations for transfer
matrices of the chiral Potts model, Int. J. Mod.
Phys. B 4 (1990) 803--870.
%
\bibitem{BPA} R. J. Baxter, J. H. H. Perk and H. Au-Yang, New solutions of the  star-triangle relations for the chiral Potts model, Phys. Lett. A 128 (1988) 138--142.
%
\bibitem{BazS} V.V. Bazhanov and Yu.G. Stroganov, Chiral
Potts model as a descendant of the six-vertex model, J.
Stat. Phys. 59 (1990) 799--817.
%
\bibitem{DJMM} E. Date, M. Jimbo, K. Miki and T. Miwa, Cyclic representations of $U_q(sl(n+1, \CZ))$ at $q^N=1$, Publ. RIMS, Kyoto Univ. 27 (1991) 347--366.
%
\bibitem{DR} E. Date
and S. S.  Roan,  The structure of quotients of the 
Onsager algebra by closed ideals, J. Phys. A:
Math. Gen. 33 (2000) 3275--3296,  math.QA/9911018; The algebraic structure of the Onsager algebra, Czech. J. Phys., Vol. 50 No. 1 (2000) 37-44; cond-mat/0002418.  
%
\bibitem{Dav} B. Davies, Onsager's algebra and
superintegrability, J. Phys. A: Math. Gen. 23 (1990)
2245--2261; Onsager's algebra and the 
Dolan-Grady condition in the non-self case, J. Math. Phys.
32 (1991) 2945--2950.
%
\bibitem{DK} C. DeConcini and V. G. Kac, Representations of quantum groups at roots of unity, in {\it Operator Algebra, Unitary Representations, Enveloping Algebras, and Invariant Theory}, Paris (1989) {\it Progress in Mathematics} 92, Birkh\"{a}user, Boston, Massachusstts (1990) 471-- 506.
%
\bibitem{DFM} T. Deguchi, K. Fabricius and B. M. McCoy, The $sl_2$ loop algebra symmetry for the six-vertex model at roots of unity, J. Stat. Phys. 102 (2001) 701--736; cond-mat/9912141. 
%
\bibitem{FM01} K. Fabricius and B. M. McCoy, Evaluation parameters and Bethe roots for the six vertex model at roots of unity, {\it Progress in Mathematical Physics} Vol 23, eds. M. Kashiwara and T. Miwa,  Birkh\"{a}user Boston (2002), 119--144; cond-mat/0108057.
%
\bibitem{Fad} L. D. Faddeev, How algebraic Bethe
Ansatz works for integrable models, eds. A.
Connes, K. Gawedzki and J. Zinn-Justin, {\it
Quantum symmetries/ Symmetries quantiques},
Proceedings of the Les Houches summer school,
Session LXIV, Les Houches, France, August 1-
September 8, 1995, North-Holland (1998),  149--219.
%
\bibitem{GR} G. von Gehlen and V. Rittenberg,   
$Z_n$-symmetric quantum chains with infinite set of
conserved charges and $Z_n$ zero modes, Nucl. Phys. B 257
(1985) 351--370.
%
\bibitem{IG} N. Iorgov, V. Shadura, Tu. Tykhyy, S. Pakuliak and G. von Gehlen, Spin operator matrix elements in the superintegrable chiral Potts quantum chain, arXiv.0912.5027.
%
\bibitem{KBI} V. E. Korepin, N. M. Bogoliubov, and A. G. Izegin, Quantum inverse scattering method and correlation functions, Cambridge Univ. Press, Cambridge, 1993.
%
\bibitem{KiR} A. N. Kirillov and N. Yu. Reshetikhin, Exact solution of the integrable XXZ Heisenberg model with arbitrary spin: I. The ground state and the excitation spectrum, J. Phys. A: Math. Gen. 20 (1987) 1565 -- 1595.
%
\bibitem{KS} P. P. Kulish and E. K. Sklyanin, Quantum spectral transform method. Recent
developments, eds. J. Hietarinta and C. Montonen,
Lecture Notes in Physics 151 Springer (1982),
61--119.
%
\bibitem{MaS} V. B. Matveev and A. O. Simnov, Some comments on the solvable chiral Potts model, Lett. Math. Phys. 19 (1990) 179--185.
%
\bibitem{MPTS} B. M. McCoy,  
J. H. H. Perk, S. Tang and C. H. Sah, Commuting transfer matrices for the four-state self-dual chiral Potts model with a genus-three uniformizing Fermat curve, Phys. Lett. A 125 (1987) 9--14.
%
\bibitem{MR} B. M. McCoy and S. S. Roan, Excitation spectrum and phase structure of the chiral Potts model. Phys. Lett. A 150 (1990) 347--354.
%
\bibitem{NiD} A. Nishino and T. Deguchi, The $L(sl_2)$ symmetry of the Bazhanov-Stroganov model associated with the superintegrable chiral Potts model, Phys. Lett. A 356 (2006) 366--370
; cond-mat/0605551.
%
\bibitem{ND08} A. Nishino and T. Deguchi, An algebraic derivation of the eigenspaces associated with an Ising-like spectrum of superintegrable chiral Potts model, J. Stat. Phys. 133 (2008) 587-615; arXiv 0806.1268.
%
\bibitem{Per} J. H. H. Perk, Star-triangle equations, quantum Lax pairs, and higher genus curves, Proc. Symp. Pure Math. 49 I (1989) 341--354.
%
\bibitem{R91} S. S. Roan, Onsager's algebra, loop algebra and chiral Potts model, Preprint Max-Planck-Inst. f\"{u}r Math.,
Bonn, MPI 91-70, 1991.
%
\bibitem{R05o} S. S. Roan, The Onsager algebra symmetry of $\tau^{(j)}$-matrices in the superintegrable chiral Potts model, J. Stat. Mech. (2005) P09007; cond-mat/0505698.
%
\bibitem{R06Q} S. S. Roan, The Q-operator for root-of-unity symmetry in six vertex model, J. Phys. A: Math. Gen. 39 (2006) 12303-12325; cond-mat/0602375.
%
\bibitem{R06F} S. S. Roan, Fusion operators in the generalized $\tau^{(2)}$-model and root-of-unity symmetry of the XXZ spin chain of higher spin, J. Phys. A: Math. Theor. 40 (2007) 1481-1511; cond-mat/0607258.
%
\bibitem{R075} S. S. Roan, The transfer matrix of superintegrable chiral Potts model as the Q-operator of root-of-unity XXZ chain with cyclic representation of $U_q(sl_2)$, J. Stat. Mech. (2007) P09021; arXiv: 0705.2856.
%
\bibitem{R0710} S. S. Roan, On the equivalent theory of the generalized $\tau^{(2)}$-model and the chiral Potts model with two alternating vertical rapidities, arXiv: 0710.2764.
%
\bibitem{R0805} S. S. Roan, Bethe equation of $\tau^{(2)}$-model and eigenvalues of finite-size transfer matrix of chiral Potts model with alternating rapidities, J. Stat. Mech. (2008) P10001; arXiv:0805.1585.
%
\bibitem{R0806} S. S. Roan, On $\tau^{(2)}$-model in chiral Potts model and cyclic representation of quantum group $U_q(sl_2)$, J. Phys. A: Math. Theor. 42 (2009) 072003; arXiv:0806.0216.
%
\bibitem{R09} S. S. Roan, Duality and symmetry in chiral Potts model, J. Stat. Mech. (2009) P08012; arXiv:0905.1924.

%
\bibitem{Tar} V. O. Tarasov, Transfer matrix of the superintegrable  chiral Potts model, Bethe ansatz spectrum,  Phys. Lett. A 147 (1990) 487--490. 
\end{thebibliography}
\end{document}